\documentclass[aps,superscriptaddress,prd,
onecolumn,
floatfix, 
nofootinbib,
amsmath,amssymb,amsfonts,longbibliography]{revtex4-2}

\usepackage{mathrsfs,amsmath}
\usepackage{mathtools}
\usepackage{empheq}
\usepackage[pdftex]{graphicx}
\usepackage[dvipsnames]{xcolor}
\usepackage{float}
\usepackage{physics}
\usepackage{comment}
\usepackage[normalem]{ulem}
\usepackage{caption}
\usepackage{subcaption}
\usepackage[colorlinks,linkcolor=blue,anchorcolor=violet,citecolor=red]{hyperref}

\usepackage{bm}

\newcommand{\souteq}[1]{\hbox{}}

\newcommand{\schrodinger}{Schr\"odinger}

\newcommand{\X}{\bm{X}}

\newcommand{\x}{\bm{x}}

\begin{document}

\title{
Gravitational entanglement witness through Einstein ring image
%Gravitational lensing on superposed curved spacetime
}

\author{Youka Kaku}
\email{kaku.yuka.g4@s.mail.nagoya-u.ac.jp}
\author{Yasusada Nambu}
\email{nambu.yasusada.e5@f.mail.nagoya-u.ac.jp}
\affiliation{Department of Physics, Graduate School of Science, Nagoya University, Chikusa, Nagoya 464-8602, Japan}

%%%
\date{\today}

%%%%%%%%%%%%%%%%%%%%%%%%%%%%%%%%%%%%%%%%%%%%%%%%%%%%%%%%%%
\begin{abstract}
We investigate the interplay between quantum theory and gravity by exploring gravitational lensing and Einstein ring images in a weak gravitational field induced by a mass source in spatial quantum superposition.
We analyze a quantum massless scalar field propagating in two distinct models of gravity: the first quantized Newtonian gravity (QG) model, which generates quantum entanglement between the mass source and other systems, and the \schrodinger-Newton (SN) gravity model, which does not produce entanglement.
Visualizing the two-point correlation function of the scalar field, we find that the QG model produces a composition of multiple Einstein rings, reflecting the spatial superposition of the mass source. By contrast, the SN model yields a single deformed ring image, representing a classical spacetime configuration. 
Furthermore, we introduce a specific quantity named the which-path information indicator and visualize its image. The QG model again reveals multiple Einstein rings, while the image intensity in the SN model notably vanishes.
Our findings provide a visual approach to witness gravity-induced entanglement through distinct features in Einstein ring images.
This study advances our understanding of quantum effects in general relativistic contexts and establishes a foundation for future studies of other relativistic phenomena.
\end{abstract}

\maketitle

%%%%%%%%%%%%%%%%%%%%%%%%%%%%%%%%%%%%%%%%%%%%%%%%%%%

%%%%%----------------------------------------------
\section{Introduction}

Building a quantum gravity theory remains one of the foremost challenges in theoretical physics. One of the main challenges is the lack of experimental evidence to test the quantum aspects of gravity.
Colella \textit{et al.} conducted the pioneering experiment that bridged quantum physics and gravity~\cite{Colella1975}, which measured the phase shift in the quantum state of neutrons within the gravitational field of the Earth. This foundational work has gained significant attention for investigating quantum phenomena influenced by gravity and for inspiring several related experiments over the years~\cite{Asenbaum2017,Overstreet2022,Schlippert2014}. Despite examining the quantum characteristics of the probe system in a classical gravitational field, they offer only limited insight into the quantum nature of gravity itself.

To test the quantum nature of gravity, Feynmann proposed a thought experiment to examine how a gravitational field behaves when its source is placed in a quantum superposition~\cite{Feynmann}.
This idea inspired the experimental proposal by Bose, Marletto, Vederal, \textit{et al.}, which aimed to verify whether the gravitational interaction induced by spatially superposed massive objects can generate quantum entanglement between them~\cite{Bose2017,Marletto2017}, collectively referred to as the BMV papers. Their idea was based on a fundamental concept from quantum information theory: local operations and classical communication (LOCC) cannot create quantum entanglement between two systems. In their proposal, two massive objects obey quantum mechanics and interact solely through Newtonian gravity. If the initially separable state of the two masses evolves into an entangled state over time, this would indicate that their mutual gravitational interaction is no longer LOCC.

Let us consider a setup in which a quantum test particle with mass $m$ and position operator $\hat\x$ is influenced by a gravitational mass source with mass $M$ and position operator $\hat \X$ to illustrate the idea behind the BMV papers. The evolution of the test particle is governed by the Newtonian gravitational potential created by the mass source. 
Although a fully verified quantum-compatible theory of gravity remains unestablished, let us introduce some models of Newtonian gravity in the context of quantum mechanics as specific examples.
As a first model, we suppose that the gravitational interaction can be represented by the following first-quantized form of the classical Newtonian potential
\begin{align}\label{eq:intro_QG}
    m\Phi(\hat\x,\hat\X)=\frac{-GmM}{|\hat\x-\hat \X|}.
\end{align}
In this study, we refer to this first-quantized Newtonian gravity as the quantized gravity (QG) model. 
This interaction depends on the test particle position operator and the mass source position operator, suggesting that the gravitational potential inherits the quantum fluctuation of the mass source position. Moreover, the interaction couples the operators of the test particle and the mass source, resulting in the quantum entanglement between the two systems. 
As a counterpart to the QG model, let us also consider alternative gravity models, often referred to as semi-classical gravity theories, which predict the absence of such an entanglement~\cite{Bahrami2014,Anastopoulos2014,Ruffini1969,Kibble1978,Kibble1980,Diosi1989,Diosi2011,Penrose1996,Penrose2014,Kafri2014,Tilloy2016,Bassi2017,Carney2023,Oppenheim2023}. In these models, gravity remains fundamentally classical, although the gravitational sources themselves are treated quantum mechanically. A notable example is the \schrodinger-Newton (SN) gravity model, one of the simplest semi-classical models. In this framework, the gravitational interaction experienced by the test particle from the mass source object is given by the following:
\begin{align}\label{eq:intro_SN}
    m\Phi(\hat\x)=\left\langle \frac{-GmM}{|\hat\x-\hat \X|} \right\rangle_\text{mass source}.
\end{align}
Taking the expectation value over the mass source system in the SN gravity model eliminates the direct dependence on the mass source position operator. As such, the gravitational potential does not inherit the quantum fluctuation of the mass source position. In addition, the operators between the two systems are not directly coupled; thus, no entanglement can arise in this gravity model\cite{Aleksander2024}.
Although the SN model has some theoretical problems, such as enabling superluminal signals and inconsistencies with the Newtonian limit of the semi-classical Einstein equation~\cite{Bahrami2014,Anastopoulos2014}, it still serves as a simple counterpoint to the QG model, specifically in terms of whether gravitational interaction can induce entanglement. 
Therefore, the experimental framework proposed in the BMV papers provides a way to test whether gravitational interaction operates as LOCC and thus distinguishes between various models, including QG and SN.

The experimental proposal of the BMV papers is now considered feasible in the near future owing to recent advances in quantum technology, potentially providing the first robust evidence for the quantum nature of gravity. Indeed, techniques have been developed to maintain quantum coherence even in mesoscopic-scale objects~\cite{Panda2024,Bild2023,Fein2019}, and precise measurements of gravitational interactions in microscopic systems have also been achieved~\cite{Westphal2021,Lee2020}.
Various ideas have also been proposed to improve the feasibility of detecting gravity-induced entanglement by optimizing the experimental setup~\cite{Balushi2018,Miao2020,Matsumura2020,Miki2024,Krisnanda2020,Kaku2023,Fujita2023}.
Further advances in experimental techniques and ingeniously designed setups are also expected to enable table-top tests of gravitational behavior in a quantum system with masses approaching the Planck scale.

Notably, although the BMV proposal presents a significant guideline for experimental proposals of the quantum aspects of Newtonian gravity, it does not provide sufficient evidence of the quantization of the gravitational field or the existence of gravitons. This is because the experiment is designed to test Newtonian gravity, which does not involve the dynamical degrees of freedom of the gravitational field. 
Some interesting studies have explored the relationship between gravity-induced entanglement and the existence of gravitons~\cite{Carney2022,Martin2023,Belenchia2018,Danielson2022,Hidaka2022,Sugiyama2023,Nandi2024}; however, the primary goal of the BMV papers is to determine whether non-dynamical Newtonian gravity can be classified as LOCC or not.

Building on the thought experiment by Feynman and the BMV papers, recent efforts have focused on relativistic extensions to explore the quantum nature of gravity beyond the Newtonian regime~\cite{Zych2018,Chen2023,Giacomini2019,Kaku2022,Foo2023relativity,Foo2021,Foo2022,Foo2023minkowski,Kabel2022}. 
Specifically, these studies investigate the behavior of non-dynamical curved spacetime induced by a gravitational mass source in quantum superposition, aiming to understand the quantum properties of gravity in the relativistic regime.
Although probing quantum phenomena of relativistic gravity is more challenging than in the Newtonian case, this research direction is motivated by the theoretical interest in uncovering quantum features unique to gravity and bridging the idea of the BMV proposals toward the ultimate goal of constructing the quantum gravity theory.
In addition, an experimental motivation exists: in the study of relativistic regimes, we expect to see differences between the quantum effects of gravity and those of other quantum interactions, such as electromagnetic interaction. Such distinctions could help rule out potential loopholes in gravity-induced entanglement experiments, thereby ensuring that observed quantum signals are indeed due to gravity.
Various arguments regarding the quantum nature of relativistic gravity have been established based on these motivations. These include studies on quantum mechanical phenomena~\cite{Kaku2022,Foo2023relativity} and quantum field theoretical phenomena~\cite{Foo2021,Foo2022,Foo2023minkowski,Kabel2022} in non-dynamical curved spacetimes induced by a superposed mass source.
As an extension into further relativistic regimes, some studies have investigated the quantum phenomena of the dynamical gravitational field in the linear regime~\cite{Kanno2021,Mehdi2023,Biswas2023}, and the non-linear regime, corresponding to quantum gravity theory.

This study explores the quantum effect of general relativistic phenomena beyond the Newtonian regime, specifically examining the gravitational lensing of quantum fields in non-dynamical curved spacetimes induced by the superposed mass source treated within the framework of quantum mechanics.
Gravitational lensing, a well-known phenomenon in general relativity, refers to the light scattering caused by the gravitational field~\cite{Carroll2019,Schneider1999}. 
When light emitted from a light source passes near a massive object, it is attracted by the gravitational force of the object, causing the light to scatter behind the mass following the geodesics along the curvature of spacetime. An observer situated on the opposite side of the light source across the massive object will perceive an image of the light altered by this gravitational lensing. Specifically, when the light source, the massive object, and the observer are aligned in a line, the observer sees a ring-shaped image of the light, known as the Einstein ring, which has been confirmed in several observations~\cite{Hewitt1988,EHT2019,EHT2022}.
In addition, the scattering angle in gravitational lensing differs by a factor of two compared to the predictions from Newtonian gravity, highlighting its unique nature as a relativistic effect.
Gravitational lensing has also been extensively studied in the context of wave optics and classical massless field theory~\cite{Nakamura1999,Schneider1999,Nambu2013,Kanai2013,Nambu2019}. Moreover, several studies have explored the gravitational lensing of quantum fields, particularly in relation to Hawking radiation scattering in classical curved spacetime~\cite{Nambu2022,Caribe2023}. A recent study~\cite{Lee2024} has further examined the gravitational lensing of quantum fields in the context of specific quantum gravity theories proposed by Verlinde and Zurek~\cite{Verlinde2021}. Furthermore, Biswas \textit{ et al.}~\cite{Biswas2023} studied tree-level photon-matter scattering via graviton exchange. Although our focus is not on gravitons, our study adopts a similar approach to that of Biswas \textit{et al.} in exploring the gravitational coupling between light and matter.

This paper investigates the Einstein ring image in a curved spacetime induced by spatially superposed mass sources. For practical calculations, we assume the background spacetime is a weak gravitational field, with the Newtonian potential incorporated into the metric being described by either the QG model in Eq.~\eqref{eq:intro_QG} or the SN gravity model in Eq.~\eqref{eq:intro_SN}. We calculate the evolution of the quantum massless scalar field as a toy model to study light propagation in this background spacetime. In addition, we assume that the observer interacts with the system via the Unruh-DeWitt (UDW) detectors~\cite{Unruh1976} that are coupled to the scalar field. 
We introduce two kinds of observables accessible through the UDW detectors: the two-point correlation function of the scalar fields, which is commonly examined in the context of quantum field theory and wave optical formalism of gravitational lensing, and a newly defined quantity named the which-path information indicator, useful for witnessing gravity-induced entanglement. 
Then, we construct the Einstein ring image from these two observables. 
When visualizing the two-point correlation function, we observe a composition of multiple Einstein rings in the QG model, reflecting the spatial quantum superposition of the mass source. By contrast, the SN model produces a single deformed ring image, indicating a classical spacetime configuration.
In addition, visualizing the which-path information indicator shows multiple Einstein rings in the QG model, while in the SN model, the image intensity notably disappears.
Building on the foundational work of Feynmann and the BMV papers, this research serves as a relativistic extension, emphasizing the quantum nature of gravity beyond the Newton regime. From a perspective of gravitational lensing research, this study will clarify the quantum effects observable in the Einstein ring image.
Although it presents significant challenges to verify these effects through actual observation, we believe this theoretical exploration is valuable, as it provides a practical formulation of the quantum field theory in non-dynamical curved spacetime induced by a superposed gravitational mass source and serves as a concrete example of how the quantum effect of gravity arises in general relativistic phenomena. 

The remainder of this paper is organized as follows:
In Section~\ref{sec:Time_evolution}, we detail our setup and calculate the time evolution of the total system, which includes the mass source of the background spacetime, the scalar field and the UDW detectors. In Section~\ref{sec:Gravity_induced_entanglement}, we evaluate the quantum entanglement between the mass source and other systems resulting from the QG interaction. Section~\ref{sec:observer_accessible_quantity} provides the reduced density matrix of the UDW detectors and introduces two specific quantities measurable by these detectors: the two-point correlation function of the scalar field and the which-path information indicator. In Section~\ref{sec:imaging}, we present the results of the Einstein ring image for both the QG and SN models. Section~\ref{sec:discussion} addresses advanced topics, including the experimental feasibility of our proposal, the discussions of an alternative semi-classical gravity model that induces gravitational state collapse, and connections to other research, particularly quantum reference frames~\cite{Zych2018,Chen2023,Giacomini2019,Foo2020,Foo2021}.
Finally, we summarize our findings in Section~\ref{sec:conclusion}.

%%%%%----------------------------------------------
\section{Scalar field and the Unruh-DeWitt detector in a curved spacetime induced by a superposed mass source}
\label{sec:Time_evolution}

\begin{figure}[htbp]
    \centering
    \includegraphics[width=0.65\linewidth]{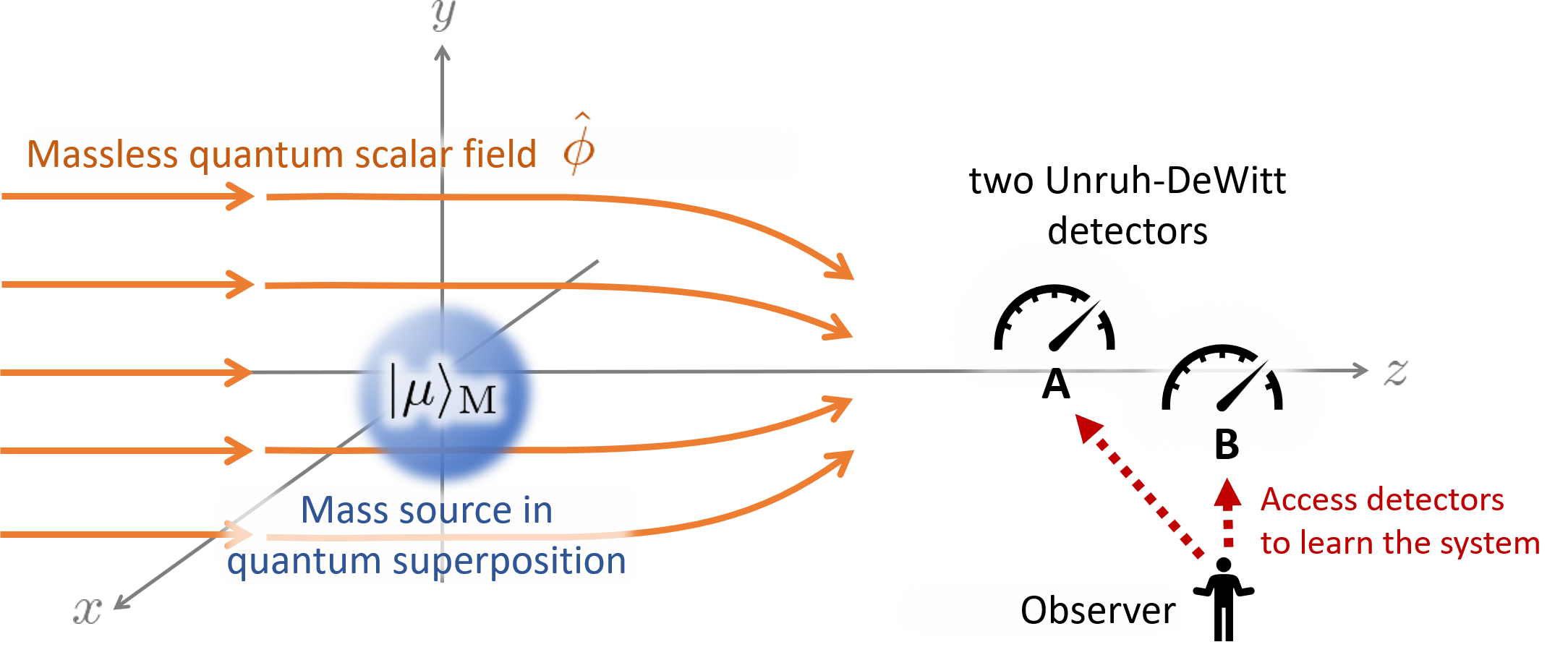}
    \caption{Setup of our proposal: The background spacetime is a weak gravitational field generated by a mass source in spatial quantum superposition. 
    A massless quantum scalar field propagates in this spacetime, and two Unruh-DeWitt detectors are coupled to the scalar field. The observer in the setup can only access the measurements obtained from these detectors.
    }
    \label{fig:setup}
\end{figure}

This section describes our setup and derives a time-evolved state of the total system. The schematic picture of our setup is shown in Fig.~\ref{fig:setup}.
We aim to investigate the gravitational lensing of light on a curved spacetime induced by a mass source in spatially quantum superposition. We do so by first considering the background spacetime induced by the first quantized mass source, described by the quantum state $|\mu\rangle_\text{M}$. We also suppose that the massless quantum scalar field $\hat \phi$ is a test field corresponding to the toy model of light propagating in the background spacetime. Furthermore, we clarify the measurement process of the scalar field by introducing two Unruh-DeWitt (UDW) detectors that couple to the scalar field and are placed sufficiently far from the mass source to ensure decoupling from the mass source system. In addition, we assume that an observer only has access to the measurement outcomes of these detectors.
Therefore, our setup comprises four subsystems: the mass source system, the scalar field system, and two UDW detector systems named A and B. 
We first describe the background metric, the Hamiltonian, and the initial state, and finally, we show the time-evolved state of the total system.

The background metric is supposed to be a weak gravitational field of a quantum point mass source with its mass $M$. If we denote the spacetime coordinate with $(t,x,y,z)$ and the mass source position operator with $\hat\X$, the background metric is given as follows:
\begin{align}\label{eq:metric}
    ds^2=g_{\mu\nu}dx^\mu dx^\nu
    =-\left(1+\frac{2\Phi^{(\text{QG})/(\text{SN})}}{c^2}\right)c^2dt^2+\left(1-\frac{2\Phi^{(\text{QG})/(\text{SN})}}{c^2}\right)d\x^2+\mathcal{O}(1/c^4)
    .
\end{align}
%%%
Here, $\Phi^{(\text{QG})},~\Phi^{(\text{SN})}$ are the Newtonian potentials for the quantized gravity model (QG) and the \schrodinger-Newton gravity model (SN) respectively:
\begin{align}
    &\Phi^{(\text{QG})}(\x,\hat\X)=\frac{-GM}{|\x-\hat \X|}, \label{eq:QG}\\
    &\Phi^{(\text{SN})}(\x)=\langle\mu| \frac{-GM}{|\x-\hat \X|} |\mu\rangle_\text{M} 
    =\int d^3 X |\mu(\X)|^2 \frac{-GM}{|\x-\X|} ,
    \label{eq:SN}
\end{align}
where $\mu(\X)=\langle \X|\mu\rangle_\text{M}$ is the wave function of the mass source system.
Note that, unlike in Eqs.~\eqref{eq:intro_QG} and \eqref{eq:intro_SN}, the spatial coordinate describing the spacetime, $\x$, here is not in the operator form. This is because our setup considers a quantum scalar field $\hat\phi(t,\x)$ as the probe system, adopting a second quantization, while the original BMV papers~\cite{Bose2017,Marletto2017} employ a first-quantized particle as the probe system, represented as $\hat\x$. Although our treatment of the probe differs slightly from the BMV approach, we will later confirm that this setup also produces gravity-induced entanglement between the scalar field probe and the mass source system.
Next, let us provide a comparison of the physical interpretations of these two gravity models.
The QG potential contains the position operator of the mass source system. This indicates that the spacetime inherits the quantum fluctuation of the mass source position, and that the spacetime itself is also quantum fluctuating. 
In the QG case, the mass source state $|\mu\rangle_\text{M}$ is not yet present at this stage, but will appear later when we compute the time-evolved state and various expectation values.
However, the SN potential is given by the expectation value of the QG potential and does not directly depend on the mass source system operator, resulting in the metric being a c-number, unlike the QG metric.
In other words, the SN potential can be viewed as a single classical entity represented by the ensemble-averaged form of the QG potential over various mass source positions $\X$, as indicated in the second equality of Eq.~\eqref{eq:SN}.
Later, we will show the calculation for the QG model in section~\ref{sec:time_evolution_QG} and the SN model in section~\ref{sec:time_evolution_SN}, respectively.

On the background spacetime given in Eq.~\eqref{eq:metric}, we consider the dynamics of the scalar field and two UDW detector systems.
The UDW detector is an idealized model of an atomic detector, described by a two-level qubit that changes its energy level owing to the interaction with the quantum field \cite{Unruh1976}. This model helps to provide an operational way of measuring field quanta and is often analyzed in the context of the entanglement harvesting of the quantum field.
The two UDW detectors in our setup are each assumed to couple independently to the massless quantum scalar field. The measurement process with UDW detectors clarifies the observable quantities in our setup, outlining specific steps for constructing the Einstein ring image from the observer-accessible quantities.

Now, the Hamiltonian of the detectors and the scalar field is given as follows:
\begin{align}
    \hat H_{\text{tot}}
    =\hat H_0+\hat H_{\text{int}},\quad
    \hat H_0
    =\hat H_\text{A}+\hat H_\text{B}+\hat H_{\text{S}}
    .
\end{align}
$\hat H_0$ describes the free evolution part and $\hat H_\text{int}$ describes the interaction between two detectors and the scalar field. $\hat H_0$ contains the free evolution Hamiltonian of each detector $\hat H_\text{A}$, $\hat H_\text{B}$, and the scalar field $\hat H_\text{S}$, whose details are given below.
The free evolution Hamiltonian of detector systems A and B is given by
\begin{align}
    \hat H_\text{A}=\frac{\hbar \omega_\text{D}}{2}\hat \sigma_{z,\text{A}},\quad
    \hat H_\text{B}=\frac{\hbar \omega_\text{D}}{2}\hat \sigma_{z,\text{B}},
\end{align}
where $\hbar\omega_\text{D}$ is the internal energy gap of the UDW detector model and $\hat \sigma_{z,\text{A/B}}$ are the $z$-component of the Pauli operators for detectors A/B. 
These Hamiltonian indicates that detectors A/B are described by two states $|0\rangle_\text{A/B}$ or $|1\rangle_\text{A/B}$, whose energy eigenvalues are $+\hbar \omega_\text{D}/2$ and $-\hbar \omega_\text{D}/2$, respectively.

Let us also introduce the free evolution Hamiltonian of the scalar field $\hat\phi$.
In general, when the background metric is expressed in a form that separates time and space, such as $ds^2=g_{00} dt^2 + g_{ij} dx^i dx^j$, the free scalar field Hamiltonian is given using the energy-momentum tensor $\hat T_{\mu\nu}$ derived from the scalar field action $S_\text{S}$~\cite{Kibble1980,Mashkevich2007,Smith2019,Birrell1984} as follows:%``Introduction to Quantum Fields in Curved Spacetime and the Hawking Effect" by Jacobson
\begin{gather}
    \hat H_{\text{S}}
    = \int d^3 x \sqrt{g^{(3)}}\, \hat{T}^0_{~0},\\
    %%%
    \hat{T}^{\mu\nu}
    =\frac{2}{\sqrt{-g}}\frac{\delta S_\text{S}}{\delta g^{\mu\nu}},
    \qquad
    S_\text{S}
    = \int d^4 x \sqrt{-g} \,
    \frac{1}{2}g^{\mu\nu}\partial_\mu \hat\phi \, \partial_\nu \hat\phi,
\end{gather}
where $g^{(3)}$ is the determinant of an induced metric of three-dimensional space.
From the above formula, the specific form of the scalar field Hamiltonian is rewritten as follows:
\begin{align}\label{eq:Hamiltonian_scalar}
    \hat H_{\text{S}}
    =\int d^3x \sqrt{g^{(3)}}\,\left[
    -\frac{1}{2g^{(3)}g^{00}}\hat\pi^2
    +\frac{1}{2}g^{ij} \nabla_i \hat\phi \, \nabla_j \hat\phi
    \right],
\end{align}
where $\hat\pi=-\sqrt{g^{(3)}}g^{00}\dot{\hat\phi}$ denotes the conjugate momentum of $\hat\phi$.
We can see that the scalar field operator couples to the background spacetime. This indicates that the scalar field evolution differs depending on the gravity model QG or SN, as we will see later.

Finally, let us introduce the interaction Hamiltonian between two detectors and the scalar field systems
\begin{align}\label{eq:Hamiltonian_int}
    \hat H_{\text{int}}
    =\lambda\, \chi(t) \sum_{j=\text{A},\text{B}}
    \hat\sigma_{x,j}\,\hat\phi(\x_j).
\end{align}
Here, $\lambda$ is the coupling constant of the scalar field and detector interaction, $\chi(t)$ is a real function describing the switching of the UDW detector, and $\hat\sigma_{x,\text{A}/\text{B}}$ are the $x$-component of the Pauli matrix for detectors A/B. 

We solve the time evolution of the total system in the interaction picture using the Hamiltonian given above. We show the formalism for the QG case in the next section and the SN case in Section \ref{sec:time_evolution_SN}.

%%%%%%
\subsection{Time evolution for the quantized gravity model}
\label{sec:time_evolution_QG}

In this section, let us solve the time evolution for the QG model, whose potential $\Phi^{(\text{QG})}(\x,\hat \X)$ is given in Eq.~\eqref{eq:QG}. Recalling that the QG spacetime depends on the mass source position operator $\hat\X$, we can see that the various physical quantities on this spacetime also depend on $\hat\X$. To emphasize this, we specify the mass source position dependence of each quantity in the QG case. For example, the scalar field Hamiltonian given in Eq.~\eqref{eq:Hamiltonian_scalar} is revisited as follows
\begin{align}\label{eq:Hamiltonian_scalar_QG}
    \hat H_{\text{S}}(\hat \X)
    =\int d^3x \sqrt{g^{(3)}}\,\left[
    -\frac{1}{2}g^{00}(\hat \X)(\partial_0 \hat\phi)^2
    +\frac{1}{2}g^{ij}(\hat \X) \nabla_i \hat\phi \, \nabla_j \hat\phi
    \right].
\end{align}
Note that the scalar field Hamiltonian depends on the mass source position operator $\hat \X$ inside the background spacetime, which induces the quantum entanglement between the mass source and the scalar field systems. 
This is nothing but gravity-induced entanglement in a framework of quantum field theory in curved spacetime, as advocated in the BMV papers~\cite{Bose2017,Marletto2017}.

We move on to the interaction picture by considering $\hat H_0(\hat\X)=\hat H_\text{A}+\hat H_\text{B}+\hat H_{\text{S}}(\hat \X)$ as the solvable part and $\hat H_\text{int}$ as the interaction part of the total Hamiltonian.
Then, the time-evolved state of the total system is written as
\begin{align}
    |\Psi^{(\text{QG})}(t)\rangle = \hat U_I^{(\hat \X)}(t) |\Psi_\text{ini}\rangle,
\end{align}
where the time evolution operator in the interaction picture is given as
\begin{align}\label{eq:time_evolution_operator}
    \hat U_I^{(\hat \X)}(t)
    =\mathcal{T}\exp\left[-\frac{i}{\hbar}\int dt \hat H_I^{(\hat \X)}(t) \right],
\end{align}
and the time-dependent interaction Hamiltonian in the interaction picture is 
\begin{align}\label{eq:Hamiltonian_I}
    \hat H_I^{(\hat \X)}(t)
    = e^{i\hat H_0(\hat\X) t/\hbar}\,\hat H_{\text{int}}\,e^{-i\hat H_0(\hat\X) t/\hbar}
    =\lambda\, \chi(t) 
    \sum_{j=\text{A},\text{B}}
    \hat\sigma_{x,j}(t)\,\hat\phi^{(\hat \X)}(t,\x_j).
\end{align}
The time-dependent operators of the detectors $\hat\sigma_{x,\text{A}/\text{B}}(t)$ are driven straightforwardly as follows:
\begin{align}\label{eq:Heisenberg_detector_operator}
    \hat\sigma_{x,j}(t)
    = \exp\left[\frac{i}{\hbar}\,\hat H_j\,t\right]\hat \sigma_{x,j}\exp\left[-\frac{i}{\hbar}\,\hat H_j\,t\right]
    =e^{-i\omega_\text{D} t}|0\rangle_j \,{}_j\langle 1|
    +e^{i\omega_\text{D} t}|1\rangle_j \,{}_j\langle 0|.
\end{align}

Let us also derive the expression of the scalar field operator in the interaction picture $\hat\phi^{(\hat \X)}(t,\x)$.
Although written earnestly as $\hat\phi^{(\hat \X)}(t,\x_j)=e^{i\hat H_0(\hat\X) t/\hbar}\,\hat\phi(\x_j)\,e^{-i\hat H_0(\hat\X) t/\hbar}$, it is better to solve the Klein-Gordon equation of the free scalar field directly.
We do so by first performing an eigendecomposition of the scalar field operator $\hat \phi^{(\hat \X)}(x)$ for the mass source position operator $\hat \X$ as follows
\begin{align}\label{eq:scalar_decomposition1}
    \hat \phi^{(\hat \X)}(x)
    =\int d^3X |\X\rangle\langle \X| \hat \phi^{(\X)}(x)
    .
\end{align}
The decomposed scalar field operator $\hat \phi^{(\X)}(x)$ depends not on the operator $\hat \X$, but on the c-number position $\X$. We can obtain its solution by solving the Klein-Gordon equation on a classical curved spacetime $g_{\mu\nu}(\X)$ with the mass source at a c-number position $\X$, following the familiar strategy of quantum field theory in curved spacetime:
\begin{align}\label{eq:KG_eq}
    \Box_{g_{\mu\nu}(\X)} \, \hat \phi^{(\X)}(x)=0
    .
\end{align}
Here, we denote the classical background spacetime in the subscript of the d'Alembert operator.
We first deal with the time evolution in the classical curved spacetime as in Eq.~\eqref{eq:KG_eq} and then sum up its solution to obtain the time evolution in the QG spacetime as in Eq.~\eqref{eq:scalar_decomposition1}.
Now, we perform a mode function decomposition to obtain the solution of Eq.~\eqref{eq:KG_eq} as follows:
\begin{align}\label{eq:scalar_decomposition2}
    \hat \phi^{(\X)}(x)
    =\int_0^\infty d\omega \left(\hat a_{\omega} e^{-i\omega t} \varphi^{(\X)}(\omega,\x)+\hat a_{\omega}^\dagger e^{i\omega t} \varphi^{(\X)}(\omega,\x)^*\right)
    .
\end{align}
By substituting this form into Eq.~\eqref{eq:KG_eq}, we obtain the equation of motion for the mode function as follows:
\begin{align}\label{eq:Schrodinger_eq}
    \left[
    \left(\frac{\partial}{\partial x}\right)^2+\left(\frac{\partial}{\partial y}\right)^2+\left(\frac{\partial}{\partial z}\right)^2
    + \frac{\omega^2}{c^2}\left(1-\frac{4\Phi^{(\text{QG})}(\x,\X)}{c^2}\right) \right]
    \varphi^{(\X)}(\omega,\x)=0
    .
\end{align}
We also adopt the following boundary condition so that the mode function flows from far south into the vicinity of the origin, as shown in Fig.~\ref{fig:setup}
\begin{align}\label{eq:boundary_condition}
    \varphi^{(\X)}(\omega,\x)\sim e^{i\omega z/c}
    \quad
    \text{for}\quad
    z<0,~|\x|\to \infty
    .
\end{align}
Here, we implicitly assume that the mass source position $\X$ is localized around the origin, ensuring it does not affect the boundary condition imposed at infinity. Notably, this boundary condition indicates that an external source of the scalar field exsits at $z\to-\infty$, generating the incoming flux.
The solution of Eq.~\eqref{eq:Schrodinger_eq} with the boundary condition Eq.~\eqref{eq:boundary_condition} is given by
\begin{align}\label{eq:mode_function_solution}
    \varphi^{(\X)}(\omega,\x)
    =e^{\pi|\gamma|/2+i\omega (z-Z)/c}\,
    \Gamma\!\left[1-i \gamma\right]\,
    {}_1 F_1\!\left[i\gamma\,;1\,;\frac{i\omega}{c}\left\{|\x-\X|-(z-Z)\right\}\right]
    ,
\end{align}
where $\gamma:=2GM\omega/c^3$ is a dimensionless gravitational coupling constant and ${}_1 F_1\!\left[a\,;b\,;z\right]$ is the confluent hypergeometric function of the first kind.  This solution is the Coulomb wave function, which describes the charged particle attracted and scattered by the Coulomb potential in quantum mechanics~\cite{Drake2007,Landau2013}. The derivation and further details of the solution in Eq.~\eqref{eq:mode_function_solution} are given in Appendix.~\ref{apdx:Coulomb_wave_function}.
In Fig.~\ref{fig:waveoptics}, we showed the real part of the mode function for the mass source at the origin, $\mathrm{Re}\left[\varphi^{(\X=0)}(\omega,\x)\right]$. The horizontal axis shows a $z$ direction, while the vertical axis shows a $\sqrt{x^2+y^2}$ direction, both normalized by a length scale $c/\omega$. We set the gravitational coupling constant as $\gamma=10$. We can see that the plane wave comes from the left side, which is consistent with the imposed boundary condition Eq.~\eqref{eq:boundary_condition}, is scattered by the mass source at the origin and produces interference fringe as it propagates.
\begin{figure}[htbp]
    \centering
    \includegraphics[width=0.4\linewidth]{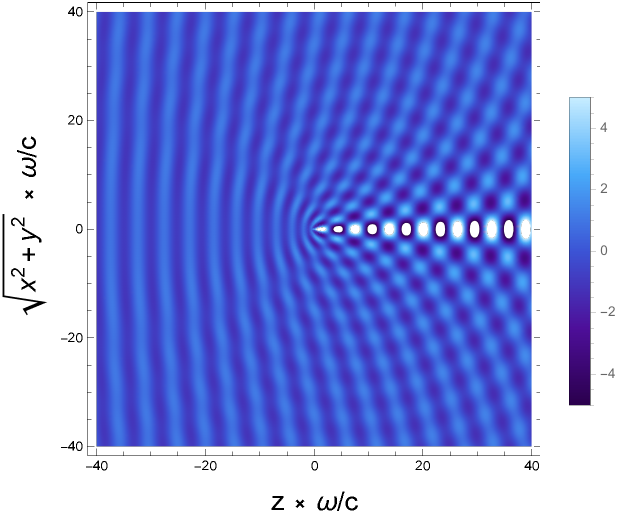}
    \caption{The real part of the mode function  for the mass source at the origin $\mathrm{Re}\left[\varphi^{(\X=0)}(\omega,\x)\right]$ in a 2-dimensional space. The horizontal and the vertical axis show the $z$ and $\sqrt{x^2+y^2}$ direction, rescaled by $\omega/c$, respectively. We set the gravitational coupling constant $\gamma = 10$.}
    \label{fig:waveoptics}
\end{figure}
Finally, the time-dependent scalar field operator is explicitly provided from Eqs.~\eqref{eq:scalar_decomposition1},~\eqref{eq:scalar_decomposition2} and \eqref{eq:mode_function_solution}.

We also mention the mode function property under translation transformation for later use. 
According to Eq.~\eqref{eq:mode_function_solution}, the mode function solution depends only on the relative distance between its position and the mass source position, $|\x-\X|$ and $z-Z$. Therefore, the solution stays the same even if we apply the translational transformation to the whole system. This results in the following relational formula
\begin{align}\label{eq:translation}
    \varphi^{(\X)}(\omega,\x)
    =\varphi^{(\X+\bm{\Delta})}(\omega,\x+\bm{\Delta}),
\end{align}
where $\bm{\Delta}$ is an arbitrary real vector of the translational transformation.
%The detailed explanation of this property is also given in Appendix~\ref{apdx:Coulomb_wave_function} and in Fig.~\ref{fig:translation}.

We obtain the explicit form of the time-dependent operators of the detectors and the scalar field systems appearing in Eq.~\eqref{eq:Hamiltonian_I}. Now, we suppose the following initially separable state  to calculate the time-evolved state:
\begin{align}\label{eq:initial_state}
    |\Psi_\text{ini}\rangle = |\mu\rangle_\text{M} \otimes |0^{(\text{QG})}\rangle_\text{S} \otimes |0\rangle_\text{A} \otimes |0\rangle_\text{B}.
\end{align}
$|\mu\rangle_\text{M}$ is the quantum state of the mass source, whose specific form is given later.
$|0^{(\text{QG})}\rangle_\text{S}$ represents the ground state of the scalar field
\footnote{
The ground state of the scalar field in our paper $|0^{(\text{QG})}\rangle_\text{S}$ is the lowest energy state satisfying the boundary condition given in Eq.~\eqref{eq:boundary_condition}; this state has preferred direction along with its boundary condition. If we assume a homogeneous boundary condition instead, the corresponding ground state no longer has a preferred direction and differs from our ground state $|0^{(\text{QG})}\rangle_\text{S}$. Furthermore, our ground state is considered to be an excited state rather than a vacuum state for the creation and annihilation operators of the homogeneous solution.
}
vanishing for the annihilation operator introduced in Eq.~\eqref{eq:scalar_decomposition2}. 
$|0\rangle_{\text{A}/\text{B}}$ are the ground states of the UDW detectors A/B.
By applying the time evolution operator to this initial state, the time evolved state is eventually given by
\begin{align}\label{eq:time_evolved_state}
    |\Psi^{(\text{QG})}(t)\rangle
    = \int d^3 X ~ \mu(\X)|X\rangle_\text{M}\,\otimes |\psi^{(\X)}(t)\rangle_\text{SAB},
\end{align}
where $|\psi^{(\X)}(t)\rangle_\text{SAB}$ is the time-evolved state of the scalar field and two detectors with the mass source at a classical position $\X$, approximately given by
\begin{align}
    &|\psi^{(\X)}(t)\rangle_\text{SAB}
    \simeq 
    \left[
    \left\{
    1
    -\frac{\lambda^2}{\hbar^2} \int dt dt'\,
    \theta(t-t')e^{i\omega_\text{D}(t-t')}\,\chi(t)\chi(t') 
    \sum_{j=\text{A,B}}
    \hat \phi^{(\X)}(t,\x_j)\hat \phi^{(\X)}(t',\x_j)
    \right\}
    |0\rangle_\text{A} |0\rangle_\text{B}
    \right.\notag\\
    &\hspace{20mm}\left.
    -\frac{i\lambda}{\hbar} \int dt \, e^{-i\omega_\text{D}t}\,\chi(t)\left(
    \hat \phi^{(\X)}(t,\x_\text{A})
    |1\rangle_\text{A} |0\rangle_\text{B}
    +\hat \phi^{(\X)}(t,\x_\text{B})
    |0\rangle_\text{A} |1\rangle_\text{B}
    \right)\right.\notag\\
    &\hspace{20mm}\left.
    -\frac{\lambda^2}{\hbar^2} \int dt dt'\,\theta(t-t')e^{i\omega_\text{D}(t+t')}\,\chi(t)\chi(t') 
    \left\{
    \hat \phi^{(\X)}(t,\x_\text{A}),\hat \phi^{(\X)}(t',\x_\text{B}))
    \right\}
    |1\rangle_\text{A} |1\rangle_\text{B}
    \right]|0^{(\text{QG})}\rangle_\text{S}
    +\mathcal{O}(\lambda^3).
\end{align}
We consider up to the second order of the scalar field-detector coupling constant $\lambda$. The derivation is given in Appendix~\ref{apdx:Time_evolved_state}. 
As indicated in Eq.~\eqref{eq:Hamiltonian_scalar_QG}, the mass source and scalar field systems evolve to the entangled state owing to gravitational interaction, although they were initially prepared as a separable state. The detectors and the mass source systems also couple indirectly through gravitational and scalar field-detector interaction and evolve into an entangled state.

\begin{figure}[htbp]
    \centering
    \begin{minipage}[h]{0.49\linewidth}
        \centering
        \includegraphics[scale=0.3]{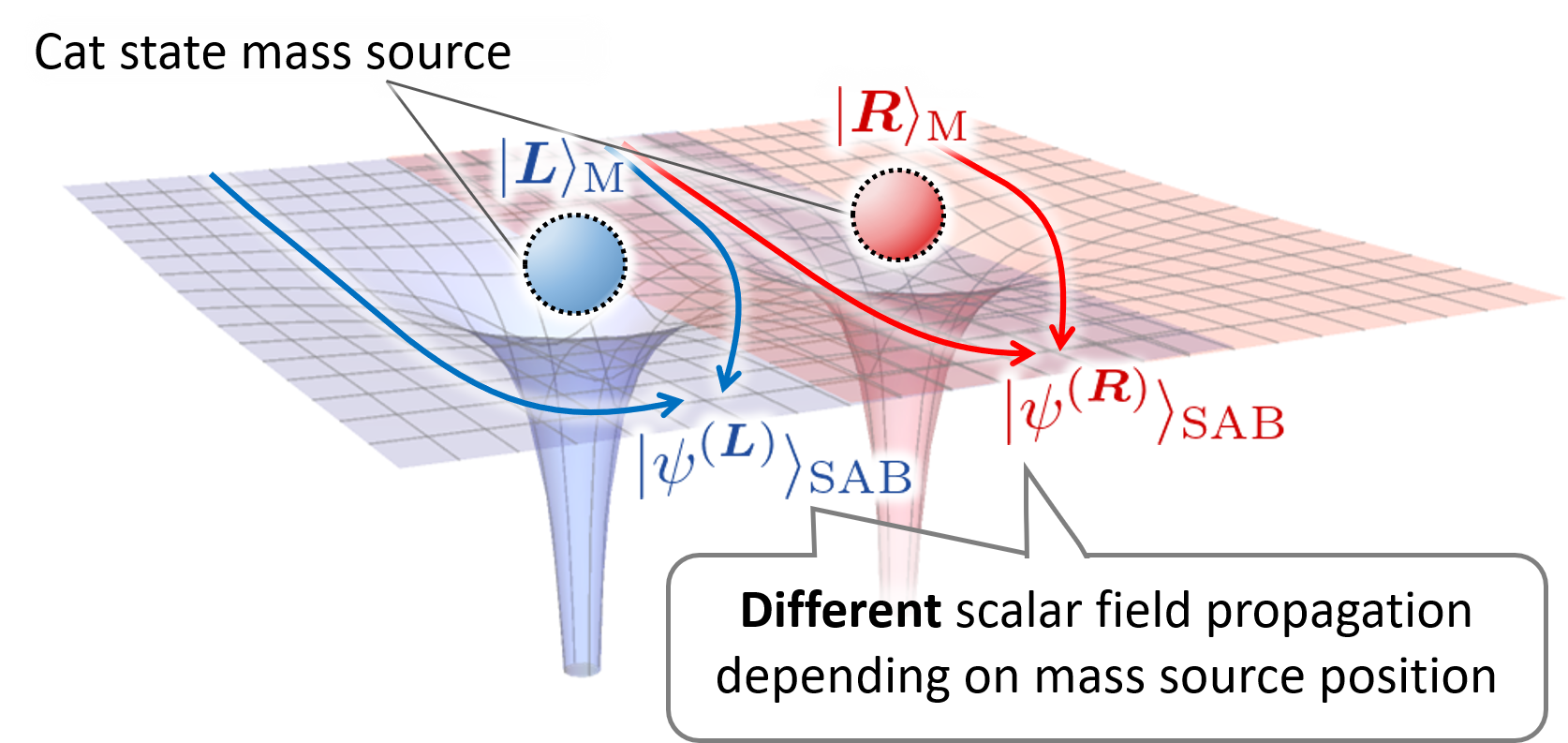}
        \subcaption{Time evolution in the QG spacetime.}
        \label{fig:QGevolution}
    \end{minipage}
    \begin{minipage}[h]{0.49\linewidth}
        \centering
        \includegraphics[scale=0.3]{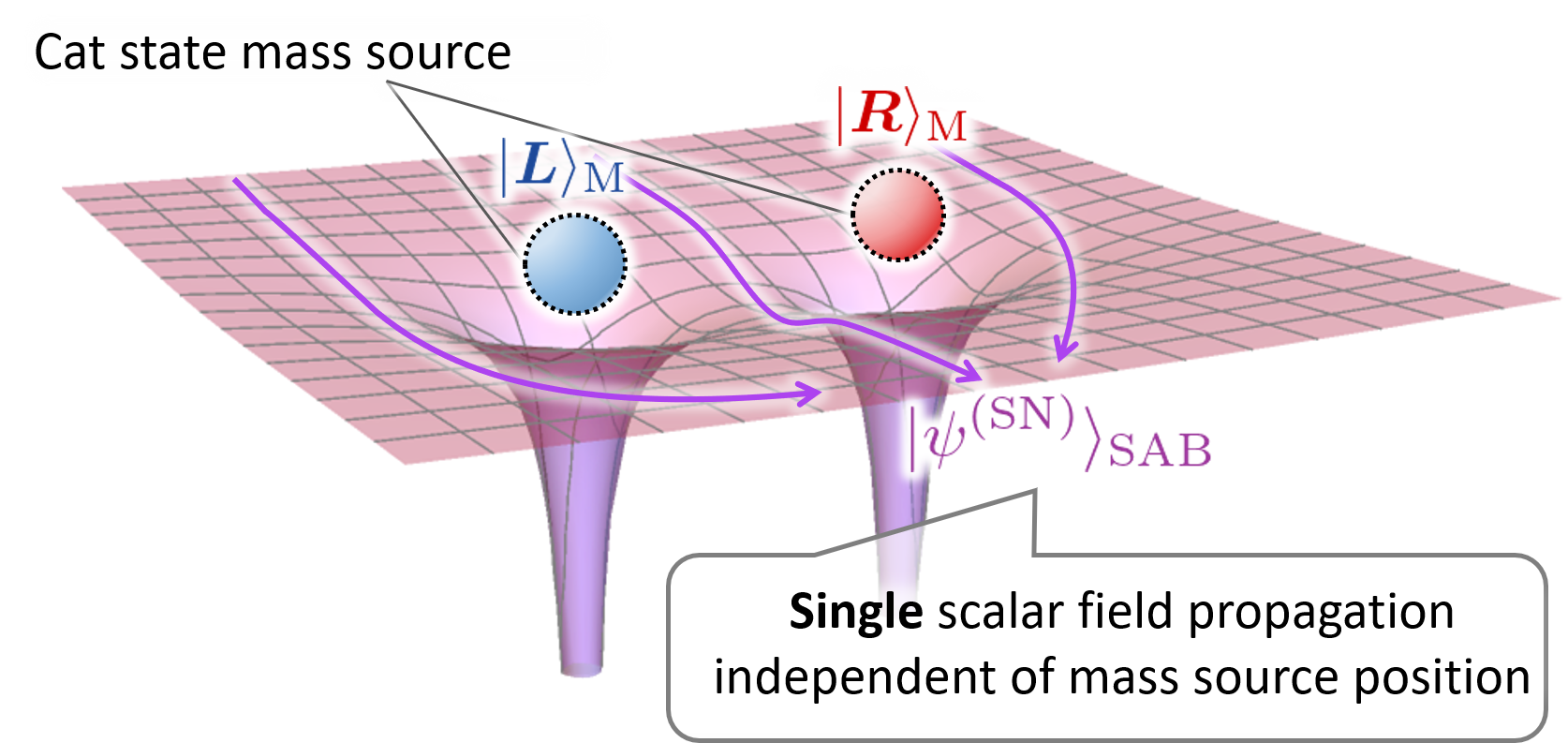}
        \subcaption{Time evolution in the SN spacetime.}
        \label{fig:SNevolution}
    \end{minipage}
    \caption{
    Schematic of time evolution in the curved spacetime induced by a superposed mass source. The mass source is in a \schrodinger~cat state with two spatially localized states as in Eq.~\eqref{eq:BMV_mass_source}. The left panel shows the QG spacetime, in which the scalar field and detectors propagate differently depending on the mass source position. By contrast, the right panel illustrates the SN spacetime, showing a single propagation of the scalar field and detectors, independent of the mass source position.
    }
\end{figure}

Following the scenario in the BMV papers ~\cite{Bose2017,Marletto2017}, let us consider the case when the mass source is in a \schrodinger~cat state at two spatially localized states:
\begin{align}\label{eq:BMV_mass_source}
    |\mu\rangle_\text{M}
    =\frac{1}{\sqrt{2}}\left(
    |\bm{L}\rangle_\text{M} + |\bm{R}\rangle_\text{M}
    \right)
    .
\end{align}
Here, $|\bm{L}\rangle_\text{M},~|\bm{R}\rangle_\text{M}$ are eigenstates of the mass source position operator $\hat\X$, which are orthogonal to each other ${}_\text{M}\langle \bm{L}|\bm{R}\rangle_\text{M}=0$. 
In this case, the time-evolved state reduces to the following form
\begin{align}\label{eq:time_evolved_state_BMV}
    |\Psi^{(\text{QG})}(t)\rangle
    =\frac{1}{\sqrt{2}}
    \left(
    |\bm{L}\rangle_\text{M} \, |\psi^{(\bm{L})}(t)\rangle_\text{SAB}
    + 
    |\bm{R}\rangle_\text{M} \, |\psi^{(\bm{R})}(t)\rangle_\text{SAB}
    \right).
\end{align}
This can be regarded as a relativistic extension of the gravitationally entangled state of two systems in the BMV papers~\cite{Bose2017,Marletto2017}
\footnote{
Our paper focuses on a static spacetime, where the mass source is continuously kept in a superposed state. In contrast, the BMV proposals consider a Stern-Gerlach experiment involving a dynamical mass source, which is initially localized and gradually splits into a superposition of two localized states. In such a scenario, the scalar field becomes excited due to the dynamical background spacetime. Although such excitation modifies the time-evolved state quantitatively, gravity-induced entanglement still occurs exclusively in the QG spacetime, regardless of the excitation.
}. 
In Fig.~\ref{fig:QGevolution}, we sketch the time evolution in the QG spacetime, where the mass source is in a \schrodinger-cat state with two spatially localized states, as in Eq.~\eqref{eq:BMV_mass_source}. 
The background spacetime of the QG model changes based on the mass source position, shown as blue and red curved slices representing the spacetime for the mass source states $|\bm{L}\rangle_\text{M}$ and $|\bm{R}\rangle_\text{M}$, respectively. The scalar field and detectors evolve differently depending on the mass source position and, consequently, the background spacetime; the system propagation for the mass source states $|\bm{L}\rangle_\text{M}$ and $|\bm{R}\rangle_\text{M}$ is represented by blue and red arrows in the figure, respectively.

%%%%%%
\subsection{Time evolution for the \schrodinger-Newton gravity model}
\label{sec:time_evolution_SN}

This section considers the time evolution of the total system for the SN model, whose potential $\Phi^{(\text{SN})}(\x)$ is given in Eq.~\eqref{eq:SN}. In contrast to the QG case, the SN spacetime does not directly depend on the mass source position operator. For instance, the scalar field Hamiltonian in Eq.~\eqref{eq:Hamiltonian_scalar} is revisited as follows:
\begin{align}\label{eq:Hamiltonian_scalar_SN}
    \hat H_{\text{S}}^{(\text{SN})}
    =\int d^3x \sqrt{g^{(3)}}\,\left[
    -\frac{1}{2}g^{(\text{SN})00}(\partial_0 \hat\phi)^2
    +\frac{1}{2}g^{(\text{SN})ij} \nabla_i \hat\phi \, \nabla_j \hat\phi
    \right]
    .
\end{align}
Here, we specify ``(SN)" as a subscript of each quantity emphasizing that we are adhering to the SN background spacetime. We observe that no coupling exists between the scalar field and the mass source position operator. Consequently, unlike in the QG case, no gravity-induced entanglement occurs between the scalar field and the mass source systems.

Now, we move on to the interaction picture to obtain the time-evolved state by considering $\hat H_0^{(\text{SN})}=\hat H_\text{A}+\hat H_\text{B}+\hat H_{\text{S}}^{(\text{SN})}$ as the solvable part and $\hat H_\text{int}$ as the interaction part of the total Hamiltonian. 
The calculation proceeds almost exactly as in the QG case, except for the dependence on the mass source position operator.
The time-evolved state of the total system in the interaction picture is given by
\begin{align}\label{eq:time_evolved}
    |\Psi^{(\text{SN})}(t)\rangle = \hat U_I^{(\text{SN})}(t) |\Psi_\text{ini}\rangle,
\end{align}
where the time evolution operator in the interaction picture is given as
\begin{align}
    &\hat U_I^{(\text{SN})}(t)
    =\mathcal{T}\exp\left[-\frac{i}{\hbar}\int dt \hat H_I^{(\text{SN})}(t) \right],\\
    %%%
    \label{eq:Hamiltonian_I_SN}
    &\hat H_I^{(\text{SN})}(t)
    = e^{i\hat H_0^{(\text{SN})} t/\hbar}\,\hat H_{\text{int}}\,e^{-i\hat H_0^{(\text{SN})} t/\hbar}
    =\lambda\, \chi(t) 
    \sum_{j=\text{A},\text{B}}
    \hat\sigma_{x,j}(t)\,\hat\phi^{(\text{SN})}(t,\x_j).
\end{align}
Since the time-dependent operators of the detectors $\hat\sigma_{x,\text{A}/\text{B}}(t)$ do not depend on the background spacetime, they are given in the same form as in Eq.~\eqref{eq:Heisenberg_detector_operator}.

Let us derive the expression of the scalar field operator in the interaction picture $\hat\phi^{(\text{SN})}(t,\x)$.
As in the QG case, we derive its explicit form by directly solving the Klein-Gordon equation of the free scalar field on the SN spacetime
\begin{align}\label{eq:KG_eq_SN}
    \Box_{g^{(\text{SN})}_{\mu\nu}} \, \hat \phi^{(\text{SN})}(x)=0
\end{align}
where we specify the SN background spacetime $g^{(\text{SN})}_{\mu\nu}$, given in Eq.~\eqref{eq:metric} and \eqref{eq:SN}, in the subscript of the d'Alembert operator.
We obtain the solution by first performing a mode function decomposition of the scalar field operator:
\begin{align}\label{eq:scalar_decomposition_SN}
    \hat \phi^{(\text{SN})}(x)
    =\int_0^\infty d\omega \left(\hat a_{\omega} e^{-i\omega t} \varphi^{(\text{SN})}(\omega,\x)+\hat a_{\omega}^\dagger e^{i\omega t} \varphi^{(\text{SN})}(\omega,\x)^*\right)
    .
\end{align}
By substituting this into Eq.~\eqref{eq:KG_eq_SN}, we obtain the equation of motion for the mode function as follows:
\begin{align}\label{eq:Schrodinger_eq_SN}
    \left[
    \left(\frac{\partial}{\partial x}\right)^2+\left(\frac{\partial}{\partial y}\right)^2+\left(\frac{\partial}{\partial z}\right)^2
    + \frac{\omega^2}{c^2}\left(1-\frac{4\Phi^{(\text{SN})}(\x)}{c^2}\right) \right]
    \varphi^{(\text{SN})}(\omega,\x)=0,
\end{align}
where $\Phi^{(\text{SN})}(\x)$ is given by Eq.~\eqref{eq:SN}.
We also adopt the boundary condition so that the mode function flows from the far south into the vicinity of the origin, which is the same as in the QG case
\begin{align}\label{eq:boundary_condition_SN}
    \varphi^{(\text{SN})}(\omega,\x)\sim e^{i\omega z/c}
    \quad
    \text{for}\quad
    z<0,~|\x|\to \infty.
\end{align}

Although we obtained the exact solution for the mode function in the QG case, as presented in Eq.~\eqref{eq:mode_function_solution}, deriving an exact solution for the SN case proves challenging owing to its Newtonian potential form in Eq.~\eqref{eq:SN}, which is generally no longer the central potential.
Consequently, we employ several approximations to address this. As discussed below, we propose different assumptions depending on the mass source configuration; however, broadly speaking, the physical assumption posed is that the scattering of the mode function occurs primarily at the slice $z=\langle \hat Z\rangle_\text{M}$. 

First, let us consider the case where the mass source is localized in the $x$ and $y$ directions, specifically at $(X_\text{M},Y_\text{M})$, and extended only along the $z$-direction
\begin{align}
    \left|\mu(\X)\right|^2 
    =\delta(X-X_\text{M})\delta(Y-Y_\text{M})
    \left|\mu_Z(Z)\right|^2.
\end{align}
In this scenario, we approximate the mass source system using the monopole approximation as 
\begin{align}
    \left|\mu(\X)\right|^2 
    \simeq
    \delta(X-X_\text{M})\delta(Y-Y_\text{M})\delta(Z-\langle \hat Z\rangle_\text{M})
    +\mathcal{O}\left(\langle \hat Z^2\rangle_\text{M}\right).
\end{align}
Thus, we approximately obtain the central force induced by the monopole mass source at $(X_\text{M},Y_\text{M},\langle \hat Z\rangle_\text{M})$. Hence, we can adapt the Coulomb wave function solution as an approximate solution, given by
\begin{align}\label{eq:mode_function_solution_SN_z}
    \varphi^{(\text{SN})}(\omega,\x)
    \simeq e^{\pi|\gamma|/2+i\omega (z-\langle \hat Z\rangle_\text{M})/c}\,
    \Gamma\!\left[1-i \gamma\right]\,
    {}_1 F_1\!\left[i\gamma\,;1\,;\frac{i\omega}{c}\left\{|\x-(X_\text{M},Y_\text{M},\langle \hat Z^2\rangle_\text{M})|-(z-\langle \hat Z\rangle_\text{M})\right\}\right].
\end{align}

Secondly, if the mass source also extends in the $x$ and $y$ directions, we use the diffraction integral formula~\cite{Nakamura1999} to obtain the solution, allowing us to capture scattering effects beyond the monopole approximation. 
Here, we only present the final expression of the mode function solution:
\begin{align}
    \varphi^{(\text{SN})}(\x)
    &\simeq \frac{\omega}{2\pi i c}
    \frac{e^{i\omega z}}{z-\langle Z\rangle_\text{M}}
    \int d^2 x'_{\perp}
    \exp\left[\frac{i\omega}{c} T[\x'_\perp,\x_\perp] \right],
    \label{eq:mode_function_solution_SN_pathintegral}\\
    T[\x'_\perp,\x_\perp]
    &\simeq \frac{|\x'_\perp-\x_\perp|^2}{2(z-\langle \hat Z\rangle_\text{M})}
    -\frac{2c\gamma}{\omega}\int d^3X \left|\mu(\X)\right|^2
    \log\left|\x'_\perp-\X_\perp\right|
    .
\end{align}
Here, we denote the two-dimensional spatial vectors in the $x,y$-plane with the subscript $\perp$: $\x=(x,y,z)=(\x_\perp,z)$, $\x'=(x',y',z')=(\x'_\perp,z')$ and $\X=(X,Y,Z)=(\X_\perp,Z)$. We mainly apply two assumptions to derive this form; First, the eikonal approximation, $\omega^2 \varphi/c^2 \gg \partial_z^2\varphi$, which indicates that the solution varies on a larger scale than the wavelength. In addition, we adopt the thin-lens approximation for the gravitational lensing, assuming that the phase shift due to gravitational scattering is dominant on the $z=\langle \hat Z\rangle_\text{M}$ slice.
The detailed derivation is given in Appendix~\ref{apdx:mode_function_SN}.
For the actual calculation, we further use the stationary phase method to evaluate the $\x'_\perp$ integral in Eq.~\eqref{eq:mode_function_solution_SN_pathintegral}, which is also explained in Appendix~\ref{apdx:mode_function_SN}.

By substituting these mode function solutions into Eq.~\eqref{eq:scalar_decomposition_SN}, we obtain approximately the explicit form of the scalar field solution, which enables us to express the time evolution operator explicitly.

Now, we introduce the initial state of the total system as:
\begin{align}\label{eq:initial_state_SN}
    |\Psi_\text{ini}\rangle = |\mu\rangle_\text{M} \otimes |0^{(\text{SN})}\rangle_\text{S} \otimes |0\rangle_\text{A} \otimes |0\rangle_\text{B}
    ,
\end{align}
where $|\mu\rangle_\text{M}$ and $|0\rangle_{\text{A}/\text{B}}$ are the initial state of the mass source and UDW detectors A/B, respectively, which is the same as in the QG case. Here, $|0^{(\text{SN})}\rangle_\text{S}$ is the ground state of the scalar field vanishing for the annihilation operator introduced in Eq.~\eqref{eq:scalar_decomposition_SN}. 
Notably, the ground states in the SN case differ from those in the QG case $|0^{(\text{QG})}\rangle_\text{S}$, since the background spacetimes are different according to each gravity model. 
Nevertheless, one naturally chooses the initial states as Eq.~\eqref{eq:initial_state} and \eqref{eq:initial_state_SN} for their respective cases, since the physical setup, such as the boundary condition and the Klein-Gordon equation, is consistent across both cases aside from the choice of the fundamental gravity model.

By applying the time evolution operator to the initial state, the time evolved state in the SN spacetime is given as 
\begin{align}\label{eq:time_evolved_state_SN}
    |\Psi^{(\text{SN})}(t)\rangle
    = |\mu\rangle_\text{M} \otimes |\psi^{(\text{SN})}(t)\rangle_\text{SAB},
\end{align}
where $|\psi^{(\text{SN})}(t)\rangle_\text{SAB}$ is the time-evolved state of the scalar field and the two detectors with the mass source in the SN spacetime, approximately given by
\begin{align}
    &|\psi^{(\text{SN})}(t)\rangle_\text{SAB}
    \simeq 
    \left[
    \left\{
    1
    -\frac{\lambda^2}{\hbar^2} \int dt dt'\,
    \theta(t-t')e^{i\omega_\text{D}(t-t')}\,\chi(t)\chi(t') 
    \sum_{j=\text{A,B}}
    \hat \phi^{(\text{SN})}(t,\x_j)\hat \phi^{(\text{SN})}(t',\x_j)
    \right\}
    |0\rangle_\text{A} |0\rangle_\text{B}
    \right.\notag\\
    &\hspace{10mm}\left.
    -\frac{i\lambda}{\hbar} \int dt \, e^{-i\omega_\text{D}t}\,\chi(t)\left(
    \hat \phi^{(\text{SN})}(t,\x_\text{A})
    |1\rangle_\text{A} |0\rangle_\text{B}
    +\hat \phi^{(\text{SN})}(t,\x_\text{B})
    |0\rangle_\text{A} |1\rangle_\text{B}
    \right)\right.\notag\\
    &\hspace{10mm}\left.
    -\frac{\lambda^2}{\hbar^2} \int dt dt'\,\theta(t-t')e^{i\omega_\text{D}(t+t')}\,\chi(t)\chi(t') 
    \left\{
    \hat \phi^{(\text{SN})}(t,\x_\text{A}),\hat \phi^{(\text{SN})}(t',\x_\text{B}))
    \right\}
    |1\rangle_\text{A} |1\rangle_\text{B}
    \right]|0^{(\text{SN})}\rangle_\text{S}
    +\mathcal{O}(\lambda^3) 
    .
\end{align}
We consider up to the second order of the scalar field-detector coupling constant $\lambda$, and the leading order of the SN correction. The derivation is almost the same as in the QG case, whose details are given in Appendix~\ref{apdx:Time_evolved_state}. 
As indicated in Eq.~\eqref{eq:Hamiltonian_scalar_SN}, the mass source and the scalar field systems remain separable in the SN case. Unlike in the QG case, the two UDW detectors also remain separable from the mass source system, as no indirect coupling exists between them.

Finally, let us demonstrate the scenario in the BMV papers, where the mass source is in quantum superposition at two spatially localized states as given in Eq.~\eqref{eq:BMV_mass_source}.
In this case, the time-evolved state in the SN spacetime reduces to the following form
\begin{align}
    |\Psi^{(\text{SN})}(t)\rangle
    =\frac{1}{\sqrt{2}}
    \left(|\bm{L}\rangle_\text{M} + |\bm{R}\rangle_\text{M}\right)
    \otimes |\psi^{(\text{SN})}(t)\rangle_\text{SAB}.
\end{align}
Compared to the QG case given in Eq.~\eqref{eq:time_evolved_state_BMV}, no gravity-induced entanglement occurs in the SN case, similar to the conclusions of non-entanglement witness in the semiclassical gravity model~\cite{Bose2017,Marletto2017}.
Fig.~\ref{fig:SNevolution} illustrates a schematic of time evolution in the SN spacetime, where the mass source is in a \schrodinger-cat state with two spatially localized states as in Eq.~\eqref{eq:BMV_mass_source}. 
The background spacetime of the SN model is a classical entity, representing the ensemble average over the mass source state $|\bm{L}\rangle_\text{M}$ and $|\bm{R}\rangle_\text{M}$, and shown as a single purple curved slice in the figure.
The scalar field and detectors propagate uniformly independent of the mass source position, indicated by purple arrows.
This demonstrates that in the SN model, mass source position variations do not influence the propagation of the scalar field and detectors.

%%%%%----------------------------------------------
\section{Gravity-induced entanglement between the mass source and other systems}
\label{sec:Gravity_induced_entanglement}

In the previous section, we calculated the time evolution of the total system for two gravity models, QG and SN. In the QG spacetime, we see that the mass source system and other systems, the scalar field and two UDW detectors, entangle through gravitational interaction, as in Eq.~\eqref{eq:time_evolved_state}.
This section quantitatively evaluates the production of the gravity-induced entanglement between the mass source and other systems for the QG case.
At the end of this section, we also mention that the entanglement indicator goes to zero if we do the same calculation for the SN case.

Specifically, we use linear entropy as an indicator of quantum entanglement.
First, let us define the linear entropy for a general setup, denoting an arbitrary subsystem as S and its complement system as $\Bar{\text{S}}$. The total system is represented by the combination of two systems S and $\Bar{\text{S}}$, and we assume that it is a pure state.
The linear entropy of the subsystem S is defined as
\begin{align}
    \mathcal{E}_S := 1-\mathrm{Tr}\left[ \hat \rho_\text{S}^2 \right],
\end{align}
where $\hat \rho_\text{S} := \mathrm{Tr}_{\Bar{\text{S}}}\left[\hat \rho_\text{tot}\right]$ is a reduced density matrix of the subsystem S calculated from the density matrix of the total system $\rho_\text{tot}$.
The second term on the right-hand side is the same as the purity.
The linear entropy $\mathcal{E}_S$ is considered as an indicator of quantum entanglement between subsystem S and its complement $\Bar{\text{S}}$ based on the assumption that the total system $\rho_\text{tot}$ is a pure state. Furthermore, this satisfies $\mathcal{E}_S=\mathcal{E}_{\Bar{\text{S}}}$.
Note that the linear entropy does not indicate entanglement if the total system is in a mixed state.

Now, we evaluate the gravity-induced entanglement using the linear entropy for the QG model. Since we suppose the total initial state given in Eq.~\eqref{eq:initial_state} is a pure state, and the time evolution of the total system is unitary, the final state in Eq.~\eqref{eq:time_evolved_state} is also a pure state. Therefore, the linear entropy works as an indicator of entanglement production in our setup. 
We estimate the linear entropy between the mass source and other systems, namely the scalar field and two UDW detectors, to capture the gravity-induced entanglement since they interact only through gravity. 
%For simplicity, we assume that 2 detectors are fixed at position $\x_\text{D}$: $\x_\text{A}=\x_\text{B}=\x_\text{D}$.
Using the reduced density matrix of the mass source system $\hat\rho_\text{M}(t)=\mathrm{Tr}_\text{SAB}\left[|\Psi^{(\text{QG})}(t)\rangle\langle \Psi^{(\text{QG})}(t)|\right]$, the linear entropy of the mass source system is given by
\begin{comment}
    \begin{align}
    \mathcal{E}(\x_\text{D})
    &:=1-\mathrm{Tr}\left[\hat\rho_\text{M}(t)^2\right] \notag\\
    &=1-\int d^3X d^3X'~
    |\mu(\X)|^2 |\mu(\X')|^2 
    \left| {}_\text{SAB}\langle\psi^{(\X')}(t)|\psi^{(\X)}(t)\rangle_\text{SAB} \right|^2 \notag\\
    &= \frac{2\lambda^2}{\hbar^2}\int d^3X d^3X'~|\mu(\X)|^2 |\mu(\X')|^2
     \int dt dt'\,\chi(t)\chi(t') e^{i\omega_\text{D}(t-t')}
     \notag\\
     &\hspace{20mm}
     \times \left\langle
     \left(\hat\phi^{(\X)}(t,\x_\text{D})-\hat\phi^{(\X')}(t,\x_\text{D})\right)
     \left(\hat\phi^{(\X)}(t',\x_\text{D})-\hat\phi^{(\X')}(t',\x_\text{D})\right)
     \right\rangle
     \label{eq:GIE}
\end{align}
\end{comment}
\begin{align}
    \mathcal{E}(\x_\text{A},\x_\text{B})
    &:=1-\mathrm{Tr}\left[\hat\rho_\text{M}(t)^2\right] 
    \label{eq:GIE_definition}\\
    &=1-\int d^3X d^3X'~
    |\mu(\X)|^2 |\mu(\X')|^2 
    \left| {}_\text{SAB}\langle\psi^{(\X')}(t)|\psi^{(\X)}(t)\rangle_\text{SAB} \right|^2 \notag\\
    &= \frac{\lambda^2}{\hbar^2}
    \int d^3X d^3X'~|\mu(\X)|^2 |\mu(\X')|^2
     \int dt dt'\,\chi(t)\chi(t') e^{i\omega_\text{D}(t-t')}
     \notag\\
     &\hspace{20mm}
     \times \sum_{j=\text{A},\text{B}}
     \left\langle
     \left(\hat\phi^{(\X)}(t,\x_j)-\hat\phi^{(\X')}(t,\x_j)\right)
     \left(\hat\phi^{(\X)}(t',\x_j)-\hat\phi^{(\X')}(t',\x_j)\right)
     \right\rangle
     .
     \label{eq:GIE}
\end{align}
We explicitly note the $\x_\text{A},\x_\text{B}$ dependence of $\mathcal{E}$ for later use in Section~\ref{sec:WPI}.

If we assume to turn on the UDW detector switch permanently as
\begin{align}
    \chi(t)=1,\notag
\end{align}
the linear entropy is further simplified as
\begin{align}\label{eq:GIE_simplified}
    \mathcal{E} (\x_\text{A},\x_\text{B})
    =\left(\int_{-\infty}^\infty dt_+\right)
    \frac{\pi\lambda^2}{\hbar^2}
    \theta(\omega_\text{D}) 
     \int d^3X d^3X'~|\mu(\X)|^2 |\mu(\X')|^2
     \sum_{j=\text{A},\text{B}}
     \left|\varphi^{(\X)}(\omega_\text{D},\x_j)-\varphi^{(\X')}(\omega_\text{D},\x_j)\right|^2,
     %\left|\varphi^{(\X)}(-\omega_\text{D},\x_j)-\varphi^{(\X')}(-\omega_\text{D},\x_j)\right|^2,
\end{align}
where $t_+=t+t'$.
From Eqs.~\eqref{eq:GIE} or \eqref{eq:GIE_simplified}, we can see that gravity-induced entanglement is described by the difference between two scalar fields, $\hat\phi^{(\X)}(t,\x_j)-\hat\phi^{(\X')}(t,\x_j)$ or $\varphi^{(\X)}(\omega_\text{D},\x_j)-\varphi^{(\X')}(\omega_\text{D},\x_j)$; one is scattered by the mass source at $\X$, and the other is scattered by the mass source at $\X'$. This indicates how much the scalar field evolves differently according to the mass source position, which is referred to as the which-path information of the mass source system obtained by the scalar field.

Finally, let us evaluate the linear entropy of the SN case by substituting the reduced density matrix of the mass source system $\hat\rho_\text{M}(t)=\mathrm{Tr}_\text{SAB}\left[|\Psi^{(\text{SN})}(t)\rangle\langle \Psi^{(\text{SN})}(t)|\right]$ into Eq.~\eqref{eq:GIE_definition}. 
Here, we will discuss the disappearance of the which-path information by deliberately expressing the linear entropy of the SN case in a form similar to Eq.~\eqref{eq:GIE_simplified} by assuming $\chi(t)=1$:
\begin{align}
    \mathcal{E}(\x_\text{A},\x_\text{B})
    &=1-\langle\mu|\mu\rangle_\text{M}
    \left| {}_\text{SAB}\langle\psi^{(\text{SN})}(t)|\psi^{(\text{SN})}(t)\rangle_\text{SAB} \right|^2 \notag\\
    &=\left(\int_{-\infty}^\infty dt_+\right)
    \frac{\pi\lambda^2}{\hbar^2}
    \theta(\omega_\text{D}) ~
    \langle\mu|\mu\rangle_\text{M}
     \sum_{j=\text{A},\text{B}}
     \left|\varphi^{(\text{SN})}(\omega_\text{D},\x_j)-\varphi^{(\text{SN})}(\omega_\text{D},\x_j)\right|^2
     %\left|\varphi^{(\text{SN})}(-\omega_\text{D},\x_j)-\varphi^{(\text{SN})}(-\omega_\text{D},\x_j)\right|^2 
     %\label{eq:GIE_simplified_SN}
     \notag\\
     &=0.
     \label{eq:GIE_simplified_SN}
\end{align}
In the SN case, we can explicitly derive that the linear entropy goes zero, which is consistent with the discussion presented in the previous section.
The disappearance of the linear entropy can also be interpreted as the absence of which-path information, as given in Eq.~\eqref{eq:GIE_simplified_SN}.
This implies that, unlike in the QG case, the scalar field in the SN case does not change its evolution depending on the mass source position.

\section{Observer-accessible quantities using the Unruh-DeWitt detectors}
\label{sec:observer_accessible_quantity}

In Section \ref{sec:Time_evolution}, we consider that two UDW detectors interact with the scalar field as described in Eq.~\eqref{eq:Hamiltonian_int}. Hence, we expect to gain information regarding the evolution of the scalar field from the detector observables. 
We suppose that the observer can only access the detector observables to clarify which quantity is accessible to the observer in our setup.
In this section, we describe the reduced density matrix and the observables of two UDW detectors, especially in Section~\ref{sec:UDW_state}. 
Then, by using the expectation values of the UDW detectors, we introduce two observer-accessible quantities; a two-point correlation function of the scalar field given in Section \ref{sec:CF}, and a which-path information indicator given in Section \ref{sec:WPI}.

%%%%%
\subsection{Reduced density matrix and the observables of the UDW detectors}
\label{sec:UDW_state}

Here, let us show the explicit form of the reduced density matrix of two UDW detectors A and B.
We will first discuss the QG case, followed by a brief examination of the SN case.

Using the time-evolved state for the QG case in Eq.~\eqref{eq:time_evolved_state}, the reduced density matrix of two detectors are given by
\begin{align}
    \hat\rho_\text{AB}(t,\x_A,\x_B)
    &=\mathrm{Tr}_\text{MS}\left[
    |\Psi^{(\text{QG})}(t)\rangle\langle \Psi^{(\text{QG})}(t)|
    \right] \notag\\
    &=\int d^3X \, |\mu(\X)|^2
    \times\sum_{m,n=1}^4\left(\mathcal{M}_\text{AB}^{(\X)}\right)_{mn}
    |e_m\rangle \langle e_n|
    .
\end{align}
Here, the detector basis are chosen to be
\begin{align}\label{eq:detector_basis}
    |\bm{e}\rangle =\left\{
    |0\rangle_\text{A} |0\rangle_\text{B},\,
    |0\rangle_\text{A} |1\rangle_\text{B},\,
    |1\rangle_\text{A} |0\rangle_\text{B},\,
    |1\rangle_\text{A} |1\rangle_\text{B}\right\}^T,
\end{align}
and the corresponding matrix components are given by the following:
\begin{align}
    &\mathcal{M}_\text{AB}^{(\X)}=
    \begin{pmatrix}
        1-\sum_{j=A,B} E^{(\X)}(\x_j,\x_j) & 0 & 0 & X^{(\X)}(\x_\text{A},\x_\text{B})\\
        0 & E^{(\X)}(\x_\text{B},\x_\text{B}) & E^{(\X)}(\x_\text{A},\x_\text{B}) & 0\\
        0 & E^{(\X)}(\x_\text{B},\x_\text{A}) & E^{(\X)}(\x_\text{A},\x_\text{A}) & 0\\
        X^{(\X)}(\x_\text{A},\x_\text{B})^* & 0 & 0 & 0
    \end{pmatrix}
    , \label{eq:detector_matrix}\\
    %%%
    &E^{(\X)}(\x_i,\x_j)
    =\frac{\lambda^2}{\hbar^2} \iint dt dt'\,
    e^{i\omega_\text{D}(t-t')}\,\chi(t)\chi(t')\,
    \left\langle
    \hat\phi^{(\X)}(t,\x_i)\hat\phi^{(\X)}(t,\x_j)
    \right\rangle
    ,\\
    %%%
    &X^{(\X)}(\x_i,\x_j)
    =-\frac{\lambda^2}{\hbar^2} \iint dt dt'\,
    e^{i\omega_\text{D}(t+t')}\,\chi(t)\chi(t') \notag\\
    &\hspace{30mm}\times\left\{
    \theta(t'-t) \left\langle
    \hat\phi^{(\X)}(t,\x_i)\hat\phi^{(\X)}(t',\x_j)
    \right\rangle
    + \theta(t-t') \left\langle
    \hat\phi^{(\X)}(t',\x_j)\hat\phi^{(\X)}(t,\x_i)
    \right\rangle
    \right\}
    .
\end{align}

Similarly, we obtain the reduced density matrix of two detectors for the SN case as follows:
\begin{align}
    \hat\rho_\text{AB}(t,\x_A,\x_B)
    =\mathrm{Tr}_\text{MS}\left[
    |\Psi^{(\text{SN})}(t)\rangle\langle \Psi^{(\text{SN})}(t)|
    \right]
    =\sum_{m,n=1}^4\left(\mathcal{M}_\text{AB}^{(\text{SN})}\right)_{mn}
    |e_m\rangle \langle e_n|
    .
\end{align}
Here, the detector bases are the same as in Eq.~\eqref{eq:detector_basis}, and the matrix components also remains the same as in Eq.~\eqref{eq:detector_matrix}, except that we have replaced the scalar field operator $\hat\phi^{(\X)}$ with that corresponding to the SN case, $\hat\phi^{(\text{SN})}$.

Each matrix component of the detector state, $\mathcal{M}_\text{AB}^{(\X)}$ for the QG case or $\mathcal{M}_\text{AB}^{(\text{SN})}$ for the SN case, is accessible to the observer by measuring the expectation value of the Pauli operators of detectors A and B.
For instance, let us introduce the following detector operator 
\begin{align}\label{eq:observable}
    \hat{\mathcal{O}}_\text{AB}
    :=\frac{1}{4}\left(\hat\sigma_{x,\text{A}}+i\hat\sigma_{y,\text{A}}\right)
    \otimes\left(\hat\sigma_{x,\text{B}}-i\hat\sigma_{y,\text{B}}\right)
    =|0\rangle_\text{A}\,{}_\text{A}\langle 1| \otimes |1\rangle_\text{B}\,{}_\text{B}\langle 0|.
\end{align}
Although $\hat{\mathcal{O}}_\text{AB}$ is not Hermitian, we refer to this as the detector observable in the sense that it is assembled from the Hermitian Pauli operators.
By assuming $\chi(t)=1$, its expectation value is explicitly given by the following with respect to each gravity model:
\begin{align}
    &\mathrm{Tr}\left[ \hat{\mathcal{O}}_\text{AB} \, \hat\rho_\text{AB}(t,\x_A,\x_B) \right]\notag\\
    %&=\int d^3X \, |\mu(\X)|^2 \, E^{(\X)}(\x_\text{A},\x_\text{B})\notag\\
    &=\left(\int_{-\infty}^\infty dt_+\right)
    \frac{\pi\lambda^2}{\hbar^2}
    \theta(\omega_\text{D}) 
    ~\times~
    \begin{dcases}
        \quad
        \int d^3X |\mu(\X)|^2 
        \varphi^{(\X)}(\omega_\text{D},\x_\text{A}) \varphi^{(\X)}(\omega_\text{D},\x_\text{B})^*
        %\varphi^{(\X)}(-\omega_\text{D},\x_\text{A})^* \varphi^{(\X)}(-\omega_\text{D},\x_\text{B})
        & \quad\text{(The QG case)},\\
        \quad
        \varphi^{(\text{SN})}(\omega_\text{D},\x_\text{A}) \varphi^{(\text{SN})}(\omega_\text{D},\x_\text{B})^*
        %\varphi^{(\text{SN})}(-\omega_\text{D},\x_\text{A})^* \varphi^{(\text{SN})}(-\omega_\text{D},\x_\text{B})
        & \quad\text{(The SN case)}.
    \end{dcases}
    \label{eq:observable_expectation_value}
\end{align}
This is one of the observer-accessible quantities since it is given in combination with the Pauli operator expectation values for the reduced density matrix of the detector systems.
Using the detector observable $\hat{\mathcal{O}}_\text{AB}$, we can construct various observer-accessible quantities. Specifically, we focus on two observer-accessible quantities; a two-point correlation function of the scalar field $\mathcal{Q}_\text{CF}$ in the next section and a which-path information indicator $\mathcal{Q}_\text{WPI}$ given in Section \ref{sec:WPI}.

%%%%%
\subsection{Two-point correlation function of the scalar field as an observer-accessible quantity}
\label{sec:CF}

The scalar field correlation function(CF) is often used to analyze the wave optical image in curved spacetime. 
In previous study~\cite{Kanai2013,Nambu2013,Nambu2016}, the authors analyzed the Einstein ring image using a Fourier analysis of the classical scalar field solution, corresponding to the mode function of the scalar field operator in our case.
In addition, in a previous study~\cite{Nambu2022}, the authors investigated the wave optical image of Hawking radiation by calculating the quantum scalar field emitted from a classical black hole and applying a Fourier analysis to its two-point CF. 
In this section, we also introduce the two-point CF of the scalar field operator, which is instrumental in studying the wave optical image, eventually revealing the Einstein ring and providing rich information about the background spacetime. 
We also show that the introduced CF is accessible to the observer.

We define the scalar field CF as follows:
\begin{align}
    &\mathcal{Q}_\text{CF}(\x_\text{A},\x_\text{B})
    :=\frac{1}{2\pi\,\theta(\omega_\text{D})}\int_{-\infty}^\infty d(t-t')~ 
    e^{i\omega_\text{D} (t-t')}
    \langle\Psi_\text{ini}|
    \hat\phi(t,\x_\text{A})\,
    \hat\phi(t',\x_\text{B})
    |\Psi_\text{ini}\rangle
    .
    \label{eq:CF}
\end{align}
Using the mode expansion of the scalar field solution, given in Eqs.~\eqref{eq:scalar_decomposition1},~\eqref{eq:scalar_decomposition2} for the QG case and Eq.~\eqref{eq:scalar_decomposition_SN} for the SN case, we can derive the explicit expression of the scalar field CF. 
Furthermore, by assuming $\chi(t)=1$, the CF can be simplified as follows:
\begin{align}
    \mathcal{Q}_\text{CF}(\x_\text{A},\x_\text{B})
    =
    \begin{dcases}
        \quad
        \int d^3X |\mu(\X)|^2 
        \varphi^{(\X)}(\omega_\text{D},\x_\text{A}) \varphi^{(\X)}(\omega_\text{D},\x_\text{B})^*
        %\varphi^{(\X)}(-\omega_\text{D},\x_\text{A})^* \varphi^{(\X)}(-\omega_\text{D},\x_\text{B})
        & \quad\text{(The QG case)}
        ,\\
        \quad
        \varphi^{(\text{SN})}(\omega_\text{D},\x_\text{A}) \varphi^{(\text{SN})}(\omega_\text{D},\x_\text{B})^*
        %\varphi^{(\text{SN})}(-\omega_\text{D},\x_\text{A})^* \varphi^{(\text{SN})}(-\omega_\text{D},\x_\text{B})
        & \quad\text{(The SN case)}
        .
    \end{dcases}
    \label{eq:CF_QG_SN}
\end{align}
This shows the correlation between the scalar fields at different spatial positions $\x_\text{A}$ and $\x_\text{B}$. 
Compared with Eq.~\eqref{eq:observable_expectation_value}, CF can be rewritten using the expectation value of the detector observable $\hat{\mathcal{O}}_\text{AB}$
\begin{align}
    \mathcal{Q}_\text{CF}(\x_\text{A},\x_\text{B})
    &=\left(\left(\int_{-\infty}^\infty dt_+\right)
    \frac{\pi\lambda^2}{\hbar^2} \theta(\omega_\text{D})\right)^{-1}
    \times~
    \mathrm{Tr}\left[
    \hat{\mathcal{O}}_\text{AB} \,\hat \rho_\text{AB}(t,\x_\text{A},\x_\text{B})
    \right]
    .
\end{align}
Therefore, the observer can reproduce the scalar field CF by performing $\hat{\mathcal{O}}_\text{AB}$ measurement.
Subsequently, in Section~\ref{sec:imaging}, we will present the wave optical image obtained by applying the imaging process to CF, which reveals the Einstein ring containing rich information on the background spacetime.

%%%%%
\subsection{Which-path information indicator as an observer-accessible quantity}
\label{sec:WPI}

In this section, we introduce another quantity, named the which-path information (WPI) indicator:
\begin{align}\label{eq:WPI}
    \mathcal{Q}_\text{WPI}(\x_\text{A},\x_\text{B})
    :=\mathcal{Q}_\text{CF}(\x_\text{A},\x_\text{B})
    -\frac{1}{2\pi\theta(\omega_\text{D})}\int_{-\infty}^\infty d(t-t')~ 
    e^{i\omega_\text{D} (t-t')}
    \langle\Psi_\text{ini}|
    \hat\phi(t,\x_\text{A})
    |\mu\rangle_\text{M}\,{}_\text{M}\langle\mu|
    \hat\phi(t',\x_\text{B})
    |\Psi_\text{ini}\rangle
    .
\end{align}
We highlight two features motivating the use of the WPI indicator: First, the WPI indicator contains the two-point CF $\mathcal{Q}_\text{CF}(\x_\text{A},\x_\text{B})$ introduced in the previous section. As discussed in the previous section, CF is commonly used to examine the wave optical images in curved spacetime. Later, by applying a Fourier analysis to the WPI indicator, we obtain the WPI indicator image which reveals the Einstein ring, as in the CF image.

More importantly, the second feature of the WPI indicator is that it is more sensitive to the gravity model (QG or SN) than the two-point CF.
By substituting the scalar field solution obtained in Section~\ref{sec:Time_evolution} and assuming $\chi(t)=1$, we obtain the explicit form of the WPI indicator. 
For the QG case, it simplifies to 
\begin{align}\label{eq:WPI_QG}
    \mathcal{Q}_\text{WPI}(\x_\text{A},\x_\text{B})
    &=\frac{1}{2}\int d^3X\,d^3X' |\mu(\X)|^2 |\mu(\X')|^2\notag\\
    &\hspace{20mm}\times
    \left(\varphi^{(\X)}(\omega_\text{D},\x_\text{A})-\varphi^{(\X')}(\omega_\text{D},\x_\text{A})\right)
    \left(\varphi^{(\X)}(\omega_\text{D},\x_\text{B})^*-\varphi^{(\X')}(\omega_\text{D},\x_\text{B})^*\right),
        %\left(\varphi^{(\X)}(-\omega_\text{D},\x_\text{A})^*-\varphi^{(\X')}(-\omega_\text{D},\x_\text{A})^*\right)
        %\left(\varphi^{(\X)}(-\omega_\text{D},\x_\text{B})-\varphi^{(\X')}(-\omega_\text{D},\x_\text{B})\right),
\end{align}
while for the SN case, we obtain
\begin{align}\label{eq:WPI_SN}
    \mathcal{Q}_\text{WPI}(\x_\text{A},\x_\text{B})=0
    .
\end{align}
Hence, the WPI indicator becomes zero in the SN case, which is an interesting feature not seen in the CF.

Now, let us explain that the WPI indicator is closely related to the which-path information, as its name suggests, and is consequently linked to the gravity-induced entanglement.
According to the WPI indicator for the QG case given in Eq.~\eqref{eq:WPI_QG}, it considers the difference between two scalar field mode functions $\varphi^{(\X)}(\omega_\text{D},\x_\text{A/B})-\varphi^{(\X')}(\omega_\text{D},\x_\text{A/B})$; one is scattered by the mass source at $\X$, and the other by the mass source at $\X'$. This implies how much the scalar field changes in its time evolution depending on the mass source position. Hence, this quantity was named the WPI indicator. 
However, in the SN case, the scalar field evolves uniformly, unaffected by the mass source position. Consequently, the field does not acquire which-path information about the mass source, consistent with Eq.~\eqref{eq:WPI_SN}.

In particular, when the mass source is in a \schrodinger~cat state as in Eq.~\eqref{eq:BMV_mass_source}, the time evolution in the QG and SN spacetime follows the patterns sketched in Figs.~\ref{fig:QGevolution} and \ref{fig:SNevolution}, respectively. 
In the QG case, the scalar field evolves differently depending on whether the mass source is on the right or the left. The blue and red arrows represent the distinct propagations of the scalar field with respect to the position of the mass source, and the WPI indicator, given in Eq.~\eqref{eq:WPI_QG}, quantifies the difference between these two propagations.
Conversely, in the SN case, the background spacetime is represented by a single ensemble-averaged configuration, regardless of the specific position of the mass source. Consequently, the propagation of the scalar field follows a single pattern, as depicted by the purple arrows, and the WPI indicator becomes trivial, as in Eq.~\eqref{eq:WPI_SN}.

Now, let us discuss the connection between the WPI indicator and gravity-induced entanglement, which is evaluated using the linear entropy of the mass source system, as described in Eq.~\eqref{eq:GIE_simplified}. 
We first propose the following logical expression linking the linear entropy and WPI indicator:
\begin{align}\label{eq:logical_E_Q}
    \mathcal{E}(\x_\text{A},\x_\text{B})=0
    \quad\rightarrow\quad
    \mathcal{Q}_\text{WPI}(\x_\text{A},\x_\text{B})=0
    .
\end{align}
The derivation is given in Appendix~\ref{apdx:WPI}.
This expression indicates that, if no gravity-induced entanglement occurs, for example, when the mass source is not in the spatial superposition or when the SN model is applied, the WPI indicator goes 0. 
Taking the contrapositive, the non-zero value of the WPI indicator directly confirms the existence of gravity-induced entanglement.

\begin{figure}[t]
    \centering
    \begin{minipage}[h]{0.49\linewidth}
        \centering
        \includegraphics[scale=0.3]{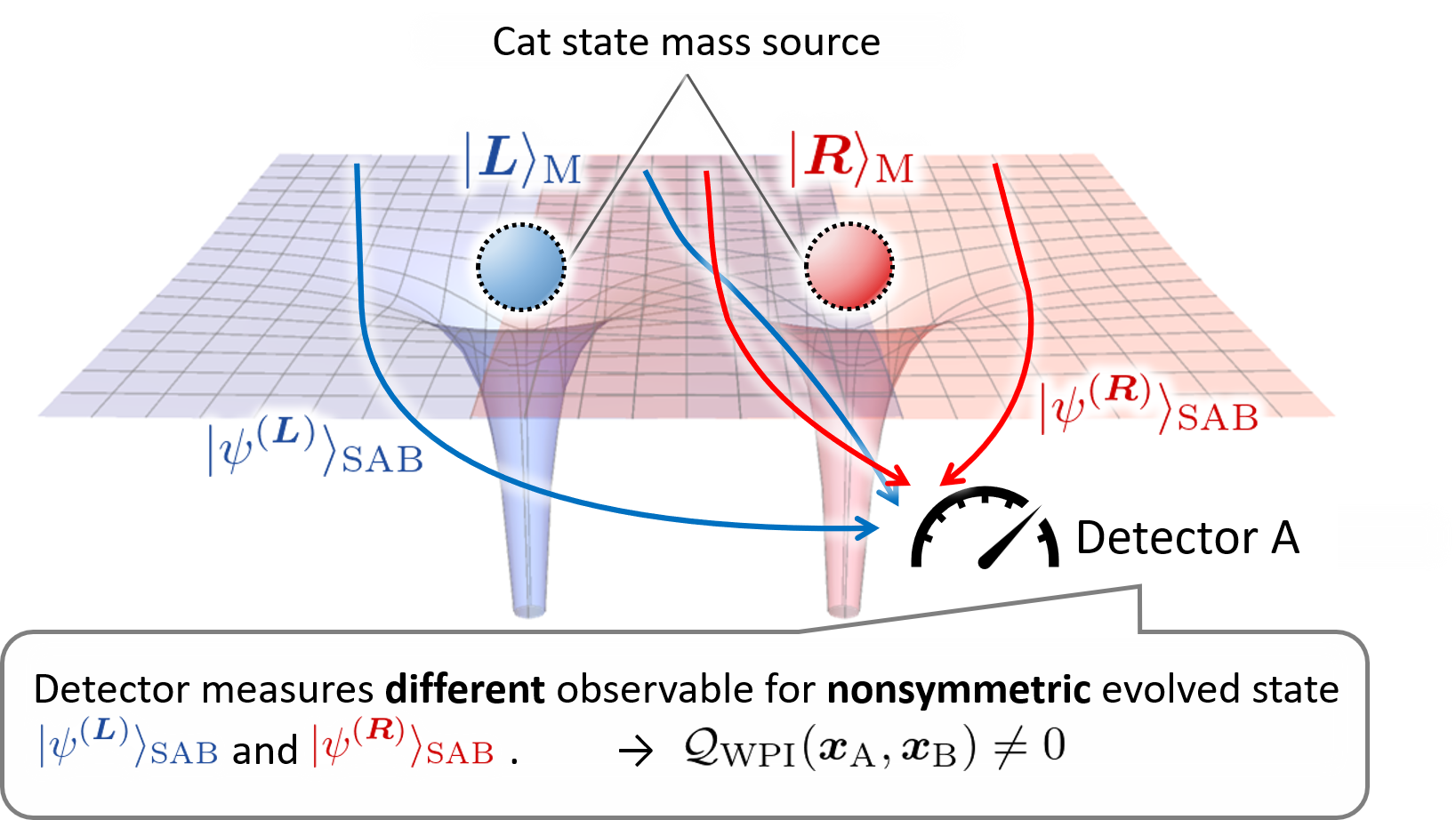}
        \subcaption{
        Nonsymmetric detector positioning resulting in a non-zero value of the WPI indicator.}
        \label{fig:non-zeroWPI}
    \end{minipage}
    \begin{minipage}[h]{0.49\linewidth}
        \centering
        \includegraphics[scale=0.3]{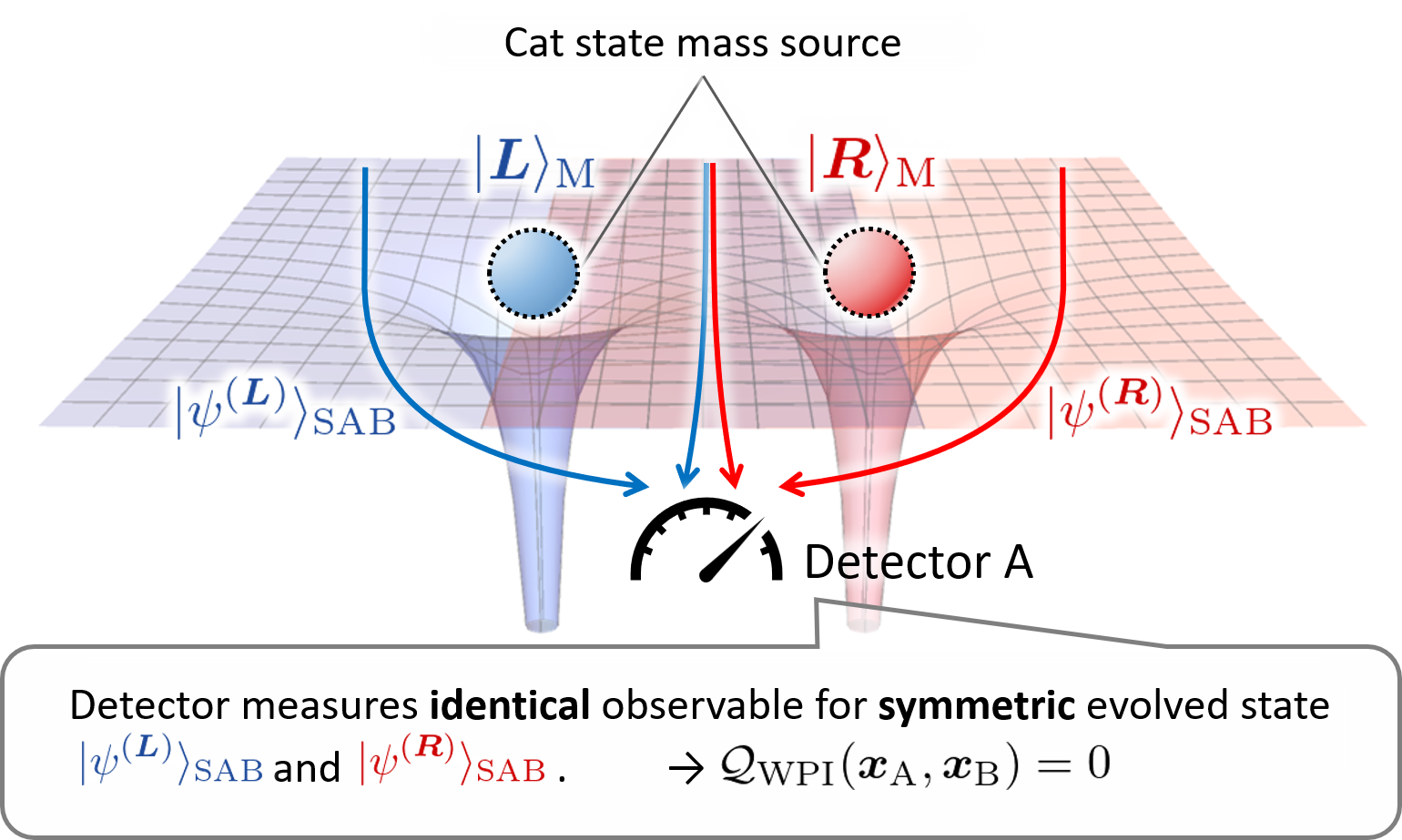}
        \subcaption{Symmetric detector positioning that results in a vanishing WPI indicator.}
        \label{fig:zeroWPI}
    \end{minipage}
    \caption{
    The existence of the WPI indicator in the QG spacetime depends on the detector arrangement.
    In most cases, the WPI indicator gives a non-zero value implying the existence of gravity-induced entanglement, as shown in the left panel. However, for a specific detector arrangement, the WPI indicator vanishes as shown in the right panel.
    }
\end{figure}

Note that for a specific case, the WPI indicator vanishes even if no gravity-induced entanglement is produced; the reverse of the logical expression given in Eq.~\eqref{eq:logical_E_Q} does not necessarily hold. 
Let us show the example for the mass source in a \schrodinger~cat state of the left and right positions as in Eq.~\eqref{eq:BMV_mass_source}.
Gravity-induced entanglement is generally produced in the QG case. In most cases, the detectors observe different scalar field propagations depending on the position of the mass source, leading to a non-zero value of the WPI indicator, as illustrated in Fig.~\ref{fig:non-zeroWPI}.
However, if detector A is particularly positioned at the same distance from the left and right mass source states, specifically satisfying $|\x_\text{A}-\X_\text{L}|=|\x_\text{A}-\X_\text{R}|$ and $z_\text{A}-Z_\text{L}=z_\text{A}-Z_\text{R}$, the WPI indicator disappears, as shown in Fig.~\eqref{fig:zeroWPI}. Two propagations of the scalar field are observed, one scattered by the left mass source (blue arrow) and another scattered by the right mass source (red arrow), which appear identical except for a parity inversion from the perspective of detector A. Then, detector A measures the scalar field locally, the value of which is identical for the two different field propagations. 
Therefore, cases may exist in which the WPI indicator vanishes, even though gravity-induced entanglement is generally produced, depending on the detector arrangement.

Moreover, if we consider the linear entropy for the same detector positions $\x_\text{A}=\x_\text{B}$, we obtain the following logical expression:
\begin{align}\label{eq:logical_E_Q_2}
    \mathcal{E}(\x_\text{A},\x_\text{A})=0
    \quad\leftrightarrow\quad
    \mathcal{Q}_\text{WPI}(\x_\text{A},\x_\text{B})=0
    \quad\text{for}~~\forall\,\x_\text{B}
    .
\end{align}
This derivation is also given in Appendix~\ref{apdx:WPI}.
The expression suggests that observing the zero/non-zero value of the WPI $\x_B$ serves as proof of the non-existence/existence of gravity-induced entanglement, at least for the same detector position $\x_\text{A}=\x_\text{B}$. 

\begin{comment}
    Moreover when $\x_\text{A}=\x_\text{B}$, the WPI indicator is proportional to the linear entropy.
\begin{align}
    \mathcal{Q}_\text{WPI}(\x_\text{A},\x_\text{B}=\x_\text{A})
    =\left(\left(\int_{-\infty}^\infty dt_+\right)
    \frac{\pi\lambda^2}{\hbar^2} \theta(\omega_\text{D})\right)^{-1}
    \times~
    \frac{1}{2}\mathcal{E}(\x_\text{A},\x_\text{B}=\x_\text{A})
\end{align}
Hence, unless the overall factor $\left(\left(\int_{-\infty}^\infty dt_+\right)
    \frac{\pi\lambda^2}{\hbar^2} \theta(\omega_\text{D})\right)^{-1}$ is zero, we get
\begin{align}
    \mathcal{Q}_\text{WPI}(\x_\text{A},\x_\text{B}=\x_\text{A})=0
    \quad\leftrightarrow\quad
    \mathcal{E}(\x_\text{A},\x_\text{B}=\x_\text{A})=0
\end{align}    
\end{comment}

In summary, the two features of the WPI indicator show that it can be used to generate wave optical images, providing insights into the background curved spacetime, while also serving as a witness of gravity-induced entanglement. Later, in Section \ref{sec:imaging}, we will perform a Fourier analysis of the WPI indicator to generate its wave optical image, allowing us to visualize the entanglement production induced by gravity as if projected onto the celestial sphere of the observer.

Finally, we show how the WPI indicator can be rewritten in terms of the detector observable and accessible to the observer.
Here, we show the driven relational formula of the WPI indicator and the detector observable holding for the QG and SN cases in the following:
\begin{align}
    \mathcal{Q}_\text{WPI}(\x_A,\x_B)
    &\simeq\left(\left(\int_{-\infty}^\infty dt_+\right)
    \frac{\pi\lambda^2}{\hbar^2} \theta(\omega_\text{D})\right)^{-1}\notag\\
    &\hspace{10mm}
    \times~
    \left(
    \mathrm{Tr}\left[\hat{\mathcal{O}}_\text{AB}\, \hat\rho(t,\x_\text{A},\x_\text{B})\right]
    -\int d^3 X |\mu(\X)|^2 \,
    \mathrm{Tr}\left[\hat{\mathcal{O}}_\text{AB}\, \hat\rho(t,\x_\text{A},\x_\text{B}^{(\text{falling})}-\X)\right]
    \right).
    \label{eq:WPI_observable}
\end{align}
Here, $\x_\text{B}^{(\text{falling})}$ is a detector position that feels the background field and changes its evolution depending on the gravity model:
\begin{align}\label{eq:xB_falling}
    \x_\text{B}^{(\text{falling})}
    =
    \begin{dcases}
        \x_\text{B}+\hat \X & \text{(The QG case)}
        ,\\
        \x_\text{B}+\langle\hat \X\rangle_\text{M} & \text{(The SN case)}
        .
    \end{dcases}
    .
\end{align}
We used the translational covariance property of the scalar field mode function given in Eq.~\eqref{eq:translation} to derive Eq.~\eqref{eq:WPI_observable} for the QG case. In addition, for both the QG and SN cases, we used the expectation value of the detector observable $\hat{\mathcal{O}}_\text{AB}$ provided in Eq.~\eqref{eq:observable_expectation_value}. 
For the SN case, we assumed that the higher-order position variances of the mass source $\mathcal{O}\left(\langle\hat\X^2\rangle_\text{M}\right)$ are negligible. The detailed derivation is outlined in Appendix~\ref{apdx:WPI}.
Based on this relational formula, we clarify what the observer should know to reproduce the WPI indicator. The first term $\mathrm{Tr}\left[\hat{\mathcal{O}}_\text{AB}\, \hat\rho(t,\x_\text{A},\x_\text{B})\right]$ in Eq.~\eqref{eq:WPI_observable} can be obtained by several measurements on the detector observable $\hat{\mathcal{O}}_\text{AB}$ and evaluate its expectation value. The second term comprises several factors: First, the observer needs to know the mass source wave function $\mu(\X)$, which can be accessed in principle for the table-top experiment, as proposed in the BMV papers~\cite{Bose2017,Marletto2017}. Second, the observer should gain the expectation value of $\hat{\mathcal{O}}_\text{AB}$, where detector A is fixed at a constant position $\x_\text{A}$, while detector B sits at $\x_\text{B}^{(\text{falling})}$ given in Eq.~\eqref{eq:xB_falling}. Equation~\eqref{eq:xB_falling} indicates that the displacement of detector B depends on the gravity model. Such a scenario could be realized, for example, by trapping the detector in an external shallow potential sensitive to gravity so that the displacement of the detector position depends on the background spacetime.

%%%%%----------------------------------------------
\section{Einstein ring image in the weak gravitational field induced by the superposed mass source}
\label{sec:imaging}

In this section, we generate Einstein ring images in the weak gravitational field induced by the spatially superposed mass source by applying an imaging process to the observer-accessible quantities, namely the two-point correlation function introduced in Section~\ref{sec:CF} and the which-path information indicator in Section~\ref{sec:WPI}. In the next section, we first explain the procedures for obtaining the image from the given data of the observer-accessible quantities. Then, in Section \ref{sec:classical_spacetime_image}, we show the image results on a classical curved spacetime and discuss what we can learn about the background spacetime from the resulting Einstein ring image.
Finally, in Section~\ref{sec:superposed_spacetime_image}, we show the image results on the curved spacetime induced by the superposed mass source, the main study findings, and discuss how they reveal differently depending on the gravity model, QG or SN.

%%%%%
\subsection{Imaging process}
\label{sec:image_formation}

In the previous section, we considered the observer who can only access the UDW detectors coupled to the propagating scalar field and introduced the observer-accessible quantities, the 
two-point CF of the scalar field $\mathcal{Q}_\text{CF}(\x_\text{A},\x_\text{B})$ and the WPI indicator $\mathcal{Q}_\text{WPI}(\x_\text{A},\x_\text{B})$.
In this section, we collectively denote such observer-accessible quantities as $\mathcal{Q}(\x_\text{A},\x_\text{B})$.

Recall that this study aims to observe the Einstein ring image on a curved spacetime induced by a superposed mass source.
For this purpose, let us explain how the observer builds the image using the given data for these quantities $\mathcal{Q}(\x_\text{A},\x_\text{B})$. The imaging process involves two primary steps.
\begin{enumerate}
    \item The observer prepares the two-dimensional array data of the observer-accessible quantity by repeating the measurement while changing the detector position.
    \item The Fourier analysis of the prepared two-dimensional array data is performed, and two-dimensional image intensity data is generated.
\end{enumerate}

\begin{figure}[htbp]
    \centering
    \includegraphics[width=0.45\linewidth]{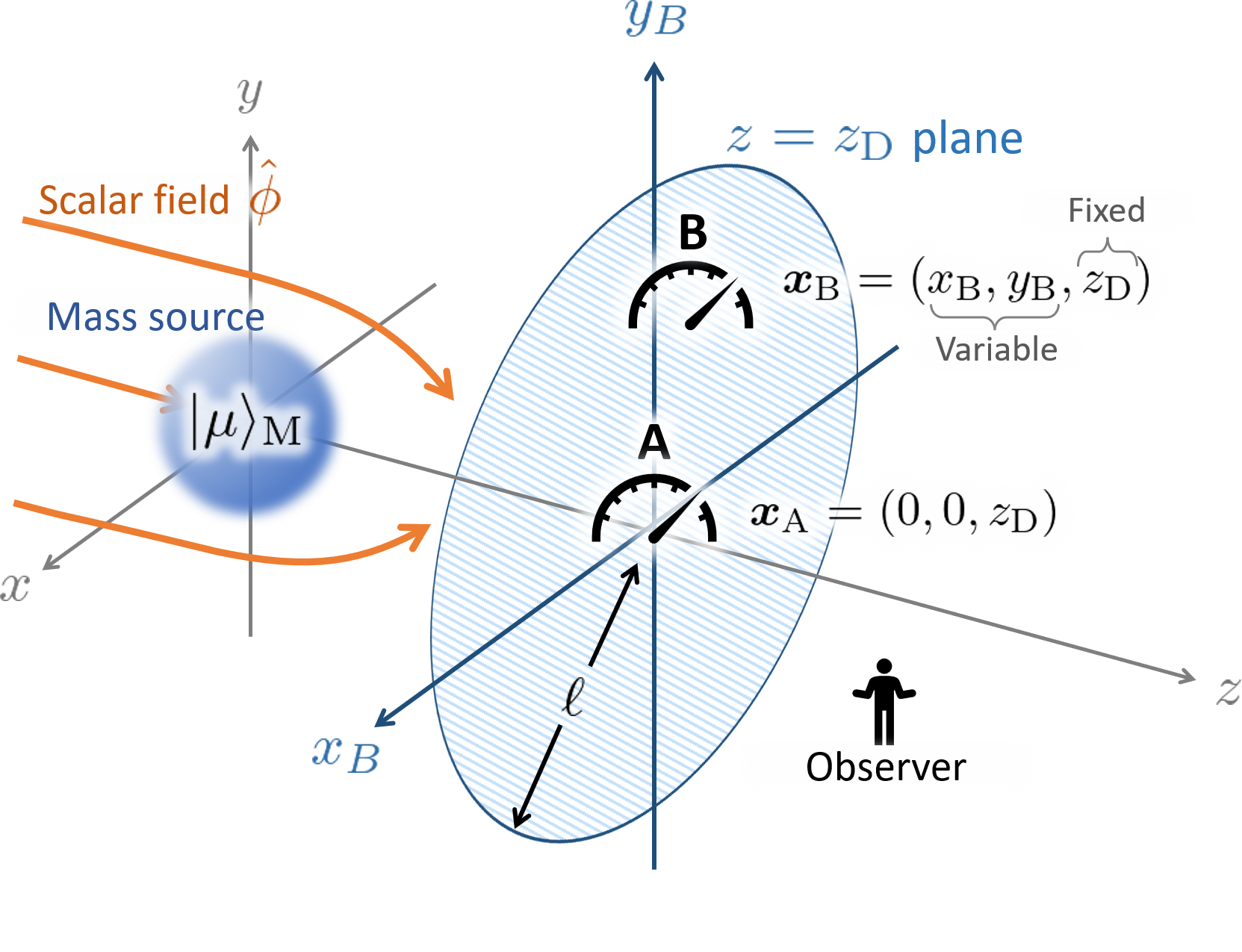}
    \caption{The imaging process setup using two UDW detectors.}
    \label{fig:imaging_process}
\end{figure}

We explain the details of each step as follows:
As a first step, let us demonstrate how the observer prepares the two-dimensional array data of the observer-accessible quantity $Q(\x_\text{A},\x_\text{B})$.
As depicted in Fig.~\ref{fig:imaging_process}, we consider keeping detector A fixed on the $z$-axis, while moving the detector B in the $x,y$-direction
\begin{align}\label{eq:detector_position}
    \x_\text{A}=(0,0,z_\text{D}),
    \quad
    \x_\text{B}=(x_\text{B},y_\text{B},z_\text{D})
    .
\end{align}
Note that $z_\text{D}$ is a fixed constant. Repetition of the measurement on $Q(\x_\text{A},\x_\text{B})$ for various positions of $x_\text{B}$ and $y_\text{B}$, the observer can collect data of the two-dimensional array of $Q(\x_\text{A},\x_\text{B})$ in the $(x_\text{B},y_\text{B})$ plane.
Specifically, in this study, we consider moving  detector B within the disc area on the $(x,\,y)$-plane centered at $(0,0,z_\text{D})$ and having radius $\ell$ as shown by a light blue disk in Fig.~\ref{fig:imaging_process}.

As a second step, let us perform the Fourier analysis of the two-dimensional array data of $Q(\x_\text{A},\x_\text{B})$ obtained in the first step and build its image
\begin{align}\label{eq:imaging_formula}
    \tilde{\mathcal{Q}}(\x_I)
    &:= \int_{|\x_\text{B}-\x_\text{A}|\leq \ell}
    d^2(\x_\text{B}-\x_\text{A}) ~
    e^{i\omega_\text{D}(\x_\text{B}-\x_\text{A}) \cdot \x_I/(cf)} \,
    \mathcal{Q}(\x_\text{A},\x_\text{B}) \\
    &= \iint_{\sqrt{x_\text{B}^2+y_\text{B}^2}\leq \ell} dx_\text{B} \, dy_\text{B} ~
    e^{i\omega_\text{D}(x_\text{B}x_I+y_\text{B}y_I)/(cf)} \,
    \mathcal{Q}(\x_\text{A},\x_\text{B})
    .
\end{align}
Here, $\x_I=(x_I,y_I)$ is the two-dimensional coordinate of the Fourier domain, and $f$ is a constant in length units.
We refer to the quantity $\tilde{\mathcal{Q}}(\x_I)$, the spectrum of the given data $\mathcal{Q}(\x_\text{A},\x_\text{B})$, as an image intensity. Let us mention why the Fourier transformation procedure in Eq.~\eqref{eq:imaging_formula} explains the imaging process and the spectrum $\tilde{\mathcal{Q}}(\x_I)$ can be interpreted as an image intensity. The operation in Eq.~\eqref{eq:imaging_formula} is motivated by the lens diffraction\footnote{
Note that the lens in the context of wave optics is unrelated to gravitational lensing.
}
 formula used in wave optics, which describes how an incident wave forms an image when passing through an optical lens, such as in a telescope~\cite{Hecht2012,Goodman2005,Sharma2006}. 
In wave optics, the Fourier transformation is used to explain the behavior of the wave as it is diffracted by an optical lens and projected onto a screen at the focal point. In this context, $f$ denotes the focal length of the lens, $\ell$ represents the lens aperture, and $(x_I,\,y_I)$ corresponds to the coordinates on the screen. 
Therefore, the operation in Eq.~\eqref{eq:imaging_formula} can be interpreted as representing how the incident wave $\mathcal{Q}(\x_\text{A},\x_\text{B})$ would appear as an image when observed through a telescope. If we observe the incident wave $\mathcal{Q}(\x_\text{A},\x_\text{B})$ directly without performing this operation, we cannot confirm a clear image like the Einstein ring.
Further details about the image formation method in wave optics are given in Appendix~\ref{apdx:image_formation}. 

By substituting each observer-accessible quantity, such as the CF and the WPI indicator, into Eq.~\eqref{eq:imaging_formula}, we obtain the corresponding image intensity. The Fourier integral in Eq.~\eqref{eq:imaging_formula} is performed numerically.
In the next section, we will show the numerical result of the image intensity for each observer-accessible quantity. Especially, in Section~\ref{sec:classical_spacetime_image}, we demonstrate the results for the classical spacetime to describe the fundamental feature of the image. 
In Section~\ref{sec:superposed_spacetime_image}, we finally show our main results for the spacetime induced by the superposed mass source, specifically the QG and SN spacetime.

%%%%%
\subsection{Images in the classical curved spacetime}
\label{sec:classical_spacetime_image}

In this section, we show the image intensity results on a classical curved spacetime. 
Let us suppose that the quantum state of the point mass source $|\mu\rangle_\text{M}$ is localized at some constant position $\X_\text{M}$ as follows:
\begin{align}
    |\mu(\X)|^2 = \delta^{(3)}(\X-\X_\text{M})
    .
\end{align}
Then, the background spacetime is regarded as classical in the sense that the mass source has no spatial quantum fluctuation. 
Note that the QG and SN models do not need to be considered here since the mass source and background spacetime are both treated classically.

Following the imaging process given in the previous section, we obtain the CF image intensity in the classical curved spacetime as follows:
\begin{align}
    \tilde{\mathcal{Q}}_\text{CF}(\x_I)
    &:=\int_{|\x_\text{B}-\x_\text{A}|\leq \ell}
    d^2(\x_\text{B}-\x_\text{A}) ~ 
    \mathcal{Q}_\text{CF}(\x_\text{A},\x_\text{B}) \, e^{-i\frac{\omega_\text{B}}{c} \x_\text{B} \cdot \x_I/f}
    \\
    &=
    \varphi^{(\X_\text{M})}(\omega_\text{D},\x_\text{A})
        \iint_{\sqrt{x_\text{B}^2+y_\text{B}^2}\leq \ell} dx_\text{B} \, dy_\text{B} ~
        e^{i\omega_\text{D}(x_\text{B}x_I+y_\text{B}y_I)/(cf)}
        \varphi^{(\X_\text{M})}(\omega_\text{D},\x_\text{B})^*
    .
    \label{eq:CF_image_classical}
\end{align}
This shows that the CF image intensity is determined by the Fourier spectrum of the scalar field mode function $\varphi^{(\X_\text{M})}(\omega_\text{D},\x_\text{B})^*$, analogous to the Einstein ring image analysis in classical wave optics~\cite{Kanai2013,Nambu2016,Nambu2019}. 
%Note that this Fourier integral is performed %numerically to obtain the following image results.
In contrast to the CF image, the WPI indicator image vanishes trivially in the classical curved spacetime. This is owing to the absence of gravity-induced entanglement in the curved spacetime, which causes the WPI indicator to be zero, as can be seen from Eq.~\eqref{eq:logical_E_Q}.
Hereafter, we define the parameter with length dimensions, $c/\omega_\text{D}$, as 1. Other parameters, such as $\x_\text{A},~\x_\text{B}$, and so on, used in the subsequent numerical results are normalized to this unit length scale.

\begin{figure}[htbp]
    \centering
    %%%
    \begin{minipage}[h]{\linewidth}
        \centering
        \begin{minipage}[h]{0.32\linewidth}
            \centering
            \includegraphics[width=1\linewidth]{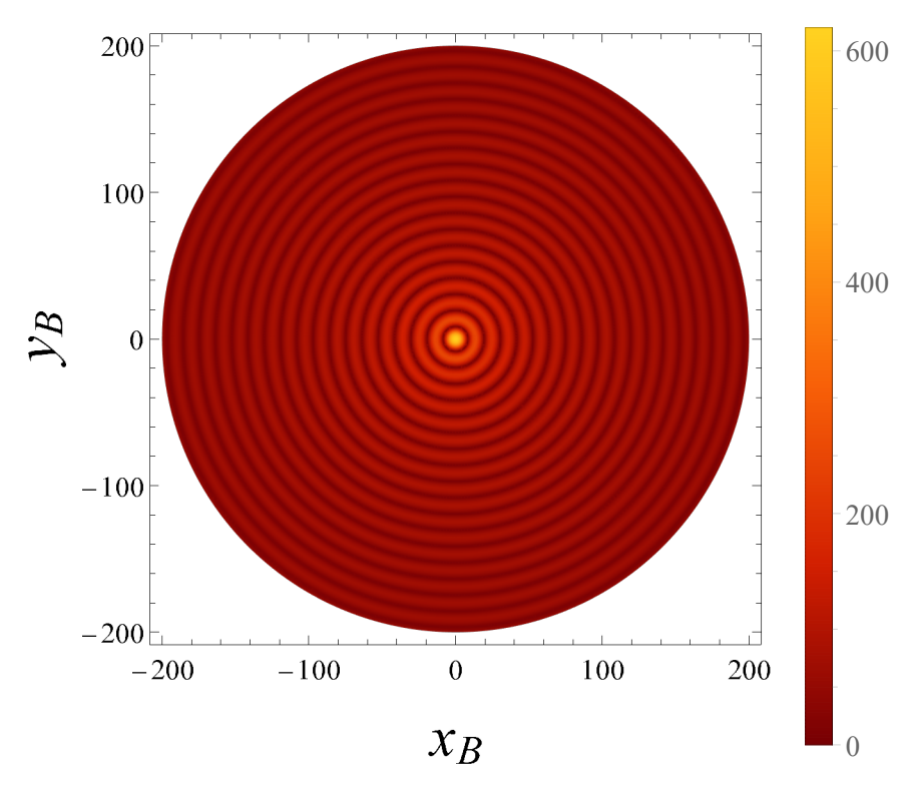}
            \subcaption{2-dim data of CF $\mathcal{Q}_\text{CF}(\x_\text{A},\x_\text{B})$.}
            \label{fig:imageCFaxclassical_CF}
        \end{minipage}
        %
        %\hspace{5mm}
        %
        \begin{minipage}[h]{0.32\linewidth}
            \centering
            \includegraphics[width=1\linewidth]{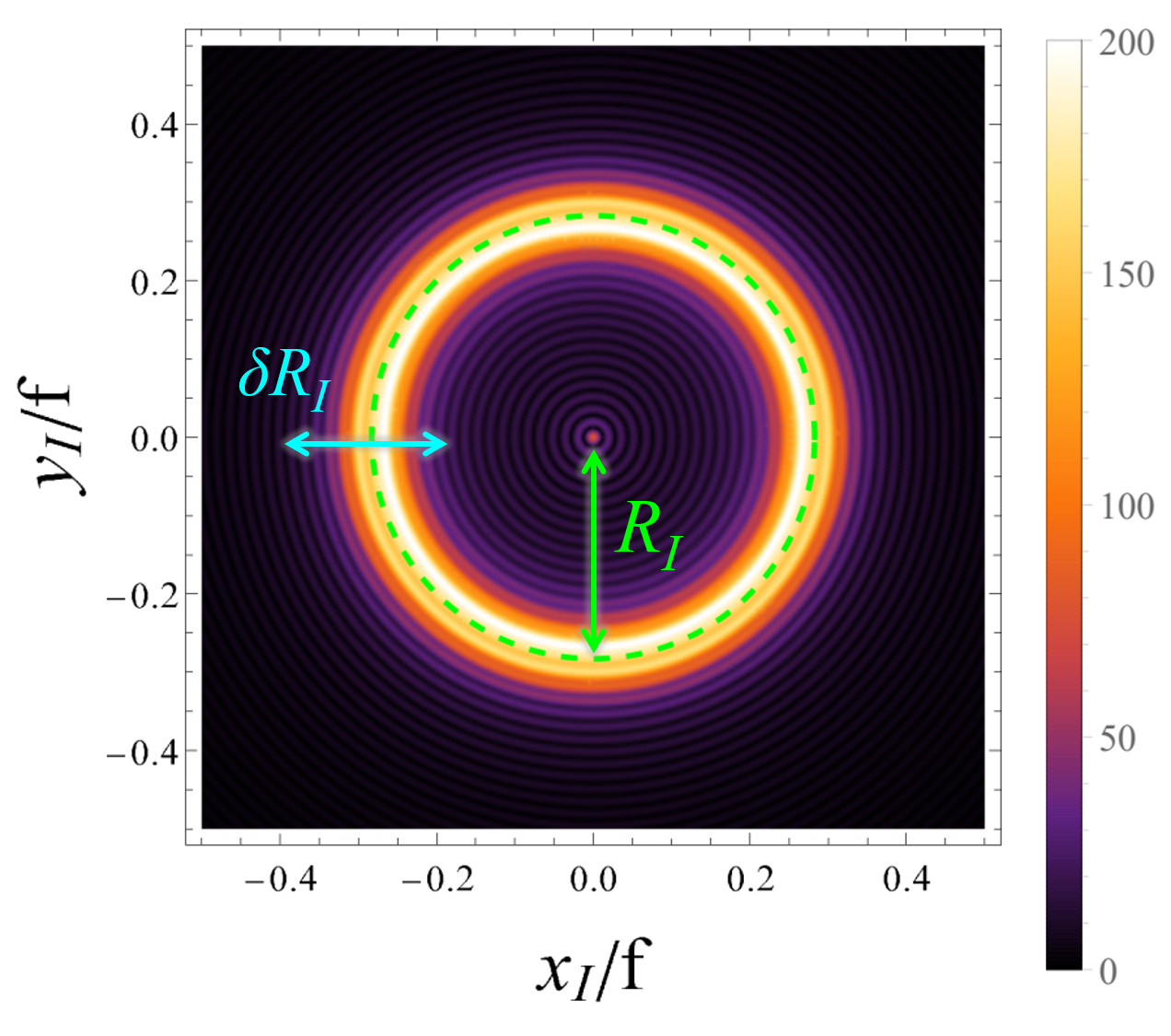}
            \subcaption{The CF image intensity $\tilde{\mathcal{Q}}_\text{CF}(\x_I)$.}
            \label{fig:imageCFaxclassical_CFimage}
        \end{minipage}
        %
        %\hspace{5mm}
        %
        \begin{minipage}[h]{0.32\linewidth}
            \centering
            \includegraphics[width=1\linewidth]{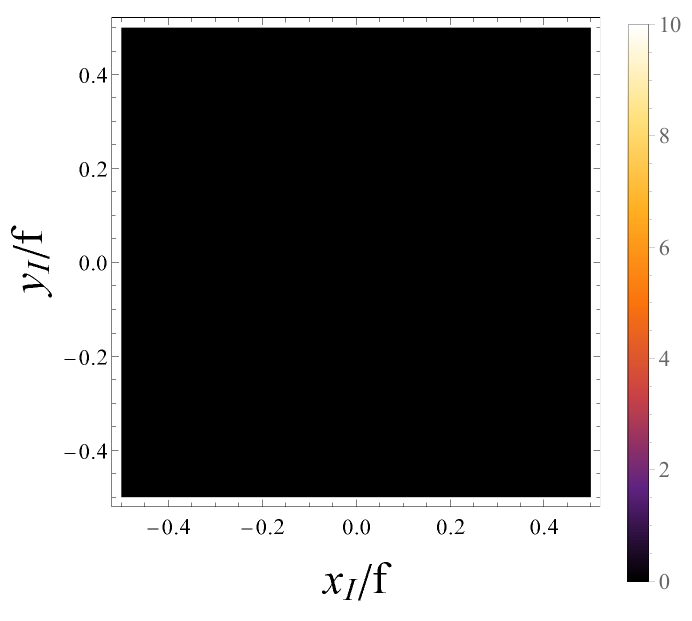}
            \subcaption{The WPI image intensity $\tilde{\mathcal{Q}}_\text{WPI}(\x_I)$.}
            \label{fig:imageCFaxclassical_imageWPIax}
        \end{minipage}
    \end{minipage}
    %%%
    \caption{ 
    We consider the classical curved spacetime induced by a point mass source located at the origin $\X_\text{M}=(0,0,0)$. Fig.~\ref{fig:imageCFaxclassical_CF} shows the two-dimensional array data of the two-point CF of the scalar field $\mathcal{Q}_\text{CF}(\x_\text{A},\x_\text{B})$ detected at various positions of detector B.
    After applying the imaging process to the CF data, we obtain the image intensity $\tilde{\mathcal{Q}}_\text{CF}(\x_I)$, as shown in Fig.~\ref{fig:imageCFaxclassical_CFimage}.
    We observe an Einstein ring image with the radius $R_I$ (indicated by a green dashed line) and the radius variance $\delta R_I$ (indicated by a blue arrow) being analytically expressed in Eq.~\eqref{eq:Einstein_ring_radius}.
    By contrast, the WPI indicator $\mathcal{Q}_\text{WPI}(\x_\text{A},\x_\text{B})$ vanishes in the classical curved spacetime, resulting in the disappearance of the image intensity, as shown in Fig.~\ref{fig:imageCFaxclassical_imageWPIax}.
    }
    \label{fig:imageCFaxclassical}
\end{figure}

We show the two-dimensional array data of CF and its image intensity in Figs.~\ref{fig:imageCFaxclassical_CF} and \ref{fig:imageCFaxclassical_CFimage}, respectively, to demonstrate how the imaging process provided in Eq.~\eqref{eq:imaging_formula} works. 
For both images, we suppose that the point mass source is localized at the origin $\X_\text{M}=(0,0,0)$. The gravitational coupling constant is also set to $\gamma=100$, and the movable radius of detector B is set to $\ell=200$. 
In Fig.~\ref{fig:imageCFaxclassical_CF}, we present the two-dimensional array data of the two-point CF of the scalar field $\mathcal{Q}_\text{CF}(\x_\text{A},\x_\text{B})$ detected at various positions of detector B. The horizontal and vertical axes represent the detector B position $x_\text{B}$ and $y_\text{B}$, respectively. Note that detector A is fixed along the $z$-axis, as described in Eq.~\eqref{eq:detector_position} and in Fig.~\ref{fig:imaging_process}, while we set $z_\text{D}=2500$. We prepared 100 data points per radius to plot Fig.~\ref{fig:imageCFaxclassical_CF}.
After applying the imaging process in Eq.~\eqref{eq:imaging_formula} to the CF data given in Fig.~\ref{fig:imageCFaxclassical_CF}, we obtain its image intensity $\tilde{\mathcal{Q}}_\text{CF}(\x_I)$ as shown in Fig.~\ref{fig:imageCFaxclassical_CFimage}. The horizontal and vertical axes show the normalized Fourier domain coordinate $x_I/f$ and $y_I/f$ respectively. We obtained 120 data points per radius for plotting this.
In the figure, we observe a ring image, corresponding to the peak of the spectrum for a typical oscillation frequency shown in Fig.~\ref{fig:imageCFaxclassical_CF}. This ring, known as the Einstein ring, arises from the diffraction of the massless scalar field caused by the curved spacetime. 
The Einstein ring radius $R_I$, indicated by a green dashed line, and its variance $\delta R_I$, shown by a blue arrow, are approximately given by
\begin{align}\label{eq:Einstein_ring_radius}
    R_I=\sqrt{\frac{2c\gamma}{(z_\text{D}-Z_\text{M})\omega_\text{D}}},
    \qquad
    \delta R_I = \frac{1}{z_\text{D}-Z_\text{M}}\left(\ell-\frac{3\pi c}{\omega_\text{D}}\right).
\end{align}
In addition, when the mass source is positioned off the $z$-axis, the center of the Einstein ring shifts correspondingly on the $(x_I/f, y_I/f)$ plane, tracking the mass source position as 
\begin{align}\label{eq:Einstein_ring_center}
    \bm{O}_I = \frac{1}{2(z_\text{D}-Z_\text{M})}\times (X_\text{M},Y_\text{M})
    .
\end{align}
Thus, we can determine the gravitational coupling constant $\gamma$ from the ring radius $R_I$, the distance from the observer to the mass source $z_\text{D}-Z_\text{M}$ from the ring radius variance $\delta R_I$, and the mass source position $(X_\text{M},Y_\text{M})$ from the center of the ring. 
Hence, the ring structure in the CF image provides rich information about the background spacetime.
In addition, the image reveals an interference fringe around the peak of the Einstein ring, of whose period is characterized by the scalar field frequency $\omega_\text{D}$, and the movable radius of detector B $\ell$ as
\begin{align}\label{eq:Einstein_ring_fringe}
    \frac{\omega_\text{D} \ell}{c} \times \frac{|\x_I|}{f} =2\pi n,
\end{align}
where $n$ is an arbitrary integer. The interference pattern arises as a consequence of wave optical formalism.
The derivations of these approximated forms are discussed in Appendix~\ref{apdx:Einstein_ring_radius}.

Figure~\ref{fig:imageCFaxclassical_imageWPIax} shows the image intensity of the WPI indicator $\tilde{\mathcal{Q}}_\text{WPI}(\x_I)$ on the classical spacetime with parameters matching those used for the CF image in Fig.~\ref{fig:imageCFaxclassical_CFimage}. As discussed below Eq.~\eqref{eq:CF_image_classical}, we observe that the image intensity clearly disappears in the classical curved spacetime.

Now, let us demonstrate how the Einstein ring images appear differently with the position of the point mass source position. In the remainder of this section, we mainly focus on the CF image, as the WPI image trivially vanishes in classical spacetime, as shown in Fig.~\ref{fig:imageCFaxclassical_imageWPIax}.

\begin{figure}[htbp]
    \centering
    \includegraphics[width=1\linewidth]{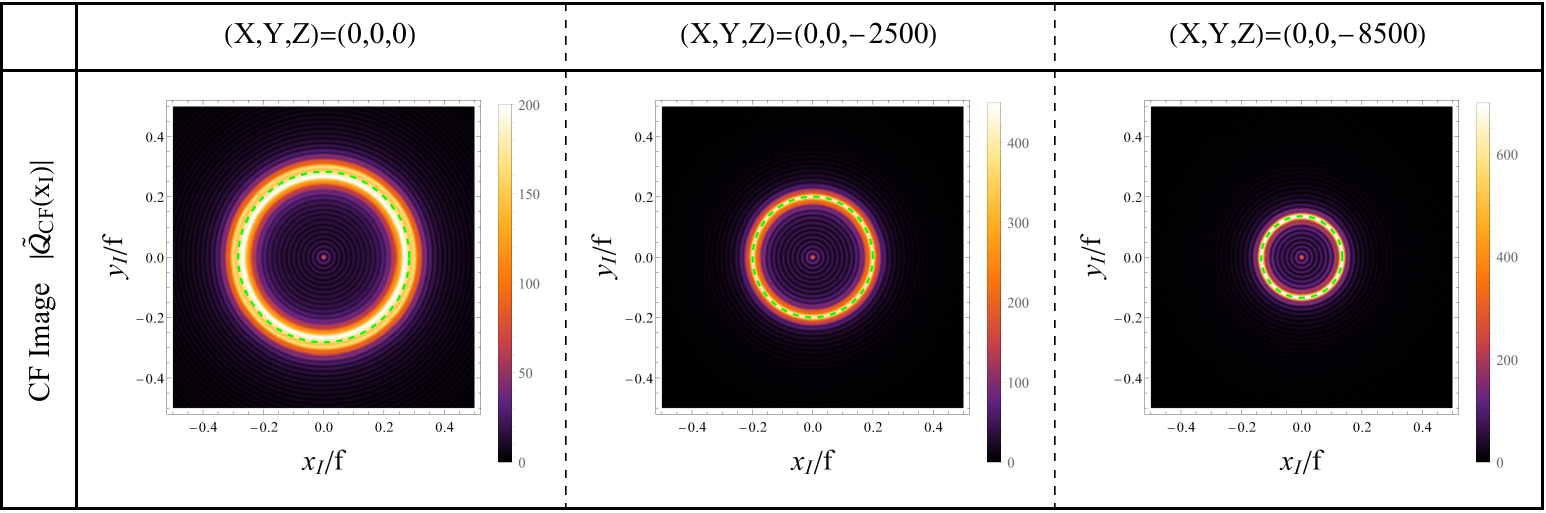}
    \caption{
    The CF images $|\tilde{\mathcal{Q}}_\text{CF}(\x_I)|$ in the classical curved spacetime where the mass source is located along the $z$-axis, corresponding to the direction of the line of sight of the observer. The left, middle, and right panels depict the CF figures for the point mass source positioned at $(X_\text{M},Y_\text{M},Z_\text{M})=(0,0,0),~(0,0,-2500)$ and $(0,0,-8500)$, respectively. The right panel shows the scenario when the mass source is farthest from the observer, resulting in a smaller ring image. The green dashed line in each panel indicates the approximate position of the Einstein ring, as given in Eq.~\eqref{eq:Einstein_ring_radius}.
    }
    \label{fig:imageCFaxclassicaltable}
\end{figure}

% Mass source at z-axis
In Fig.~\ref{fig:imageCFaxclassicaltable}, we examine how the CF image varies when the point mass source moves in the direction of the line of sight of the observer. 
We present three different CF images where the mass source is located along the $z$-axis. 
The left, middle, and right panels depict the CF figures for the point mass source positioned at $(X_\text{M},Y_\text{M},Z_\text{M})=(0,0,0),~(0,0,-2500)$ and $(0,0,-8500)$, respectively. Other parameters remain the same, specifically $\gamma=100$ and $\ell=200$, as in Fig.~\ref{fig:imageCFaxclassical}. We prepared 120 data points per radius for plotting.
The right panel shows the scenario where the mass source is farthest from the observer, resulting in a smaller ring image.
In addition, we see that the approximate expression of the Einstein ring given in Eq.~\eqref{eq:Einstein_ring_radius} and indicated by a green dashed line in each panel aligns well with each numerical result. 

\begin{figure}[htbp]
    \centering
    \includegraphics[width=1\linewidth]{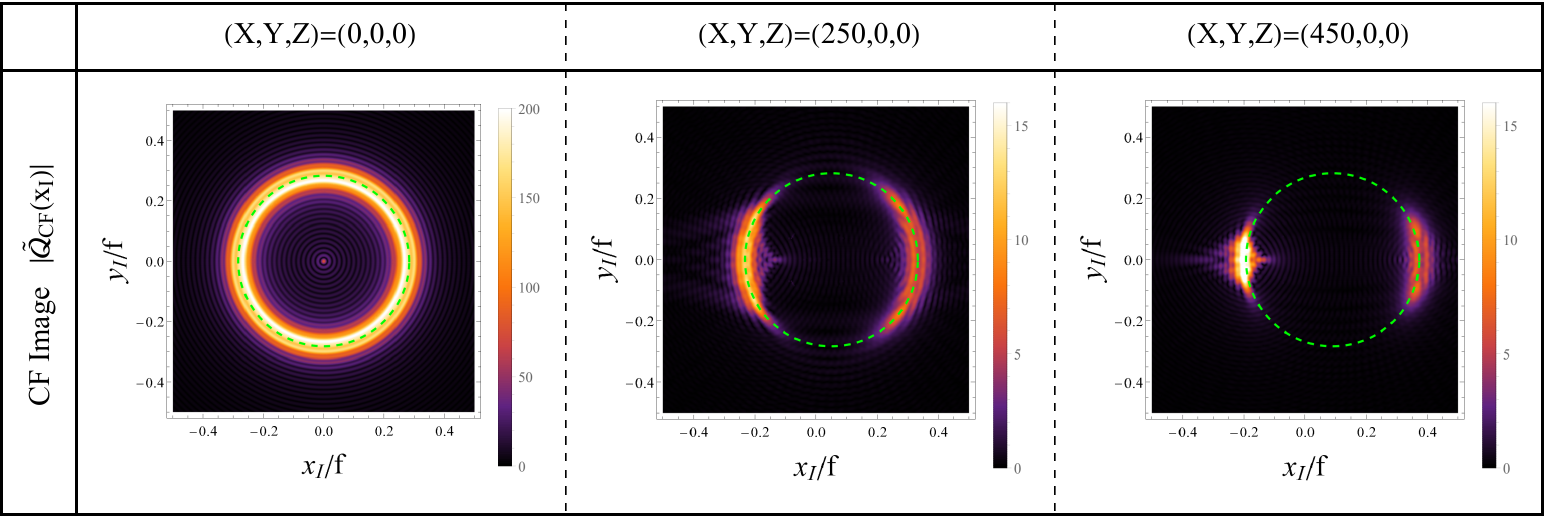}
    \caption{
    The CF images $|\tilde{\mathcal{Q}}_\text{CF}(\x_I)|$ in the classical curved spacetime where the mass source is located along the $x$-axis, corresponding to the direction perpendicular to the line of sight of the observer. The left, middle, and right panels depict the CF figures for the point mass source positioned at $(X_\text{M},Y_\text{M},Z_\text{M})=(0,0,0),~(250,0,0)$, and $(450,0,0)$, respectively. 
    The image no longer resembles a ring as the mass source is further displaced; instead, it takes the form of double arcs. Each panel shows the approximate position of the Einstein ring indicated by the green dashed line, as given in Eqs.~\eqref{eq:Einstein_ring_radius}, \eqref{eq:Einstein_ring_center}.
    }
    \label{fig:imageCFclassicaltable}
\end{figure}

% Mass source at x axis
As the final part of this section, let us examine how the CF images change when the point mass source moves in a direction perpendicular to the line of sight of the observer.
In Fig.~\ref{fig:imageCFclassicaltable}, we present three different CF images where the mass source is located along the $x$-axis. 
The left, middle, and right panels show the CF figures for the point mass source located at $(X_\text{M},Y_\text{M},Z_\text{M})=(0,0,0),~(250,0,0)$ and $(450,0,0)$, respectively. Other parameters, specifically $\gamma=100$ and $\ell=200$, remain the same as in Fig.~\ref{fig:imageCFaxclassical}. We collected $200\times200$ data points to plot the image.
In the right panel, we observe the scenario in which the mass source is most displaced in the positive $x$-direction from the perspective of the observer.
As can be seen from the middle and right panels, the image no longer resembles a ring; instead, it takes the form of double arcs. Notably, the arc on the left side becomes brighter than that on the right and gradually shifts towards the center of the figure as the mass source is further displaced. This is because the observer detects the scalar field that has traveled straight along the direction of their line of sight as the mass source moves further away.
Furthermore, we show the approximate position of the Einstein ring, provided in Eqs.~\eqref{eq:Einstein_ring_radius}, \eqref{eq:Einstein_ring_center}, with a green dashed line in each panel. These approximated expressions align well with each numerical result.

%%%%%
\subsection{Images in the curved spacetime induced by the superposed mass source}
\label{sec:superposed_spacetime_image}

This section investigates the Einstein ring image in the curved spacetime generated by the superposed mass source, examining the QG and SN spacetime, which is the primary aim of this study.
Before showing the numerical results, we first outline the explicit form of the image intensity derived from the CF data and the WPI indicator data.

Using Eqs.~\eqref{eq:CF_QG_SN}, \eqref{eq:imaging_formula}, the CF image intensity $\tilde{\mathcal{Q}}_\text{CF}(\x_I)$ is given by
\begin{align}
    &\tilde{\mathcal{Q}}_\text{CF}(\x_I)
    :=\int_{|\x_\text{B}-\x_\text{A}|\leq \ell}
    d^2(\x_\text{B}-\x_\text{A}) ~ 
    \mathcal{Q}_\text{CF}(\x_\text{A},\x_\text{B}) \, e^{-i\frac{\omega_\text{B}}{c} (\x_\text{B}-\x_\text{A}) \cdot \x_I/f}
    \notag\\
    &=
    \begin{dcases}
        ~
        \int d^3X |\mu(\X)|^2 \, \varphi^{(\X)}(\omega_\text{D},\x_\text{A})
        \iint_{\sqrt{x_\text{B}^2+y_\text{B}^2}\leq \ell} dx_\text{B} \, dy_\text{B} ~
        e^{i\omega_\text{D}(x_\text{B}x_I+y_\text{B}y_I)/(cf)}
        \varphi^{(\X)}(\omega_\text{D},\x_\text{B})^*
        & \quad\text{(The QG case)}
        ,\\
        ~
        \varphi^{(\text{SN})}(\omega_\text{D},\x_\text{A})
        \iint_{\sqrt{x_\text{B}^2+y_\text{B}^2}\leq \ell} dx_\text{B} \, dy_\text{B} ~
        e^{i\omega_\text{D}(x_\text{B}x_I+y_\text{B}y_I)/(cf)}
        \varphi^{(\text{SN})}(\omega_\text{D},\x_\text{B})^*
        & \quad\text{(The SN case)}
        .
    \end{dcases}
    \label{eq:CF_image_QG_SN}
\end{align}
For the QG case, the expression suggests that we need to sum the Fourier spectrum of $\varphi^{(\X)}(\omega_\text{D},\x_\text{B})^*$ over the different positions of the mass source $\X$, using the mass source probability distribution $|\mu(\X)|^2$ as a weighting factor. 
This signifies that the image intensity in the QG spacetime is obtained by summing multiple classical Einstein ring images for the mass source located at various positions $\X$.
More broadly, the image intensity becomes blurred owing to the quantum fluctuation of the mass source position, suggesting quantum dephasing or quantum decoherence of the image.
Conversely, the SN case involves a single Fourier spectrum of $\varphi^{(\text{SN})}(\omega_\text{D},\x_\text{B})^*$, and there is no need to sum over mass source positions, which means image decoherence does not occur, unlike the QG case.

According to Eq.~\eqref{eq:imaging_formula}, the WPI image intensity $\tilde{\mathcal{Q}}_\text{WPI}(\x_I)$ is given by 
\begin{align}
    \tilde{\mathcal{Q}}_\text{WPI}(\x_I)
    :=\int_{|\x_\text{B}-\x_\text{A}|\leq \ell}
    d^2(\x_\text{B}-\x_\text{A}) ~ 
    \mathcal{Q}_\text{WPI}(\x_\text{A},\x_\text{B}) \, e^{-i\frac{\omega_\text{B}}{c}(\x_\text{B}-\x_\text{A}) \cdot \x_I/f}.
    \label{eq:WPI_image}
\end{align}
For the QG case, owing to the WPI indicator given in Eq.~\eqref{eq:WPI_QG}, the image intensity is explicitly given by
\begin{align}
    \tilde{\mathcal{Q}}_\text{WPI}(\x_I)
    &=\int d^3X\,d^3X' |\mu(\X)|^2 |\mu(\X')|^2
        \left(\varphi^{(\X)}(\omega_\text{D},\x_\text{A})-\varphi^{(\X')}(\omega_\text{D},\x_\text{A})\right)
        \notag\\
    &\hspace{10mm}\times~
        \iint_{\sqrt{x_\text{B}^2+y_\text{B}^2}\leq \ell} dx_\text{B} \, dy_\text{B} ~
        e^{i\omega_\text{D}(x_\text{B}x_I+y_\text{B}y_I)/(cf)}
        \left(\varphi^{(\X)}(\omega_\text{D},\x_\text{B})^*-\varphi^{(\X')}(\omega_\text{D},\x_\text{B})^*\right).
    \label{eq:WPI_image_QG}
\end{align}
Similarly to the CF image intensity, the WPI image for the QG case is obtained by integrating over various mass source positions, revealing the summation of multiple Einstein ring images, as will be demonstrated later.
Conversely, for the SN case, the WPI image disappears since the WPI indicator itself goes to zero, as shown in Eq.~\eqref{eq:WPI_SN}:
\begin{align}
    \tilde{\mathcal{Q}}_\text{WPI}(\x_I)=0
    .
    \label{eq:WPI_image_SN}
\end{align}
In Section~\ref{sec:WPI}, we highlighted the following key feature of the WPI indicator: it vanishes in the SN case and is sensitive to the presence of gravity-induced entanglement. This is why the WPI image also vanishes for the SN case and is strongly influenced by gravity-induced entanglement. 
This allows us to witness the gravity-induced entanglement through the Einstein ring image constructed from the WPI indicator, the observer-accessible quantity.

From Eqs.~\eqref{eq:CF_image_QG_SN} and \eqref{eq:WPI_image_SN}, we anticipate a significant difference between the CF and WPI images in the SN case. Here, let us compare the CF and WPI images also in the QG case to demonstrate their difference.
For simplicity, we consider the weak gravitational field induced by a point mass source that exists in a \schrodinger~cat state, comprising two spatially localized states at $\X_\text{L}$ and $\X_\text{R}$, as discussed in the BMV papers~\cite{Bose2017,Marletto2017}. The corresponding probability distribution is given by
\begin{align}\label{eq:BMV_wave_function}
    |\mu(\X)|^2
    = \frac{1}{2}\left( \delta^{(3)}(\X-\X_\text{L}) +\delta^{(3)}(\X-\X_\text{R}) \right).
\end{align}
Substituting this form into Eq.~\eqref{eq:CF_image_QG_SN} and \eqref{eq:WPI_image_QG}, we derive the CF and WPI image intensities for the QG case as follows:
\begin{align}
    \tilde{\mathcal{Q}}_\text{CF}(\x_I)
    &\propto \frac{1}{2}\left(
    \tilde{\mathcal{Q}}_\text{CF}^{(\X_\text{L})}(\x_I)
    +\tilde{\mathcal{Q}}_\text{CF}^{(\X_\text{R})}(\x_I)
    \right)\\
    \tilde{\mathcal{Q}}_\text{WPI}(\x_I)
    &\propto \frac{1}{2}\left(
    \frac{\tilde{\mathcal{Q}}_\text{CF}^{(\X_\text{L})}(\x_I)}{\varphi^{(\X_\text{L})}(\x_\text{A})}
    -\frac{\tilde{\mathcal{Q}}_\text{CF}^{(\X_\text{R})}(\x_I)}{\varphi^{(\X_\text{R})}(\x_\text{A})}
    \right).
\end{align}
Here, $\tilde{\mathcal{Q}}_\text{CF}^{(\X_\text{L/R})}(\x_I)$ denotes the CF image intensity in the classical spacetime where the mass source is located at $\X_\text{L/R}$, revealing the Einstein ring image as shown in the previous section. 
The above expression suggests that the CF image is a simple summation of two classical Einstein ring images corresponding to the mass source located at $\X_\text{L}$ and $\X_\text{R}$. 
Conversely, the WPI image intensity is also composed of two classical Einstein ring images but incorporates weighting factors $\varphi^{(\X_\text{L})}(\x_\text{A})^{-1}$ and $-\varphi^{(\X_\text{R})}(\x_\text{A})^{-1}$, respectively. 
In summary, we note the following points: First, both the CF and WPI images are composed of two classical Einstein ring images. Second, the difference between the CF and WPI images becomes apparent when the weighting factors $\varphi^{(\X_\text{L})}(\x_\text{A})^{-1}$ and $-\varphi^{(\X_\text{R})}(\x_\text{A})^{-1}$ differ significantly. In such situation, the WPI image captures distinct phase information of the scalar field compared to the CF image.
In practice, the image intensity plots do not show a marked difference between the CF and WPI cases unless the parameters are finely tuned. Nonetheless, subtle difference can be observed in the relative brightness of the two superposed classical images.

Now, let us show the numerical results of the Einstein ring images in the QG and SN spacetime.
We numerically perform the Fourier integral in the imaging process to obtain the image results.
Hereafter, we define the parameter with length dimensions, $c/\omega_\text{D}$, as 1. Other parameters, such as $\x_\text{A}$ and $\x_\text{B}$, used in the subsequent numerical results are normalized to this unit length scale.

\begin{figure}[t]
    \centering
    \includegraphics[width=0.8\linewidth]{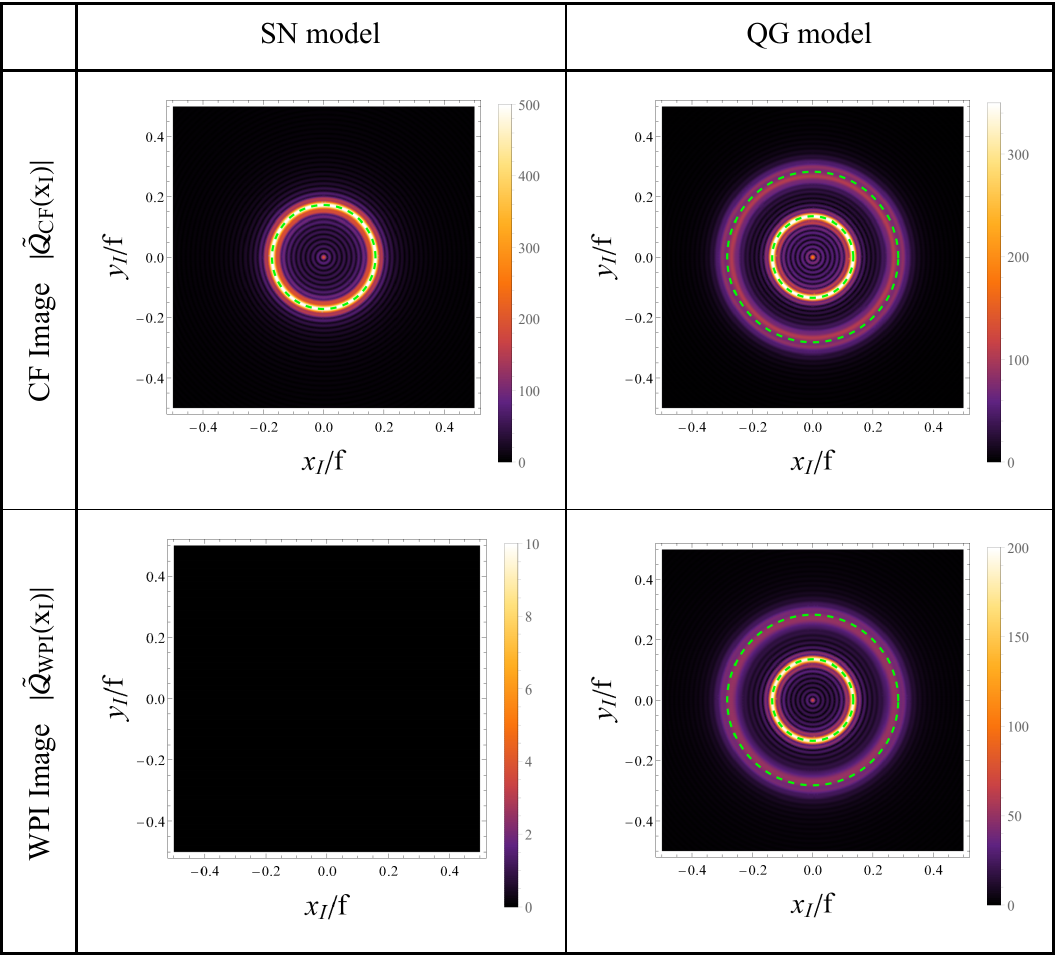}
    \caption{
    The CF images and WPI images for the SN and QG models, where the mass source is superposed at two distinct positions along the $z$-axis, $\X_\text{L}=(0,0,0)$ and $\X_\text{R}=(0,0,-8500)$. In the QG model, the CF and WPI images display the composition of two ring images, corresponding to the two Einstein rings on the curved spacetime induced by the point mass source at $\X_\text{L}$ and $\X_\text{R}$, respectively. By contrast, for the SN model, the CF image shows a single Einstein ring corresponding to the averaged mass source position $\langle \hat \X\rangle=(\X_\text{L}+\X_\text{R})/2$, while the intensity in the WPI image disappears, as predicted by Eq.~\eqref{eq:WPI_image_SN}.
    }
    \label{fig:imageaxCFWPI_QG_SN}
\end{figure}

% Mass source at z-axis
First, let us examine how the images differ based on observer-accessible quantities and the gravity model, where the point mass source is superposed along the $z$-axis, corresponding to the direction of the line of sight of the observer.
In Fig.~\ref{fig:imageaxCFWPI_QG_SN}, we present the CF and WPI images for both the SN and QG models. The upper left and right panels depict the CF images for the SN and QG models, respectively, while the lower left and right panels depict the WPI image for the SN and QG models respectively. The mass source is in quantum superposition at two distinct positions along the $z$-axis, $\X_\text{L}=(0,0,0)$ and $\X_\text{R}=(0,0,-8500)$. Other parameters, namely $\gamma=100$ and $\ell=200$, remain the same as in Fig.~\ref{fig:imageCFaxclassical}. We collected 120 data points per radius to plot the images.
The following two key features are noted this figure: First, in the QG model, both the CF and WPI images show the composition of two ring images. In particular, each ring closely matches the green dashed line, which represents the approximate position of the Einstein ring on the classical curved spacetime from Eq.~\eqref{eq:Einstein_ring_radius}. The outer and inner lines show the position of the ring for $\X_\text{L}$ and $\X_\text{R}$, respectively. Therefore, the image displays a summation of the two classical Einstein rings for the mass source located at $\X_\text{L}$ and $\X_\text{R}$, indicating the quantum dephasing of the image in the QG spacetime.
Second, in the SN model, we observe significant differences compared to the QG model for the CF and WPI images. In the CF image, we observe a single Einstein ring, which aligns well with the Einstein ring predicted for the curved spacetime induced by the mass source at the average position $\langle \hat \X\rangle=(\X_\text{L}+\X_\text{R})/2=(0,0,-4250)$, depicted by the green dashed line. 
However, the WPI image in the SN model clearly vanishes as expected from Eq.~\eqref{eq:WPI_image_SN}. 

% Mass source at x-axis
\begin{figure}[ht]
    \centering
    \includegraphics[width=0.8\linewidth]{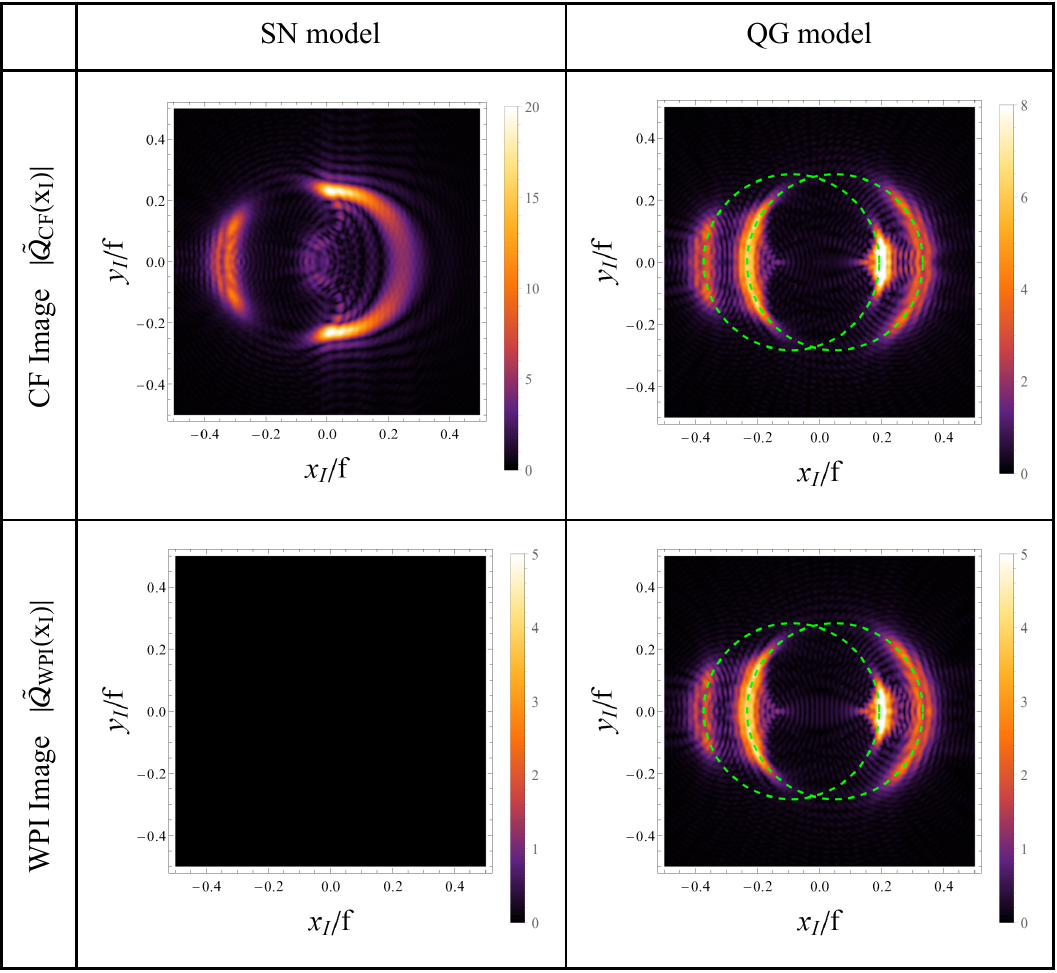}
    \caption{
    The CF images and WPI images for the SN and QG models, where the mass source is superposed at two distinct positions along the $x$-axis: $\X_\text{L}=(-450,0,0)$ and $\X_\text{R}=(250,0,0)$. In the QG model, the CF and WPI images both display the composition of the large and the small double arcs, corresponding to the image on the curved spacetime induced by the point mass source at $\X_\text{R}$ and $\X_\text{L}$, respectively. By contrast, in the SN model, the CF image shows a single image corresponding to the mass source at averaged position $\langle \hat \X\rangle=(\X_\text{L}+\X_\text{R})/2$, while the intensity in the WPI image disappears, as predicted by Eq.~\eqref{eq:WPI_image_SN}.
    }
    \label{fig:imagenonsymCFWPI_QG_SN}
\end{figure}
Let us also examine the case where the point mass source is superposed along the $x$-axis, corresponding to the perpendicular direction of the line of sight of the observer.
In Fig.~\ref{fig:imagenonsymCFWPI_QG_SN}, we present the CF and WPI images for both the SN and QG models. The upper left and right panels depict the CF images for the SN and QG models, respectively, while the lower left and right panels depict the WPI images for the SN and QG models, respectively. The mass source is in quantum superposition at two distinct positions along the $x$-axis, $\X_\text{L}=(-450,0,0)$ and $\X_\text{R}=(200,0,0)$. Other parameters, namely $\gamma=100$ and $\ell=200$, remain the same as in Fig.~\ref{fig:imageCFaxclassical}. We collected $200\times200$ data points to plot the images.
The important properties of this figure are nearly identical to those in Fig.~\ref{fig:imageaxCFWPI_QG_SN}; in the CF images, we observe a composition of classical images for the QG model, which reflects the superposition of the mass source, whereas the SN model displays only a single image reflecting the ensemble-averaged spacetime in one classical configuration.
Specifically, for the QG model, the panels depict the approximate position of the Einstein ring given by Eqs.~\eqref{eq:Einstein_ring_radius} and \eqref{eq:Einstein_ring_center} with a green dashed line. The left and right side rings correspond to the mass source at $\X_\text{L}$ and $\X_\text{R}$, respectively.
The images differ from Fig.~\ref{fig:imageaxCFWPI_QG_SN} in that each image appears as double arcs instead of a ring, which owing to the mass source configuration being separately superposed in the direction perpendicular to the line of sight of the observer.
In the QG model, the CF and WPI images both reveal the composition of two sets of double arcs. 
In the SN model, the CF image shows deformed double arcs, illustrating how light propagates through the ensemble-averaged spacetime described by a summation of spatially separated central potentials at $\X_\text{L}$ and $\X_\text{R}$ rather than a simple central potential, as depicted in Fig.~\ref{fig:SNevolution}.
In addition, in the QG model, we observe a weak vertical fringe pattern between two double arc images for the CF and WPI cases. The fringe pattern represents a beat effect caused by composing two images with a period given in Eq.~\eqref{eq:Einstein_ring_fringe}, each shifted by a distance of $\bm{O}_I|_{\X_\text{L}}$ and $\bm{O}_I|_{\X_\text{R}}$. This visualizes the interference between two propagations of the scalar field scattered by the mass source at the left and right and can be observed only within the framework of wave optics, beyond the analysis of null geodesics.
We observe a stronger interference fringe when the two states of the mass source, $|\bm{L}\rangle_\text{M}$ and $|\bm{R}\rangle_\text{M}$, are sufficiently close, as shown in Fig.~\ref{fig:imagesymCFWPI_QG_SN} in Appendix~\ref{apdx:more_images}.

Let us briefly summarize the highlights of the Einstein ring images in the curved spacetime induced by the superposed mass source, shown in Figs~\ref{fig:imageaxCFWPI_QG_SN} and \ref{fig:imagenonsymCFWPI_QG_SN}. 
The CF images vary depending on the gravity model: in the QG case, we observe the composition of two classical Einstein rings, reflecting the superposition of the mass source; however only a single Einstein ring is present in the SN case, indicating a single classical configuration of the ensemble-averaged spacetime. 
The WPI image reveals even more dramatic differences; in the QG case, it also displays the composition of two classical Einstein rings, whereas in the SN case, the image trivially vanishes. 
From a different perspective, we can also conclude that the WPI image intensity arises only when the gravitationally scattered light reaches the observer and gravity-induced entanglement is present. Thus, the WPI image serves as a useful tool for witnessing gravity-induced entanglement through the Einstein ring image, which contains substantial information about the background spacetime.

In Appendix~\ref{apdx:more_images}, we further explore how the image changes depending on the various configurations of the mass source state. 
Notably, in Fig.~\ref{fig:imagesymCFWPI_QG_SN}, we examine additional image results for the case where the mass source is symmetrically superposed in $x$-direction, such that $\X_\text{L}=-\X_\text{R}$. We highlight three interesting features observed in this case; First, the WPI image completely disappears even in the QG model. This is because detector A cannot capture the which-path information in this specific arrangement, as discussed in Fig.~\ref{fig:zeroWPI}. Second, when $\X_\text{L}$ and $\X_\text{R}$ are sufficiently close together, we observe a strong interference fringe between the two Einstein rings in the QG model. Third, the Poisson spot, which is the bright spot appearing in the center of the image, exhibits a different behavior depending on the gravity model. These features are extensively discussed in the Appendix~\ref{apdx:more_images}.

%%%%%----------------------------------------------
\section{Discussion}
\label{sec:discussion}

This section explore advanced topics relevant to our study, including the feasibility of our proposal, predictions from alternative semi-classical gravity models inducing gravitational state collapse, and the connections to quantum reference frames.

We first assess the feasibility of our proposal in two aspects: the distinguishability of the superposed Einstein rings and the value of the WPI indicator. 
To experimentally reproduce our setup, we consider injecting a high-intensity laser beam, such as the CoReLS petawatt laser~\cite{Yoon2021}, to observe gravitational scattering from a massive source object in a \schrodinger~cat state as described in Eq.~\eqref{eq:BMV_wave_function}.
The UDW detectors are configured as an Al$^+$ optical atomic clock system with a frequency of $\omega_D=1\times 10^{15}\,[Hz]$.
For the atomic clock to act as a laser photon detector, the laser wavelength must be $\lambda_\text{D}=2\pi c/\omega_\text{D}\sim 1\,[\mu m]$, which can be achieved using the CoReLS petawatt laser~\cite{Yoon2021}.
Assuming an optimistic scenario, we consider the mass source as a \schrodinger~cat state of a $16\,[\mu g]$ mechanical oscillator, as demonstrated in~\cite{Bild2023}. 
The detector is placed at a distance of $z_\text{D}=25[cm]$ from the mass source, based on prior work~\cite{Biswas2023} calculating graviton exchange between laser and matter. Furthermore, the detector is assumed to have a movable radius of $\ell=1\,[cm]$.

Subsequently, we outline two concerns for assessing experimental feasibility.
The experiment described in \cite{Bild2023} achieved superposition in the phonon mode at a high frequency of the order of $10^6\,[Hz]$; however, our setup treats the mass source as a quantum point particle with negligible time evolution, corresponding to a low frequency.
Second, we imposed the incoming wave at $z\to-\infty$ as the boundary condition of the massless scalar field in Eq.~\eqref{eq:boundary_condition}. By contrast, the experimental setup~\cite{Yoon2021} assumes that the laser is trapped and amplified within an optical cavity, leading to different boundary conditions for light.
Despite these distinctions, we use parameters from previous studies~\cite{Bild2023,Yoon2021} as a useful reference for our rough feasibility assessment.

%\blue{[In \cite{Biswas2023}, they consider the LIGO mirror as the mass source, since they need a quantum ground state, but not necessary a cat state. In contrast, our proposal a pure cat state, which cannot be realized with LIGO mirror for now.]}

First, we evaluate the required separation between the superposed mass source to distinguish the superposition of two Einstein ring images.
We consider that the mass source is represented by a cat state, as described in Eq.~\eqref{eq:BMV_wave_function}, along the $x$-axis as follows:
\begin{align}\label{eq:BMV_state_feasibility}
    \X_\text{L}=(\langle X\rangle_\text{M}-\Delta X,0,0),
    \qquad
    \X_\text{R}=(\langle X\rangle_\text{M}+\Delta X,0,0).
\end{align}
When the separation between the radii of the two Einstein rings is sufficiently larger than the variance of the ring radius, the two Einstein rings become distinguishable. This condition is expressed as
\begin{align}
    \left|\left.\bm{O}_I\right|_{\X_\text{M}=\X_\text{L}}
    -
    \left.\bm{O}_I\right|_{\X_\text{M}=\X_\text{R}}\right|
    ~\gtrsim~
    \left.\delta R_I\right|_{\X_\text{M}=\bm{0}}.
\end{align}
Using Eqs.~\eqref{eq:Einstein_ring_radius} and \eqref{eq:Einstein_ring_center}, we can rewrite the above condition in terms of the separation of the superposed mass source as follows:
\begin{align}
    \Delta X
    ~\gtrsim~
    \ell-\frac{3}{2}\lambda_D
    =
    1\,[cm]
    ~~
    \left\{
    \left(\frac{\ell}{1\,[cm]}\right)
    -\frac{3}{2}\left(\frac{\lambda_\text{D}}{1\,[\mu m]}\right)\right\}.
\end{align}
Currently, as reported in \cite{Biswas2023}, the cat state has been realized for a coherent parameter of $\alpha=1.75$, resulting in a separation of the order of $10^{-25}\,[m]$. 
Therefore, achieving the desirable value $1\,[cm]$ required to observe separable Einstein ring images is extremely challenging with current experimental status. 
Nevertheless, the realization of a \schrodinger-cat state over a large distance exceeding $1\,[cm]$ has been demonstrated in a different experimental setup~\cite{Kovachy2015}. This suggests that the feasibility of our proposal, while difficult, cannot be considered entirely out of reach.

Secondly, we evaluate the required separation of the cat state mass source necessary to observe the WPI indicator on the order of unity.
We assume that the mass source is again superposed along the $x$-axis as given in Eq.~\eqref{eq:BMV_state_feasibility}.
The WPI indicator, with detectors A and B positioned at the same point, is given by
\begin{align}
    \mathcal{Q}_\text{WPI}(\x_\text{A},\x_\text{A})
    = 
    \frac{|\alpha|^2}{4}
    \left|\varphi^{(\X_\text{L})}(\omega_\text{D},\x_\text{A})
    -
    \varphi^{(\X_\text{R})}(\omega_\text{D},\x_\text{A})\right|^2.
\end{align}
Here, we rescaled the mode function solution by the factor $|\alpha|^2$, which represents the mean photon number of the initial laser beam:
\begin{align}
    |\alpha|^2
    =\frac{4 I V_c}{\hbar \omega_\text{D} c}.
\end{align}
The volume of the optical cavity is given by $V_c \simeq 2 z_\text{D} \times \pi \tilde{w}^2$, where $\tilde{w}$ is the cavity waist. 
Using the asymptotic form of the mode function as provided in Eq.~\eqref{eq:Coulomb_wave_asymptotic} and assuming $|\Delta X/\langle X\rangle_\text{M}| \ll 1$, we obtain
\begin{align}
    \mathcal{Q}_\text{WPI}(\x_\text{A},\x_\text{A})
    = 
    \frac{2\gamma^2|\alpha|^2 \Delta X^2}{z_\text{D}}.
\end{align}
To arrive this form, we also used that $|\x_\text{A}-\X_\text{L,R}|-z_\text{D}$, $|\x_\text{A}-\X_\text{L,R}|$ and $|\X_\text{L,R}|$ are all on the order of $z_\text{D}$.
To detect the WPI indicator at order unity, specifically $\mathcal{Q}_\text{WPI}(\x_\text{A},\x_\text{A})\sim \mathcal{O}(1)$, we find that the separation of the cat state mass source should be
\begin{align}
    \Delta X \sim 2\times 10^{14}\,[m]
    \left( \frac{z_\text{D}}{25\,[cm]} \right)^{1/2}
    \left( \frac{M}{16\,[\mu g]} \right)^{-1}
    \left( \frac{\lambda_\text{D}}{1\,[\mu m]} \right)^{1/2}
    \left( \frac{\tilde{w}}{6\,[cm]} \right)^{-1}
    \left( \frac{I}{10^{-13}\,[W\cdot cm^{-2}]} \right)^{-1/2}.
\end{align}
Therefore, we need to prepare the mass source in a cat state with a separation of $2\times 10^{14}\,[m]$ to visualize the WPI image with an intensity of order unity%, where the ring radius is estimated to be $R_I\sim 1\times 10^{-17}$
. 
In contrast, the currently achievable separation is on the order of $10^{-25}\,[m]$~\cite{Biswas2023}.
Consequently, our proposal presents an incredibly challenging task, with the likelihood of observation nearly nonexistent at this stage.
This situation arises as we enter the relativistic regime of quantum gravitational interaction, which is expected to be significantly more difficult than testing the quantum entanglement induced by Newtonian gravity~\cite{Bose2017,Marletto2017}.Although further enhancements are necessary to conduct the experiment, our study remains valuable for exploring the relativistic phenomena that may arise from results validated in next-generation investigations of the BMV experiment~\cite{Bose2017,Marletto2017}.

Regarding the experimental realization, we also address the impact of environmental noise on our proposal. In realistic settings, even if the total system begins in a pure state, it generally evolves into a mixed state by interacting with the environment. For instance, light undergoing gravitational lensing could be disturbed by quantum interactions with the environment, such as photon-electron interactions. As a result, the Einstein ring image exhibits decoherence arising not only from QG interactions, but also from the environmental noise. In practice, such environmental decoherence is expected to be dominant, rendering the QG-induced decoherence indistinguishable.

Next, we discuss our expectations regarding alternative semi-classical gravity models in contrast to the \schrodinger-Newton gravity model.
In many of these alternative models, \cite{Diosi1989,Diosi2011,Tilloy2016,Penrose1996,Penrose2014,Kafri2014,Bassi2017,Carney2023,Oppenheim2023}, which do not generate gravity-induced entanglement like the SN model, the Newtonian gravitational interaction is treated as a classical noise term affecting the probe system. This interaction ultimately leads to the collapse of the probe quantum state into a mixed state in the position basis, a process often referred to as gravitational state collapse. This mechanism closely resembles the state collapse induced by measurement.
In the SN gravity model described in Section \ref{sec:time_evolution_SN}, we found that an initially pure state remains pure after time evolution. Conversely, in the alternative semi-classical gravity model that induces a gravitational state collapse, the total state evolves into a mixed state owing to the classical noise from gravitational interactions affecting the system. In this case, we naively anticipate that the time-evolved state can be expressed as follows:
\begin{align}
    \hat\rho(t)
    \sim
    \int d^3 X~
    |\mu(\X)|^2 ~
    |X\rangle_\text{M}\,{}_\text{M}\langle X|
    \otimes 
    |\psi^{(\X)}(t)\rangle_\text{SAB}\,
    {}_\text{SAB}\langle\psi^{(\X)}(t)|.
\end{align}
Here, $|\psi^{(\X)}(t)\rangle_\text{SAB}$ represents the time-evolved state of the scalar field and the two detectors, which have been gravitationally scattered by the point mass source located at $\X$. This formulation suggests that the time-evolved pure states $|\psi^{(\X)}(t)\rangle_\text{SAB}\,{}_\text{SAB}\langle\psi^{(\X)}(t)|$ for various mass source positions $\X$ are realized with classically mixed probabilities.
In alternative semi-classical gravity models, the emergence of the mixed states requires significant adjustments to our analysis, primarily regarding the following three points: First, we cannot use linear entropy to evaluate gravity-induced entanglement, as this measure is only a valid indicator for pure states.
Second, if we calculate the CF image for these semi-classical gravity models, we expect to observe a summation of multiple Einstein ring images, similar to the predictions in the QG case given in Eq.~\eqref{eq:CF_image_QG_SN}. The classical decoherence of the image in semi-classical gravity models arises from the classical mixture present in the final state.
Third, the WPI image will also display non-zero intensity, since the linear entropy of alternative semi-classical gravity models is generally non-zero, resulting in the non-zero value of the WPI indicator as presented in Eq.~\eqref{eq:logical_E_Q_2}.
Hence, we must consider an improved methodology to effectively distinguish the QG model and semi-classical gravity models that induce gravitational state collapse. For example, we could employ negativity~\cite{Vidal2002}, an entanglement indicator that is applicable to mixed states, to explore the gravity-induced entanglement between the mass source and other systems. In addition, rather than relying on the WPI indicator, we can develop an alternative quantity that is sensitive to differences among gravity models, referring to the expression of negativity.
Although these investigations are essential for examining various semi-classical gravity models, we will defer them to future work.

Finally, we discuss the connection to the concept of the quantum reference frame~\cite{Zych2018,Giacomini2019,Kabel2022}.
Quantum reference frames extend classical coordinate transformations by replacing the classical transformation parameter with a quantum operator. This approach has been applied to scenarios involving quantum superpositions of gravitational fields, as developed in recent studies~\cite{Foo2023relativity,Foo2021,Foo2022,Foo2023minkowski}. The following two scenarios are considered to be related to each other by such quantum-extended coordinate transformations: (i) a probe system in a specific state within a superposed gravitational field and (ii) a probe system in an appropriate quantum superposition within a purely classical gravitational field.
Now, let us apply the quantum reference frame transformation to our setup, reinterpreting it as a phenomenon within a purely classical gravitational field, following the discussion in \cite{Zych2018,Foo2021}. For clarity, we focus on the QG spacetime, assuming the mass source state is in a \schrodinger~cat state of two localized states as given in Eq.~\eqref{eq:BMV_mass_source}.
First, we introduce the operator $\hat T(\bm{\Delta})$, which describes a coordinate translation transformation with respect to the parameter $\Delta$, as $\x \to \x'=\x+\bm{\Delta}$. Applying this operator yields various translated quantities as follows:
\begin{align}
    &\hat T_\text{M}(\bm{\Delta})\,|\X\rangle_\text{M}
    =|\X+\bm{\Delta}\rangle_\text{M},\\
    &\hat T_\text{S}(\bm{\Delta})^\dagger\,
    \hat\phi^{(\X)}(t,\x)\,
    \hat T_\text{S}(\bm{\Delta})
    =\hat\phi^{(\X+\bm{\Delta})}(t,\x+\bm{\Delta}),\\
    &\hat T_\text{SAB}(\bm{\Delta})\,
    |\psi^{(\X)}(t;\x_\text{A},\x_\text{B})\rangle_\text{SAB}
    =|\psi^{(\X+\bm{\Delta})}(t;\x_\text{A}+\bm{\Delta},\x_\text{B}+\bm{\Delta})\rangle_\text{SAB}.
\end{align}
Here, the subscript of $\hat T(\bm{\Delta})$ indicates the quantum system to which this transformation is applied.
Now, we rewrite the time-evolved state in the QG spacetime, as given in Eq.~\eqref{eq:time_evolved_state_BMV}, as follows:
\begin{align}
    |\Psi^{(\text{QG})}(t)\rangle
    &=\frac{1}{\sqrt{2}}
    \left(
    |\bm{L}\rangle_\text{M} \, |\psi^{(\bm{L})}(t)\rangle_\text{SAB}
    + 
    |\bm{R}\rangle_\text{M} \, |\psi^{(\bm{R})}(t)\rangle_\text{SAB}
    \right)\notag\\
    &=\frac{1}{\sqrt{2}}
    \left(
    \hat{T}_\text{tot}(\bm{L})\,
    |\bm{0}\rangle_\text{M} \, 
    |\psi^{(0)}(t;\x_\text{A}-\bm{L},\x_\text{B}-\bm{L})\rangle_\text{SAB}
    + 
    \hat{T}_\text{tot}(\bm{R})\,
    |\bm{0}\rangle_\text{M} \, 
    |\psi^{(0)}(t;\x_\text{A}-\bm{R},\x_\text{B}-\bm{R})\rangle_\text{SAB}
    \right)\notag\\
    &=\hat{T}_\text{tot}(
    \x_\text{A}-\hat \x_\text{A}
    \delta \hat\x_\text{A})\,
    |0\rangle_\text{M}\otimes \frac{1}{\sqrt{2}}\left(
    |\psi^{(0)}(t;\x_\text{A}-\bm{L},\x_\text{B}-\bm{L})\rangle_\text{SAB}
    + 
    |\psi^{(0)}(t;\x_\text{A}-\bm{R},\x_\text{B}-\bm{R})\rangle_\text{SAB}
    \right).
\end{align}
In the second equality, each of the first and the second terms are rewritten using the translation operators $\hat{T}_\text{tot}(\bm{L}),~\hat{T}_\text{tot}(\bm{R})$, which act on the entire system. 
In the final step, we combine the two translation operators, $\hat{T}_\text{tot}(\bm{L})$ and $\hat{T}_\text{tot}(\bm{R})$, and collectively express them as $\hat{T}(\x_\text{A} - \hat{\x}_\text{A})$, representing the coordinate translation transformation relative to the position operator of detector A, $\hat{\x}_\text{A}$, offset by the original c-number position, $\x_\text{A}$.
This final expression represents the superposition of the UDW detectors A and B in a purely classical background spacetime induced by the point mass source sharply localized at the origin, with the entire system subject to a quantum-extended translation transformation based on $\hat\x_\text{A}$.
Thus, by applying the concept of the quantum reference frame, our analysis in the QG spacetime can be reinterpreted as a study of superposed UDW detectors in a classical background spacetime, as explored in a previous study~\cite{Foo2020}.

%%%%%----------------------------------------------
\section{Conclusion}
\label{sec:conclusion}

This study investigated gravitational lensing and the Einstein ring image resulting from a spatially superposed mass source. We considered a weak gravitational field characterized by two types of Newtonian potential: the quantized gravity (QG) model, which produces quantum entanglement between the mass source and probe system, and the \schrodinger-Newton (SN) gravity model, a semi-classical gravity model that does not produce entanglement. We treated a massless quantum scalar field as a toy model for light propagation in this background spacetime and introduced an observer utilizing two Unruh-DeWitt (UDW) detectors coupled to the scalar field (see Fig.~\ref{fig:setup}).
In Section~\ref{sec:observer_accessible_quantity}, we defined two quantities accessible to the observer through UDW detector measurements: the two-point correlation function (CF) of the scalar field and the which-path information (WPI) indicator. In particular, the WPI indicator was useful for witnessing gravitational entanglement, as shown in Eqs.~\eqref{eq:logical_E_Q} and \eqref{eq:logical_E_Q_2}.
Applying the imaging process detailed in Section~\ref{sec:image_formation} to these observer-accessible quantities, we obtained the Einstein ring image in the curved spacetime induced by the superposed mass source. The resulting images are shown in Figs.~\ref{fig:imageaxCFWPI_QG_SN} and \ref{fig:imagenonsymCFWPI_QG_SN}. In the case of the QG model, the images exhibited the composition of multiple Einstein rings, which can be interpreted as a manifestation of quantum decoherence. By contrast, in the SN model, the CF image revealed a single deformed Einstein ring, while the WPI image showed no intensity. 
Consequently, our findings offer a visual method for distinguishing the two gravity models through the Einstein ring images. In particular, the WPI image highlights a striking difference depending on the existence of gravity-induced entanglement. They reveal the composition of multiple Einstein ring images if and only if gravitational entanglement exists and otherwise vanishes.

In Section~\ref{sec:discussion}, we note that our analysis presents greater experimental challenges compared to proposals to test the entanglement produced by Newtonian gravity~\cite{Bose2017,Marletto2017}, as we explore their relativistic extensions at higher energy levels. Nevertheless, our proposal holds significant theoretical value, as it offers a practical formulation of the quantum field theory in non-dynamical curved spacetime induced by superposed mass source, and specifically presents how quantum effects of gravity arise in general relativistic phenomena.

Several avenues for future work can be considered.
First, we can explore applications to alternative semi-classical gravity models beyond the SN model, particularly those exhibiting the gravitational state collapse of the probe state, as briefly discussed in Section~\ref{sec:discussion}.
In addition, investigating applications to other general relativistic phenomena, such as Shapiro time delay and frame dragging, would be interesting.
Expanding our focus to different types of background curved spacetiems, especially strong gravitational fields like black holes spacetimes, as examined in \cite{Foo2022}, opens the door to even richer phenomena, including quasi-normal modes, photon spheres, and the Penrose process.

%%%%%%%%%%%%%%%%%%%%%%%%%%%%%%%%%%%%%%%%%%%%%%%%

\begin{acknowledgements}
We would like to thank Tomohiro Fujita, Akira Matsumura, and Igor Pikovski for fruitful discussions. Y.K. was supported by Grant-in-Aid for JSPS Fellows, and Y. N. was supported by JSPS KAKENHI (Grant No. 23K25871) and MEXT KAKENHI Grant-in-Aid for Transformative Research Areas A ``Extreme Universe" (Grant No. 24H00956). 
\end{acknowledgements}

%%%%%%%%%%%%%%%%%%%%%%%%%%%%%%%%%%%%%%%%%%%%%%%%
\appendix

%%%%%------------------------------------------
\section{Mode function solution}
\label{apdx:Coulomb_wave_function}

In this section, we show how to derive the mode function solution in the quantized gravity (QG) and \schrodinger-Newton (SN) spacetime.

%%%%%
\subsection{Coulomb wave function as the mode function solution in the QG spacetime}

Recall that the quantized gravity (QG) spacetime is described by the weak gravitational field with the following Newtonian potential, as given in Eq.~\eqref{eq:QG}:
\begin{align}
    &\Phi^{(\text{QG})}(\x,\hat\X)=\frac{-GM}{|\x-\hat \X|}
    .
\end{align}
The mode function of the scalar field operator in the QG spacetime follows the \schrodinger-like equation given in Eq.~\eqref{eq:Schrodinger_eq}.

First, let us focus on the case where the mass source is placed at the origin $\X=0$. The \schrodinger-like equation of the mode function is given by
\begin{align}\label{eq:schrodinger_equation_2}
    \left[
    \left(\frac{\partial}{\partial x}\right)^2+\left(\frac{\partial}{\partial y}\right)^2+\left(\frac{\partial}{\partial z}\right)^2
    + \frac{\omega^2}{c^2}\left(1+\frac{4GM}{c^2 |\x|}\right) \right]
    \varphi^{(0)}(\omega,\x)=0.
\end{align}
By introducing the new coordinate $\xi:=|\x|+z,~\eta=|\x|-z,~\phi=\arctan(y/x)$, the \schrodinger-like equation can be rewritten as 
\begin{align}\label{eq:schrodinger_equation_3}
    \left[
    \frac{\partial}{\partial\xi}\left(\xi\frac{\partial}{\partial \xi}\right)
    +\frac{1}{4\xi}\left(\frac{\partial}{\partial \phi}\right)^2
    +\frac{\omega^2}{4c^2} \xi 
    +\frac{\partial}{\partial\eta}\left(\eta\frac{\partial}{\partial \eta}\right)
    +\frac{1}{4\eta}\left(\frac{\partial}{\partial \phi}\right)^2
    +\frac{\omega^2}{4^2} \eta
    +\frac{\gamma \, \omega}{c}
    \right]
    \varphi^{(0)}(\omega,\x)=0,
\end{align}
where $\gamma := 2GM\omega/c^3$ is a dimensionless gravitational coupling constant. We now omit the $\phi$-derivative term due to the axisymmetry of the system. Since the $\xi$ and $\eta$ dependence in the above equation is separable, the general solution of $\varphi^{(0)}(\omega,\x)$ is given by
\begin{align}
    \varphi^{(0)}(\omega,\x)=\,f_1(\xi)f_2(\eta).
\end{align}
Substituting this form into Eq.~\eqref{eq:schrodinger_equation_3}, we obtain the equations for $f_1(\xi)$ and $f_2(\eta)$ as follows:
\begin{align}
    \left[
    \frac{d}{d\xi}\left(\xi\frac{d}{d\xi}\right)
    +\frac{\omega^2}{4c^2} \xi 
    +\frac{\gamma_1 \, \omega}{c}
    \right]f_1(\xi)
    &=0,
    \label{eq:f1EOM}\\
    \left[
    \frac{d}{d\eta}\left(\eta\frac{d}{d\eta}\right)
    +\frac{\omega^2}{4c^2} \eta
    +\frac{\gamma_2 \, \omega}{c}
    \right]f_2(\eta)
    &=0
    \label{eq:f2EOM},
\end{align}
where
\begin{align}\label{eq:gamma_consistency}
    \gamma_1+\gamma_2 &= \gamma
\end{align}
must be satisfied for consistency.

Let us also rewrite the boundary condition of the mode function presented in Eq.~\eqref{eq:boundary_condition}, where we assume an incoming plane wave for $z<0,~|\x|\to\infty$, in terms of the new coordinates $\xi$ and $\eta$\footnote{
$|\x|\to\infty$ is equivalent to the condition $\xi+\eta\to\infty$. When $z<0$, it follows that $\eta>\xi$. Therefore, $\eta\to\infty$ is sufficient for the original boundary condition.
}
\begin{align}
    \varphi^{(0)}(\omega,\x)\sim e^{i\omega (\xi-\eta)/(2c)}
    \quad
    \text{for}\quad
    \eta\to\infty\,
    .
\end{align}
Consequently, $f_1(\xi)$ and $f_2(\eta)$ must satisfy the following conditions:
\begin{align}
    f_1(\xi)&=e^{i\omega\xi/(2c)} \quad \text{for} \quad \forall\,\xi
    ,
    \label{eq:f1BC}\\
    f_2(\eta)&\sim e^{-i\omega\eta/(2c)} \quad \text{for} \quad \eta\to\infty
    .
    \label{eq:f2BC}
\end{align}
We can see that the $\xi$ dependence of the mode function, represented by $f_1(\xi)$, is determined solely by the boundary condition. 
Substituting Eq.~\eqref{eq:f1BC} into Eq.~\eqref{eq:f1EOM} yields
\begin{align}
    \gamma_1=-\frac{i}{2}
    .
\end{align}
Using Eq.~\eqref{eq:gamma_consistency}, we obtain
\begin{align}
    \gamma_2 = \frac{i}{2} +\gamma.
\end{align}
Finally, we are left with the equation for $f_2(\eta)$
\begin{align}
    \left[
    \frac{d}{d\eta}\left(\eta\frac{d}{d\eta}\right)
    +\frac{\omega^2}{4c^2} \eta
    +\left(\frac{i}{2} +\gamma\right)\frac{\omega}{c}
    \right]f_2(\eta)=0.
\end{align}
By rescaling $f_2(\eta)$ as
\begin{align}
    f_2(\eta) = e^{-i\omega\eta/(2c)} \tilde{f}_2(\eta),
\end{align}
we obtain the alternative equation to solve for the $\eta$ dependence of the mode function:
\begin{align}
    \eta \tilde{f}_2''(\eta) + \left(1-\frac{i\omega}{c}\eta\right)\tilde{f}_2'(\eta) 
    + \frac{\gamma\,\omega}{c}\tilde{f}_2(\eta)
    =0
    .
\end{align}
Choosing a solution that remains finite for all $\eta$, we arrive at the following analytic solution:
\begin{align}
    \tilde{f}_2(\eta) 
    = e^{\pi|\gamma|/2}\,\Gamma\!\left[1-i \gamma\right]
    {}_1 F_1\!\left[i\gamma\,;1\,;\frac{i\omega}{c}\eta\right],
\end{align}
where ${}_1 F_1\![a,b,z]$ denotes the confluent hypergeometric function of the first kind.
Finally, the analytic solution of the mode function when the mass source is located at the origin $\X=(0,0,0)$ is given by
\begin{align}
    \varphi^{(0)}(\omega,\x)
    =e^{\pi|\gamma|/2+i\omega z/c}\,
    \Gamma\!\left[1-i \gamma\right]\,
    {}_1 F_1\!\left[i\gamma\,;1\,;\frac{i\omega}{c}(r-z)\right].
\end{align}
Here, we normalized $\varphi^{(0)}(\omega,\x)$ to satisfy $\varphi^{(0)}(\omega,\x)=\varphi^{(0)}(-\omega,\x)^*$ and produce a plane wave with unit amplitude $e^{i\omega z/c}$ in the limit as $\gamma\to0$.

When $|r-z|\to\infty$, the asymptotic form of the solution is given by the following
\begin{align}\label{eq:Coulomb_wave_asymptotic}
    \varphi^{0}(\omega,\x)
    \simeq
    \left(1-\frac{c\gamma^2}{i\omega(r-z)}\right)
    e^{i\omega z/c-i\gamma \log\left[\omega(r-z)/c\right]}
    +\frac{\Gamma[1-i\gamma]}{\Gamma[i\gamma]}\frac{c}{i\omega(r-z)}e^{i\omega r/c+i \gamma \log\left[\omega(r-z)/c\right]}.
\end{align}
Even at sufficiently large distances from the origin, the solution 
is modified by a logarithmic correction to the simple plane wave propagating in the $z$ direction. The second term represents the outward spherical wave scattered by the point mass source.

Now, let us consider the case of an arbitrary $\X$. By applying a translation transformation with a non-zero parameter $+\X$, the mass source position is shifted to $\X$ , and the scalar field position $\x$ is transformed to $\x+\X$:
\begin{align}
    \varphi^{(\X)}(\omega,\x+\X)
    =\varphi^{(0)}(\omega,\x)
    .
\end{align}
By substituting $\x\to\x-\X$ to the above form, we obtain the following relational formula
\footnote{
This relational formula does not hold if an additional system breaks the translation symmetry. For example, in the scenario where the light is emitted from a localized light source and the boundary condition specifies outgoing flux around that source, the boundary condition would break the translation symmetry, invalidating the relational formula.
} connecting the mode function solutions for the different mass source positions:
\begin{align}
    \varphi^{(\X)}(\omega,\x)
    =\varphi^{(0)}(\omega,\x-\X)
    .
\end{align}
Consequently, this gives the mode function solution $\varphi^{(\X)}(\omega,\x)$ for the general mass source position $\X$, as explicitly provided in Eq.~\eqref{eq:mode_function_solution}.

\subsection{Mode function solution for the \schrodinger-Newton gravity model using the diffraction integral formula}
\label{apdx:mode_function_SN}

This section investigates the mode function solution for the \schrodinger-Newton (SN) gravity model, particularly when the mass source is not localized in the $x,y$-directions; the final expression is given in Eq.~\eqref{eq:mode_function_solution_SN_pathintegral}. 
The SN spacetime is given by a weak gravitational field with the following Newtonian potential, which was also given in Eq.~\eqref{eq:SN}:
\begin{align}
    \Phi^{(\text{SN})}(\x)
    =\langle\mu| \frac{-GM}{|\x-\hat \X|} |\mu\rangle_\text{M}
    =\int d^3 X |\mu(\X)|^2 \frac{-GM}{|\x-\X|}.
\end{align}
This serves as a single classical entity represented by the ensemble-averaged form of the QG potential over various mass source positions $\X$.
Unlike the QG spacetime, which can be described by a simple central potential, the SN spacetime has reduced symmetry. 
Consequently, determining the exact solution of the mode function in the SN spacetime is challenging, for which the equation of motion is given in Eq.~\eqref{eq:Schrodinger_eq_SN}.

We employ the diffraction integral formula presented in \cite{Schneider1999,Nakamura1999} to obtain the analytic solution of the mode function in the SN spacetime.
Our approach is based on two main assumptions.
First, we adopt the eikonal approximation, which asserts that $\omega^2\varphi/c^2 \gg \partial_z^2 \varphi$. This condition indicates that the length scale over which the mode function varies is sufficiently larger than the wavelength.
Second, we utilize the thin-lens approximation for gravitational lensing. This interprets that the mode function scattering occurs locally at $z\sim \langle \hat Z\rangle_\text{M}$.
Next, we demonstrate the detailed derivation of the mode function solution using the diffraction integral formula.

First, we rescale the mode function as follows:
\begin{align}
    \varphi^{(\text{SN})}(\x)=e^{i\omega z} F(\x)
    .
\end{align}
According to Eq.~\eqref{eq:Schrodinger_eq_SN}, the rescaled mode function then satisfies the following equation of motion:
\begin{align}
    \partial_z^2 F+\frac{2i\omega}{c} \partial_z F +\left(\partial_x^2+\partial_y^2\right)F
    -\frac{4\omega^2}{c^2} \Phi^{(\text{SN})}(\x) F=0
    .
\end{align}
By applying the eikonal approximation, $\omega^2 F/c^2\gg \partial_z^2 F$, we can neglect the first term of the left-hand side. This gives the approximate equation of motion as follows:
\begin{align}\label{eq:schrodinger_eq_SN_aprx}
    i\partial_z F
    \simeq \left[ -\frac{c}{2\omega}\left(\partial_x^2+\partial_y^2\right)
    +\frac{2\omega}{c} \Phi^{(\text{SN})}(\x)\right]
    F
    .
\end{align}
This equation resembles the time-dependent {\schrodinger} equation if we interpret $z$ as time, $\omega/c$ as mass, and $2\omega\Phi^{(\text{SN})}(\x)/c$ as the potential term.

Now, let us obtain the solution of Eq.~\eqref{eq:schrodinger_eq_SN_aprx} in the path integral formalism.
The ``Lagrangian" corresponding to this \schrodinger~equation is given by
\begin{align}
    L(z;\x_\perp,\dot{\x}_\perp)
    =\frac{\omega}{2c}\left|\dot{\x}_\perp\right|^2-\frac{2\omega}{c}\Phi^{(\text{SN})}(\x),
\end{align}
where the quantities with the subscript $\perp$ denotes the two-dimensional spatial vector in the $x,y$-direction: $\x_\perp=(x,y)$, $\dot{\x}_\perp=(\dot{x},\dot{y})$.
\begin{figure}[b]
    \centering
    \includegraphics[width=0.5\linewidth]{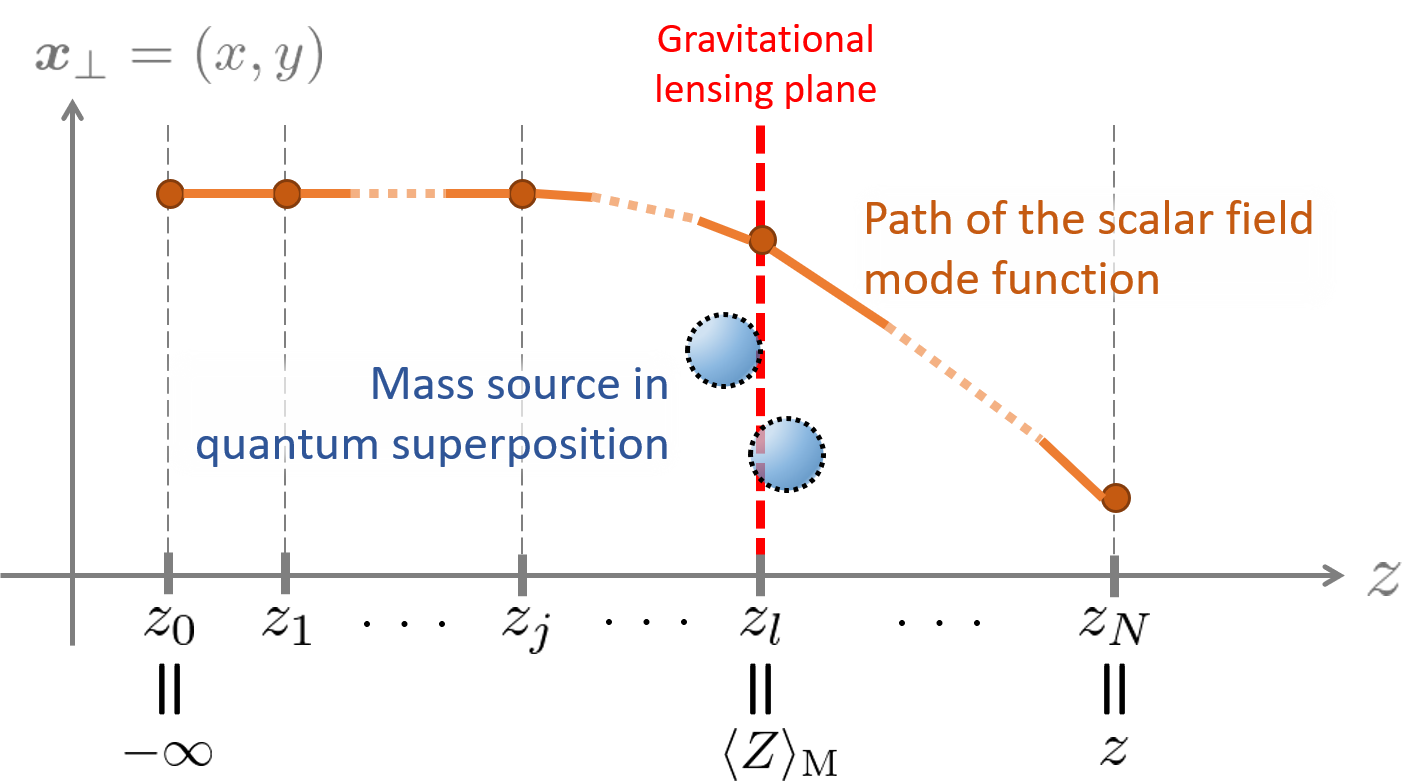}
    \caption{Path integral formalism is used to obtain the mode function solution in the SN spacetime.}
    \label{fig:path_integral}
\end{figure}
Using the path integral formalism, the solution is formally written as
\begin{align}\label{eq:path_integral}
    F(\x)
    =\int \mathcal{D}\x_\perp(z)
    \exp\left[i\int_{z_0}^{z}dz' L(z';\x_\perp,\dot{\x}_\perp)\right].
\end{align}
where $z_0$ determines the initial slice in the $z$-direction, corresponding to the position of the source of the propagating wave\footnote{
In the context of the gravitational lensing in astronomy, this usually corresponds to the position of the star emitting the light towards the earth.
}, in our case $z_0\to-\infty$, the position in which we adopt the boundary condition of incoming flux. 
Let us obtain the explicit form of Eq.~\eqref{eq:path_integral} by dividing the $z$-direction integral into $N$ small segments as depicted in Fig.~\ref{fig:path_integral}. We suppose that the initial and final slices are at $z_0=-\infty$ and $z_N=z$, respectively. In addition, we denote the dividing step in the $z$-direction as $\Delta z$. Then, Eq.~\eqref{eq:path_integral} can be rewritten as follows:
\begin{align}
    F(\x)
    &=\int \mathcal{D}\x_\perp(z)
    \exp\left[i
    \left(\int_{z_0}^{z_1}+\int_{z_1}^{z_2}+\cdots \int_{z_{N-1}}^{z_N}\right)
    dz' L(z';\x_\perp,\dot{\x}_\perp)\right]\notag\\
    &=\int \mathcal{D}\x_\perp(z)
    \lim_{N\to\infty}
    \exp\left[i
    \sum_{j=0}^{N-1}
    L(z_j;\x_{\perp,j},\dot{\x}_{\perp,j})
    \Delta z\right]\notag\\
    &=\lim_{N\to\infty}
    \left(\prod_{j=1}^{N-1}\int d^2 x_{\perp,j}\right)
    \exp\left[
    i\sum_{j=0}^{N-1}\left\{
    \frac{\omega}{2c}\left|\frac{\x_{\perp,j+1}-\x_{\perp,j}}{\Delta z}\right|^2
    -\frac{2\omega}{c}\Phi^{(\text{SN})}(\x_{\perp,j},z_j)
    \right\}\Delta z
    \right]\notag\\
    &=\lim_{N\to\infty}
    \left(\prod_{j=1}^{N-1}\int d^2 x_{\perp,j}\right)
    \exp\left[
    \frac{i\omega}{2c\Delta z}\sum_{j=0}^{N-1}
    \left|\frac{\x_{\perp,j+1}-\x_{\perp,j}}{\Delta z}\right|^2
    -\frac{2i\omega}{c}
    \int_{z_0}^z dz' \Phi^{(\text{SN})}(\x_{\perp,j},z')
    \right]
    .
\end{align}
Using the thin-lens approximation, we replace $\Phi^{(\text{SN})}(\x_{\perp,j},z')\to \Phi^{(\text{SN})}(\x_{\perp,l},z')$, where $j=l$ is the index corresponding to the lens plane of the gravitational lensing system; $z_l=\langle Z\rangle_\text{M}$. This implies that the phase shift induced by the gravitational lensing scattering is dominant at $z_l=\langle Z\rangle_\text{M}$
\begin{align}\label{eq:F}
    F(\x)
    &\simeq\lim_{N\to\infty}
    \left(\prod_{j=1}^{N-1}\int d^2 x_{\perp,j}\right)
    \exp\left[
    \frac{i\omega}{2c\Delta z}\sum_{j=0}^{N-1}
    \left|\frac{\x_{\perp,j+1}-\x_{\perp,j}}{\Delta z}\right|^2
    -\frac{2i\omega}{c}
    \int_{z_0}^z dz' \Phi^{(\text{SN})}(\x_{\perp,l},z')
    \right]
    .
\end{align}

As a preparatory step, let us first consider the case with no gravitational potential, $\Phi^{(\text{SN})}=0$. By straightforwardly performing the $\x_{\perp,j}$ integration, we derive the following expression:
\begin{align}
    F(\x)
    &\simeq\lim_{N\to\infty}
    \left(\prod_{j=1}^{N-1}\int d^2 x_{\perp,j}\right)
    \exp\left[
    \frac{i\omega}{2c\Delta z}\sum_{j=0}^{N-1}
    \left|\frac{\x_{\perp,j+1}-\x_{\perp,j}}{\Delta z}\right|^2
    \right]\notag\\
    &=\lim_{N\to\infty}
    \iint d^2x_{\perp,N-1} \cdots \iint d^2x_{\perp,2}~
    \exp\left[
    \frac{i\omega}{2c\Delta z}\sum_{j=2}^{N-1}
    \left|\frac{\x_{\perp,j+1}-\x_{\perp,j}}{\Delta z}\right|^2
    \right]\notag\\
    &\hspace{40mm}\times \iint d^2x_{\perp,1}
    \exp\left[\frac{i\omega}{2c\Delta z}\left(
    \left|\x_{\perp,2}-\x_{\perp,1}\right|^2+
    \left|\x_{\perp,1}-\x_{\perp,0}\right|^2
    \right)\right]\notag\\
    &=\lim_{N\to\infty}
    \iint d^2x_{\perp,N-1} \cdots \iint d^2x_{\perp,2}~
    \exp\left[
    \frac{i\omega}{2c\Delta z}\sum_{j=2}^{N-1}
    \left|\frac{\x_{\perp,j+1}-\x_{\perp,j}}{\Delta z}\right|^2
    \right]\notag\\
    &\hspace{40mm}\times 
    \left(\frac{-i\omega}{2\pi c\Delta z}\times 2\right)^{-1/2}
    \exp\left[
    \frac{i\omega}{4c\Delta z}
    \left|\x_{\perp,2}-\x_{\perp,0}\right|^2
    \right]\notag\\
    &=\lim_{N\to\infty}
    \left(\prod_{j=1}^{N-1}
    \frac{-i\omega}{2\pi c\Delta z}\frac{1+j}{j}\right)
    \exp\left[
    \frac{i\omega}{2 c N \Delta z}
    \left|\x_{\perp,N}-\x_{\perp,0}\right|^2
    \right]
    .
    \label{eq:F_no_potential}
\end{align}
In the third line, we performed a Gaussian integral over $\x_{\perp,1}$, and in the fourth line, we successively performed Gaussian integrals over $\x_{\perp,2},\cdots \x_{\perp,N-1}$. In addition, since the initial condition specifies an incoming plane wave traveling in the $z$-direction at $z=z_0$, we assume that the path originating from $\x_{\perp,0}$ remains parallel to the $z$-axis
and terminates at $\x_{\perp,N}$. Thus, it is natural to set $\x_{\perp,0}=\x_{\perp,N}$. Finally, we obtain
\begin{align}
    F(\x)
    &=\lim_{N\to\infty}\frac{1}{N}
    \left(\prod_{j=1}^{N-1}
    \frac{2\pi i c\Delta z}{\omega}\right).
    \label{eq:F_no_potential_2}
\end{align}

Next, before revisiting the path integral formalism in the presence of the SN potential given in Eq.~\eqref{eq:F}, we introduce the replacement as
\begin{align}
    \int d^2 x_{\perp,j}
    \to
    \int\frac{d^2 x_{\perp,j}}{2\pi i c \Delta z/\omega},
\end{align}
and rescale $F$ by a factor of $N$ as follows:
\begin{align}\label{eq:F_rescaled}
    F(\x)
    &\simeq \lim_{N\to\infty}
    N \left(\prod_{j=1}^{N-1}\int \frac{d^2 x_{\perp,j}}{2\pi i c \Delta z/\omega}\right)
    \exp\left[
    \frac{i\omega}{2c\Delta z}\sum_{j=0}^{N-1}
    \left|\frac{\x_{\perp,j+1}-\x_{\perp,j}}{\Delta z}\right|^2
    -\frac{2i\omega}{c}
    \int_{z_0}^z dz' \Phi^{(\text{SN})}(\x_{\perp,l},z')
    \right]
    .
\end{align}
This rescaling ensures that, in the absence of the SN gravitational interaction, $F$ approaches unity; thus, $\varphi^{(\text{SN})}$ becomes a plane wave with unit amplitude.

Now, let us perform the path integral in the presence of the SN potential.
Focusing on Eq.~\eqref{eq:F_rescaled}, we see that the integration over $\x_{\perp,1},\cdots , \x_{\perp,l-1}$ and $\x_{\perp,l+1},\cdots , \x_{\perp,N-1}$ do not depend on the SN potential, and thus can each be evaluated straightforwardly by performing Gaussian integrals, as in Eq.~\eqref{eq:F_no_potential}.
Specifically, the integration over $\x_{\perp,1},\cdots , \x_{\perp,l-1}$ yields a form similar to Eq.~\eqref{eq:F_no_potential_2}, while the integration over $\x_{\perp,l+1},\cdots , \x_{\perp,N-1}$ results in a form similar to the final expression in Eq.~\eqref{eq:F_no_potential}. However, to maintain consistency with the normalization introduced in Eq.~\eqref{eq:F_rescaled}, we must appropriately adjust the overall factor. The resulting form is given as follows:
\begin{align}
    F(\x)
    &\simeq \lim_{N\to\infty}
    N 
    \int \frac{d^2 x_{\perp,l}}{2\pi i c \Delta z/\omega}
    \exp\left[
    -\frac{2i\omega}{c}
    \int_{z_0}^z dz' \Phi^{(\text{SN})}(\x_{\perp,l},z')
    \right]
    \times \frac{1}{l} 
    \times \frac{1}{N-l}\exp\left[
    \frac{i\omega}{2c\Delta z}\frac{\left|\x_{\perp,N}-\x_{\perp,l}\right|^2}{N-l}
    \right]\notag\\
    &=\frac{\omega}{2\pi i c \Delta z}\frac{N}{l(N-l)}
    \int d^2 x_{\perp,l}
    \exp\left[
    \frac{i\omega}{c}\left\{
    \frac{\left|\x_{\perp,N}-\x_{\perp,l}\right|^2}{2(N-l)\Delta z}
    -2\int_{z_0}^z dz' \Phi^{(\text{SN})}(\x_{\perp,l},z')
    \right\}
    \right]
    .
\end{align}
Finally, using $l\Delta z=\langle Z\rangle_\text{M}-z_0$ and $N\Delta z=z-z_0$ with the limit $z_0\to-\infty$, we obtain the explicit form of Eq.~\eqref{eq:path_integral} as follows:
\begin{align}
    F(\x)
    &\simeq \frac{\omega}{2\pi i c}
    \frac{1}{z-\langle Z\rangle_\text{M}}
    \int d^2 x'_{\perp}
    \exp\left[\frac{i\omega}{c} T[\x'_\perp,\x_\perp] \right],\\
    T[\x'_\perp,\x_\perp]
    &=\frac{|\x'_\perp-\x_\perp|^2}{2(z-\langle \hat Z\rangle_\text{M})}
    -2\int_{z_0}^z dz' \Phi^{(\text{SN})}(\x_{\perp,l},z'),
\end{align}
where we replaced the integration variable $\x_{\perp,l}$ with $\x'_{\perp}$. 

Let us also evaluate the $z'$ integral within $T[\x'_\perp,\x_\perp]$. By substituting the explicit form of the SN potential, we can perform $z'$ as follows:
\begin{align}
    \int_{z_0}^z dz' \Phi^{(\text{SN})}(\x_{\perp,l},z')
    &=\int_{z_0}^z dz' 
    \int d^3X
    \left|\mu(\X)\right|^2
    \frac{-GM}{\sqrt{(x_l-X)^2+(y_l-Y)^2+(z'-Z)^2}}\notag\\
    &=-\frac{c \gamma}{\omega}
    \int d^3X \left|\mu(\X)\right|^2
    \log\left[\frac{|\x_\perp-\X_{\perp}|+z'-Z}{
    |\x_\perp-\X_{\perp}|-(z'-Z)}\right]_{z'=z_0}^{z'=z}
    .
\end{align}
We can expand the above expression by assuming $Z-z_0\to\infty$ and $z-Z\to\infty$. In this process, we neglect constant terms, which do not affect the overall behavior of the function, leading to the following form:
\begin{align}
    \int_{z_0}^z dz' \Phi^{(\text{SN})}(\x_{\perp,l},z')
    &\simeq
    \frac{c\gamma}{\omega}\int d^3X \left|\mu(\X)\right|^2
    \log\left|\x'_\perp-\X_\perp\right|
    .
\end{align}
Consequently, the mode function solution $\varphi(\x)$ in the SN spacetime is derived as follows:
\begin{align}
    \varphi^{(\text{SN})}(\x)
    &\simeq \frac{\omega}{2\pi i c}
    \frac{e^{i\omega z}}{z-\langle Z\rangle_\text{M}}
    \int d^2 x'_{\perp}
    \exp\left[\frac{i\omega}{c} T[\x'_\perp,\x_\perp] \right]
    ,
    \label{eq:varphi_SN_pathintegral}\\
    T[\x'_\perp,\x_\perp]
    &\simeq \frac{|\x'_\perp-\x_\perp|^2}{2(z-\langle \hat Z\rangle_\text{M})}
    -\frac{2c\gamma}{\omega}\int d^3X \left|\mu(\X)\right|^2
    \log\left|\x'_\perp-\X_\perp\right|
    .
\end{align}

In the actual calculation, we apply the stationary phase method to evaluate the $\x'_\perp$ integral in Eq.~\eqref{eq:varphi_SN_pathintegral}, assuming that $\omega$ is sufficiently large and that the integrand is highly oscillatory. First, we numerically determine the critical points $\bm{s}$ that satisfy the following conditions:
\begin{align}
    \left.\frac{\partial}{\partial x'_\perp}T[\x'_\perp,\x_\perp]\right|_{\x'_\perp=s_x}=0,
    \qquad
    \left.\frac{\partial}{\partial y'_\perp}T[\x'_\perp,\x_\perp]\right|_{\x'_\perp=s_y}=0.
\end{align}
The stationary phase approximation simplifies the mode function solution using these critical points to the following form:
\begin{align}\label{eq:varphi_SN_stationary_apprx}
    \varphi^{(\text{SN})}(\x)
    &\simeq 
    \frac{e^{i\omega z}}{i(z-\langle Z\rangle_\text{M})}
    \sum_{\bm{s}}\left|
    \det\left[\mathcal{H}[\bm{s},\x_\perp]\right]
    \right|^{-1/2}
    \exp\left[\frac{i\omega}{c} T[\bm{s},\x_\perp]
    +\frac{i\pi}{4}
    \mathrm{sgn}\left[\mathcal{H}[\bm{s},\x_\perp]\right]
    \right]
    .
\end{align}
Here, $\mathcal{H}[\x'_\perp,\x_\perp]$ is the Hessian matrix of $T[\x'_\perp,\x_\perp]$ defined as
\begin{align}
    \mathcal{H}[\x'_\perp,\x_\perp]
    =
    \begin{pmatrix}
    \partial_{x'_\perp}\partial_{x'_\perp}
    & 
    \partial_{x'_\perp}\partial_{y'_\perp}
    \\
    \partial_{y'_\perp}\partial_{x'_\perp}
    & 
    \partial_{y'_\perp}\partial_{y'_\perp}
    \end{pmatrix}
    T[\x'_\perp,\x_\perp] 
    .
\end{align}
Thus, by numerically identifying the critical points $\bm{s}$ on the two-dimensional $\x'_\perp$ plane and evaluating Eq.~\eqref{eq:varphi_SN_stationary_apprx}, we numerically obtain the amplitude of $\varphi^{(\text{SN})}(\x)$, which considers the reduced symmetry of the SN spacetime.

%%%%%------------------------------------------
\section{Calculate time-evolved state in the QG spacetime}
\label{apdx:Time_evolved_state}

Here, we demonstrate the derivation of the time-evolved state in the QG spacetime given in Eq.~\eqref{eq:time_evolved_state}:
\begin{align}
    &|\Psi(t)\rangle=\hat U_I^{(\hat\X)}(t)\,|\Psi_\text{ini}\rangle\notag\\
    &= \left(\int d^3 X ~ |X\rangle_\text{M} \, {}_\text{M}\langle X|\right)
    \hat U_I^{(\hat\X)}(t)\,|\mu\rangle_\text{M} |0\rangle_\text{S} |0\rangle_\text{A}|0\rangle_\text{B}\notag\\
    &= \int d^3 X ~ \mu(\X)|X\rangle_\text{M}\,\otimes 
    \hat U_I^{(\X)}(t)\,|\mu\rangle_\text{M} |0\rangle_\text{S} |0\rangle_\text{A}|0\rangle_\text{B}\notag\\
    &\simeq \int d^3 X ~ \mu(\X)|X\rangle_\text{M}\,\otimes 
    \left(1-\frac{i}{\hbar}\int dt \hat H_I^{(\hat \X)}(t)-\frac{1}{2\hbar^2}\int dt dt'\,
    \mathcal{T}\left[\hat H_I^{(\hat \X)}(t) \hat H_I^{(\hat \X)}(t')\right]\right)|0\rangle_\text{S} |0\rangle_\text{A}|0\rangle_\text{B}
    +\mathcal{O}(\lambda^3)\notag\\
    &\simeq \int d^3 X ~ \mu(\X)|X\rangle_\text{M}\notag\\
    &\hspace{10mm}\otimes
    \left[
    \left\{
    1
    -\frac{\lambda^2}{\hbar^2} \int dt dt'\,
    \theta(t-t')e^{i\omega_\text{D}(t-t')}\,\chi(t)\chi(t') 
    \sum_{j=\text{A,B}}
    \hat \phi^{(\X)}(t,\x_j)\hat \phi^{(\X)}(t',\x_j)
    \right\}
    |0\rangle_\text{A} |0\rangle_\text{B}
    \right.\notag\\
    &\hspace{20mm}\left.
    -\frac{i\lambda}{\hbar} \int dt \, e^{-i\omega_\text{D}t}\,\chi(t)\left(
    \hat \phi^{(\X)}(t,\x_\text{A})
    |1\rangle_\text{A} |0\rangle_\text{B}
    +\hat \phi^{(\X)}(t,\x_\text{B})
    |0\rangle_\text{A} |1\rangle_\text{B}
    \right)\right.\notag\\
    &\hspace{20mm}\left.
    -\frac{\lambda^2}{\hbar^2} \int dt dt'\,\theta(t-t')e^{i\omega_\text{D}(t+t')}\,\chi(t)\chi(t') 
    \left\{
    \hat \phi^{(\X)}(t,\x_\text{A}),\hat \phi^{(\X)}(t',\x_\text{B}))
    \right\}
    |1\rangle_\text{A} |1\rangle_\text{B}
    \right]|0\rangle_\text{S}
    +\mathcal{O}(\lambda^3)
    .
\end{align}
In the third line, we expand the time evolution operator, as described in Eq.~\eqref{eq:time_evolution_operator}, up to the second order in the scalar field-detector coupling constant $\lambda$. In the last line, we substitute the explicit form of the interaction Hamiltonian given in Eq.~\eqref{eq:Hamiltonian_I} and apply the Pauli operator of the detector, $\hat\sigma_{x,\text{A/B}}(t)$, to the detector state $|0\rangle_\text{A/B}$.

%%%%%------------------------------------------
\section{Detailed discussion about the which-path information indicator}
\label{apdx:WPI}

This section discusses the details of the which-path information (WPI) indicator $\mathcal{Q}_\text{WPI}(\x_A,\x_B)$, as introduced in Section~\ref{sec:WPI}. We will focus on two main topics: First, the proof of the logical expressions connecting the WPI indicator to gravity-induced entanglement, as presented in Eqs.~\eqref{eq:logical_E_Q} and \eqref{eq:logical_E_Q_2}, and second, the demonstration that the WPI indicator is an observer-accessible quantity, as shown in Eq.~\eqref{eq:WPI_observable}.

We begin by proving the logical expression presented in Eq.~\eqref{eq:logical_E_Q}, which states
\begin{align}
    \mathcal{E}(\x_\text{A},\x_\text{B})=0
    \quad\rightarrow\quad
    \mathcal{Q}_\text{WPI}(\x_\text{A},\x_\text{B})=0
    .
\end{align}
Since the case for the SN spacetime is trivial, we focus on the QG scenario.
Recall that the linear entropy in the QG spacetime is given by Eq.~\eqref{eq:GIE_simplified}, as follows:
\begin{align}
    \mathcal{E} (\x_\text{A},\x_\text{B})
    =\left(\int_{-\infty}^\infty dt_+\right)
    \frac{\pi\lambda^2}{\hbar^2}
    \theta(\omega_\text{D}) 
     \int d^3X d^3X'~|\mu(\X)|^2 |\mu(\X')|^2
     \sum_{j=\text{A},\text{B}}
     \left|\varphi^{(\X)}(\omega_\text{D},\x_j)-\varphi^{(\X')}(\omega_\text{D},\x_j)\right|^2
     =0
     .
\end{align}
If the overall factor $\left(\int_{-\infty}^\infty dt_+\right)\frac{\pi\lambda^2}{\hbar^2}\theta(\omega_\text{D})$ is non-zero, then the integrand must equal zero, as the integrand comprises non-negative quantities. Consequently, we have
\begin{align}
    \left|\varphi^{(\X)}(\omega_\text{D},\x_\text{A})-\varphi^{(\X')}(\omega_\text{D},\x_\text{A})\right|
    =
    \left|\varphi^{(\X)}(\omega_\text{D},\x_\text{B})-\varphi^{(\X')}(\omega_\text{D},\x_\text{B})\right|
    =0
    \qquad
    \text{for all }\X,~\X'
    .
\end{align}
Next, recalling that the WPI indicator in the QG spacetime is given in Eq.~\eqref{eq:WPI_QG}, we substitute the above results to obtain:
\begin{align}
    &\mathcal{Q}_\text{WPI}(\x_\text{A},\x_\text{B})\notag\\
    &=\frac{1}{2}\int d^3X\,d^3X' |\mu(\X)|^2 |\mu(\X')|^2
        \left(\varphi^{(\X)}(\omega_\text{D},\x_\text{A})-\varphi^{(\X')}(\omega_\text{D},\x_\text{A})\right)
        \left(\varphi^{(\X)}(\omega_\text{D},\x_\text{B})^*-\varphi^{(\X')}(\omega_\text{D},\x_\text{B})^*\right)\notag\\
    &= \frac{1}{2}\int d^3X\,d^3X' |\mu(\X)|^2 |\mu(\X')|^2
        \left|\varphi^{(\X)}(\omega_\text{D},\x_\text{A})-\varphi^{(\X')}(\omega_\text{D},\x_\text{A})\right|
        \left|\varphi^{(\X)}(\omega_\text{D},\x_\text{B})-\varphi^{(\X')}(\omega_\text{D},\x_\text{B})\right|\notag\\
    &\hspace{20mm}\times
    \exp\left[i\left(
    \arg\left[\varphi^{(\X)}(\omega_\text{D},\x_\text{A})-\varphi^{(\X')}(\omega_\text{D},\x_\text{A})\right]
    -\arg\left[\varphi^{(\X)}(\omega_\text{D},\x_\text{B})-\varphi^{(\X')}(\omega_\text{D},\x_\text{B})\right]
    \right)\right]\notag\\
    &=0
    .
\end{align}
Thus, we have successfully shown the logical expression in Eq.~\eqref{eq:logical_E_Q}.

Let us also prove the logical expression given in Eq.~\eqref{eq:logical_E_Q_2}, which states:
\begin{align}
    \mathcal{E}(\x_\text{A},\x_\text{A})=0
    \quad\leftrightarrow\quad
    \mathcal{Q}_\text{WPI}(\x_\text{A},\x_\text{B})=0
    \quad\text{for}~~\forall\,\x_\text{B}.
\end{align}
Since the SN spacetime case is trivial, we will focus on the QG case.
First, the implication from left to right can be demonstrated in a manner similar to our proof for Eq.~\eqref{eq:logical_E_Q}. Therefore, we will only present the proof of the right to left implication here.
According to Eq.~\eqref{eq:WPI_QG}, we have
\begin{align}
    &\mathcal{Q}_\text{WPI}(\x_\text{A},\x_\text{B})=0\notag\\
    &=\frac{1}{2}\int d^3X\,d^3X' |\mu(\X)|^2 |\mu(\X')|^2
        \left(\varphi^{(\X)}(\omega_\text{D},\x_\text{A})-\varphi^{(\X')}(\omega_\text{D},\x_\text{A})\right)
        \left(\varphi^{(\X)}(\omega_\text{D},\x_\text{B})^*-\varphi^{(\X')}(\omega_\text{D},\x_\text{B})^*\right).
\end{align}
Note that the integrand can take either positive or negative values depending on the choices of $\X$ and $\X'$. Therefore, we cannot conclude that the integrand must necessary be zero even if $\mathcal{Q}_\text{WPI}(\x_\text{A},\x_\text{B})=0$. However, if $\mathcal{Q}_\text{WPI}(\x_\text{A},\x_\text{B})=0$ holds for an arbitrary $\x_\text{B}$, it follows that
\begin{align}
    \varphi^{(\X)}(\omega_\text{D},\x_\text{A})-\varphi^{(\X')}(\omega_\text{D},\x_\text{A})=0.
\end{align}
Substituting this result into the expression for the linear entropy given in Eq.~\eqref{eq:GIE_simplified} with the detector positioned at $\x_\text{B}=\x_\text{A}$ yields
\begin{align}
    \mathcal{E}(\x_\text{A},\x_\text{A})=0
    .
\end{align}
Thus, we can also demonstrate the logical expression stated in Eq.~\eqref{eq:logical_E_Q_2}.

Finally, let us demonstrate how the WPI indicator us expressed in terms of observer accessible quantities, specifically $\mathrm{Tr}\left[\hat{\mathcal{O}} \, \hat\rho(t,\x_\text{A},\x_\text{B})\right]$, as shown in Eq.~\eqref{eq:WPI_observable}.
We provide the proof of the QG and SN cases in order subsequently.

First, for the QG case, the WPI indicator from Eq.~\eqref{eq:WPI_QG} can be rewritten as follows:
\begin{align}
    &\mathcal{Q}_\text{WPI}(\x_A,\x_B)\notag\\
    &=\int d^3X |\mu(\X)|^2
    \varphi^{(\X)}(\omega_\text{D},\x_\text{A})\varphi^{(\X)}(\omega_\text{D},\x_\text{B})^*\notag\\
    &\hspace{20mm}
    -\frac{1}{2}\int d^3X\,d^3X' |\mu(\X)|^2 |\mu(\X')|^2 \left(
    \varphi^{(\X)}(\omega_\text{D},\x_\text{A})\varphi^{(\X')}(\omega_\text{D},\x_\text{B})^*
    +\varphi^{(\X')}(\omega_\text{D},\x_\text{A})\varphi^{(\X)}(\omega_\text{D},\x_\text{B})^*
    \right)\notag\\
    &=\int d^3X |\mu(\X)|^2
    \varphi^{(\X)}(\omega_\text{D},\x_\text{A})\varphi^{(\X)}(\omega_\text{D},\x_\text{B})^*\notag\\
    &\hspace{20mm}
    -\frac{1}{2}\int d^3X\,d^3X' |\mu(\X)|^2 |\mu(\X')|^2 \left(
    \varphi^{(\X)}(\omega_\text{D},\x_\text{A})\varphi^{(\X)}(\omega_\text{D},\x_\text{B}+\X-\X')^*
    +\left(\X\leftrightarrow\X'\right)
    \right)\notag\\
    &=\int d^3X |\mu(\X)|^2
    \varphi^{(\X)}(\omega_\text{D},\x_\text{A})\varphi^{(\X)}(\omega_\text{D},\x_\text{B})^*\notag\\
    &\hspace{20mm}
    -\int d^3X\,d^3X' |\mu(\X)|^2 |\mu(\X')|^2 
    \varphi^{(\X')}(\omega_\text{D},\x_\text{A})\varphi^{(\X')}(\omega_\text{D},\x_\text{B}+\X'-\X)^*\notag\\
    &=\left(\left(\int_{-\infty}^\infty dt_+\right)
    \theta(\omega_\text{D}) \,
    \frac{\pi\lambda^2}{\hbar^2}\right)^{-1}
    \left(
    \mathrm{Tr}\left[\hat{\mathcal{O}} \, \hat\rho(t,\x_\text{A},\x_\text{B})\right]
    -\int d^3 X |\mu(\X)|^2 
    \mathrm{Tr}\left[\hat{\mathcal{O}} \, \hat\rho(t,\x_\text{A},\x_\text{B}+\hat \X-\X)\right]
    \right)
    .
\end{align}
Here, we start by expanding the WPI indicator for the QG case given in Eq.~\eqref{eq:WPI_QG}. Next, we used the translation property of the mode function from Eq.~\eqref{eq:translation} to rewrite the second term. In the third equality, we combined terms inside the parentheses by exchanging the integration parameters $\X$ and $\X'$. Finally, we rewrite each term in terms of observer-accessible quantities using Eq.~\eqref{eq:observable_expectation_value}.

For the SN case, we start with the final expression provided in Eq.~\eqref{eq:WPI_observable}, and show that it reduces to zero, aligning with the WPI indicator in the SN case as given in Eq.~\eqref{eq:WPI_SN}:
\begin{align}
    &\left(\left(\int_{-\infty}^\infty dt_+\right)
    \frac{\pi\lambda^2}{\hbar^2} \theta(\omega_\text{D})\right)^{-1}
    \left(
    \mathrm{Tr}\left[\hat{\mathcal{O}}_\text{AB}\, \hat\rho(t,\x_\text{A},\x_\text{B})\right]
    -\int d^3 X |\mu(\X)|^2 \,
    \mathrm{Tr}\left[\hat{\mathcal{O}}_\text{AB}\, \hat\rho(t,\x_\text{A},\x_\text{B}+\langle\hat \X\rangle_\text{M}-\X)\right]
    \right)\notag\\
    &\simeq
    \left(\left(\int_{-\infty}^\infty dt_+\right)
    \frac{\pi\lambda^2}{\hbar^2} \theta(\omega_\text{D})\right)^{-1}
    \left(\mathrm{Tr}\left[\hat{\mathcal{O}}_\text{AB}\, \hat\rho(t,\x_\text{A},\x_\text{B})\right]
    -
    \mathrm{Tr}\left[\hat{\mathcal{O}}_\text{AB}\, \hat\rho(t,\x_\text{A},\x_\text{B}+\langle\hat \X\rangle_\text{M}-\langle\hat \X\rangle_\text{M})\right]\right)
    +\mathcal{O}\left(\langle\hat \X^2\rangle_\text{M}\right)\notag\\
    &= 0 +\mathcal{O}\left(\langle\hat \X^2\rangle_\text{M}\right)\notag\\
    &=\mathcal{Q}_\text{WPI}(\x_A,\x_B) +\mathcal{O}\left(\langle\hat \X^2\rangle_\text{M}\right)
    .
\end{align}
In the second line, we transform the integral over the mass source position $\X$ into an expectation value of the mass source operator; specifically, we use the relation $\int d^3 X |\mu(\X)|^2 f(\X)=\langle f(\hat \X)\rangle_\text{M}$, where $f$ is an arbitrary function. We then apply the expansion of the expectation value as $\langle f(\hat \X)\rangle_\text{M} \simeq f(\langle\hat \X\rangle_\text{M}) + \mathcal{O}\left(\langle\hat \X^2\rangle_\text{M}\right)$. The third line can be demonstrated explicitly. In the final line, we express the result in terms of the WPI indicator for the SN case, using Eq.~\eqref{eq:WPI_SN}.
Thus, we have proven that the relational expression given in Eq.~\eqref{eq:WPI_observable} holds approximately for the SN case, by following the above equation manipulation in the reverse order.

In conclusion, the relational formula of the WPI indicator and the observer accessible quantities for both the QG and SN cases can be uniformly expressed as in Eq.~\eqref{eq:WPI_observable}.

%%%%%
\section{Imaging process using an optical convex lens in wave optics}
\label{apdx:image_formation}

\begin{figure}[htbp]
    \centering
    \includegraphics[width=0.45\linewidth]{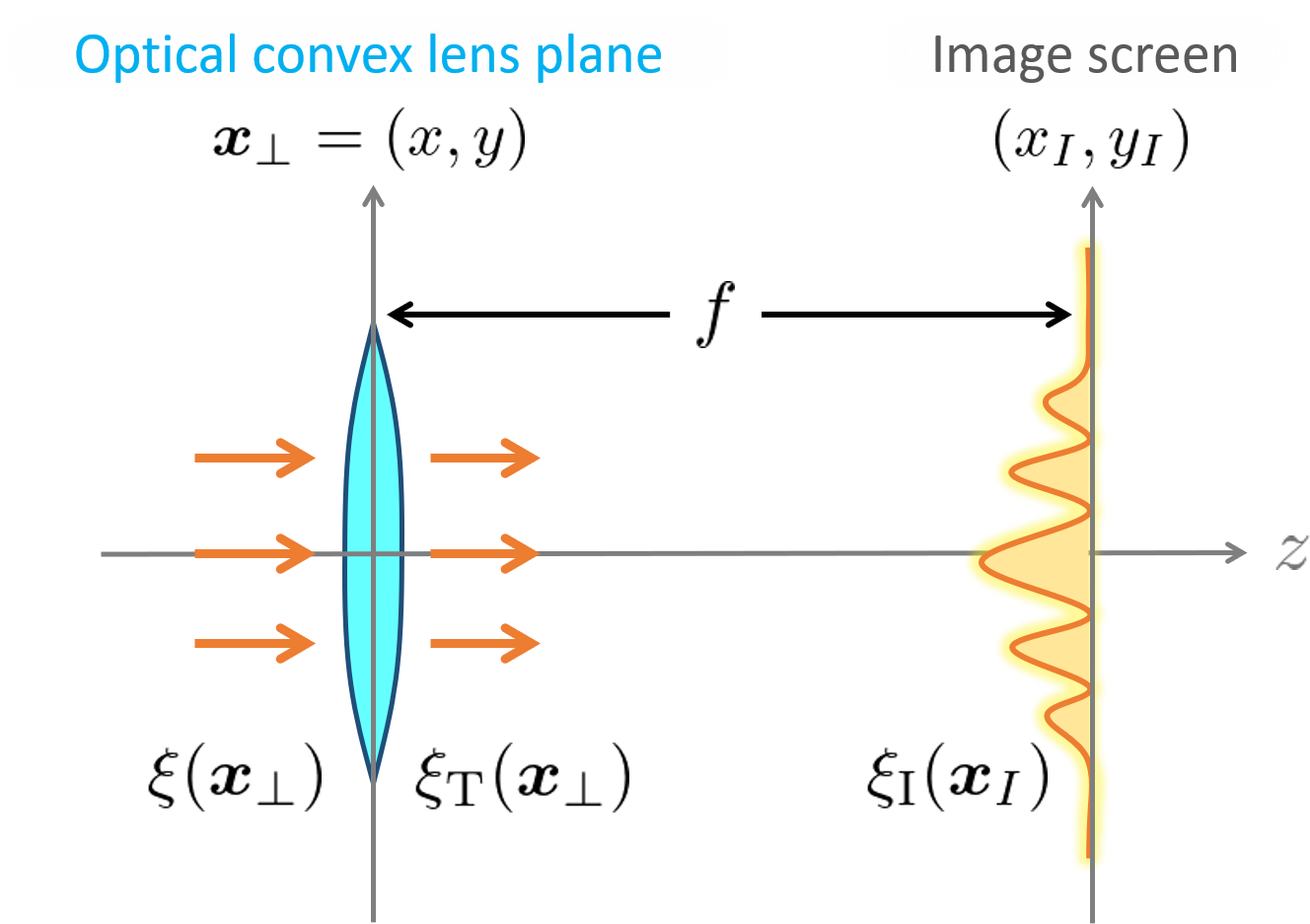}
    \caption{Imaging process using an optical convex lens.}
    \label{fig:optical_lens}
\end{figure}

In this section, we briefly review the imaging process using an optical convex lens, a concept well known in wave optics theory~\cite{Sharma2006,Hecht2012,Goodman2005,Kanai2013}, and commonly applied to explain image formation in telescopes or the human eye. The imaging process discussed in the main context in Section~\ref{sec:imaging} is inspired by this wave optical framework.
Specifically, we describe how an arbitrary incident wave with amplitude $\xi(x,y)$ is diffracted by an optical convex lens and how this diffraction leads to the image intensity $\xi_\text{I}(x_I,y_I)$ on an image screen, as depicted in Fig.~\ref{fig:optical_lens}. Note that in this section, we do not consider any gravitational mass sources; the lens discussed here refers to an optical convex lens and is unrelated to gravitational lensing.

First, we assume that the incident wave is propagating in the $z$ direction and is entering the optical convex lens perpendicularly. We denote the amplitude of the incident wave as $\xi(\x_\perp)$, where $\x_\perp=(x,y)$ represents the transverse coordinate. After diffraction by the lens, the amplitude of the transmitted wave, $\xi_\text{T}(\x_\perp)$ , is given by the following expression:
\begin{align}\label{eq:lens_Diffraction}
    \xi_\text{T}(\x_\perp)=e^{-i \omega |\x_\perp|^2/(2cf)} \xi(\x_\perp),
\end{align}
where $f$ is the focal length of the lens.
Next, we consider that the image screen is located at a distance $f$ from the lens, and the plane of the screen is characterized by the two-dimensional coordinate $\x_I=(x_I,y_I)$. According to the Fresnel-Kirchhoff's diffraction formula~\cite{Sharma2006}, the wave amplitude on the image screen, $\xi_\text{I}(\x_I)$, is given by
\begin{align}\label{eq:Fresnel-Kirchhoff}
    \xi_\text{I}(\x_I)
    \propto \int_\Sigma d^2 x_\perp \,
    \frac{e^{-i\omega L(\x_\perp,\x_I)}/c}{L(\x_\perp,\x_I)} \,
    \xi_\text{T}(\x_\perp),
\end{align}
where $L(\x_\perp,\x_I):=\sqrt{|\x_\perp-\x_I|^2+f^2}$ represents the distance between a point on the lens and a point on the image screen. The integral region $\Sigma$ denotes the finite region of the lens.
Substituting Eq.~\eqref{eq:lens_Diffraction} into Eq.~\eqref{eq:Fresnel-Kirchhoff} and applying the far-field approximation, $f\gg |\x_\perp-\x_I|$, we obtain
\begin{align}\label{eq:imaging_formula_optics}
    \xi_\text{I}(\x_I)
    \simeq \int_\Sigma d^3 x \,
    e^{-i\omega \x_\perp\cdot \x_I/(c f)} \,
    \xi(\x).
\end{align}
Here, the overall constant factor has been omitted. 
Hence, the image intensity observed through the optical convex lens can be evaluated based on the Fourier transformation of the incident wave.

In Section~\ref{sec:image_formation}, we discussed how to construct the Einstein ring image using observer-accessible quantities, such as the scalar field CF and the WPI indicator. Specifically, Eq.~\eqref{eq:imaging_formula} describes an imaging process that can be performed via the Fourier analysis of observer-accessible quantities. This process is similar to that used in optical convex lens imaging, as described in Eq.~\eqref{eq:imaging_formula_optics}.
Thus, the operation in Eq.~\eqref{eq:imaging_formula} represents how the incident wave, corresponding to the observer-accessible quantities, would appear as an image when observed through a convex lens.

Here, note that in Eq.~\eqref{eq:imaging_formula}, we use the opposite sign in the Fourier exponent compared to Eq.~\eqref{eq:imaging_formula_optics}. The minus sign in the Fourier exponent in Eq.~\eqref{eq:imaging_formula_optics} indicates that the image is inverted through the convex lens. However, we neglect this image inversion in the main context for simplicity.

%%%%%------------------------------------------
\section{Analytic expressions for the Einstein ring image in the classical spacetime}
\label{apdx:Einstein_ring_radius}

This section focuses on the Einstein ring image within classical spacetime and provides analytical formulas for the position, variance, and period of interference fringes of the Einstein ring.

Let us begin by considering the axisymmetric case, where the point mass source is positioned along the $z$-axis at $\X_\text{M}=(0,0,Z_\text{M})$.
According to Eq.~\eqref{eq:CF_image_classical}, the CF image intensity is explicitly given as follows:
\begin{align}
    \tilde{\mathcal{Q}}_\text{CF}(\x_I)
    &\propto 
        \iint_{\sqrt{x_\text{B}^2+y_\text{B}^2}\leq \ell} dx_\text{B} \, dy_\text{B} ~
        e^{i\omega_\text{D}(x_\text{B}x_I+y_\text{B}y_I)/(cf)}\,
        {}_1 F_1\!\left[-i\gamma\,;1\,;-\frac{i\omega_\text{D}}{c}\left\{|\x_\text{B}-(0,0,Z_\text{M})|-(z_\text{D}-Z_\text{M})\right\}\right]
        .
        \label{eq:CF_image_classical_1}
\end{align}
Next, we introduce polar coordinates for the integration variables as
\begin{align}
    &r_\text{B}=\sqrt{x_\text{B}^2+y_\text{B}^2},
    \qquad
    \phi_\text{B}=\arctan\left(\frac{y_\text{B}}{x_\text{B}}\right),\\
    &r_I=\sqrt{x_I^2+y_I^2},
    \qquad
    \phi_I=\arctan\left(\frac{y_I}{x_I}\right),
\end{align}
This transformation allows us to explicitly perform the integration over $\phi_\text{B}$, yielding the following expression:
\begin{align}
    \tilde{\mathcal{Q}}_\text{CF}(\x_I)
    &\propto 
    \int_0^\ell dr_\text{B} \int_0^{2\pi} d\phi_\text{B} ~
    r_\text{B}\,
    e^{i\omega_\text{D} r_\text{B} r_I \cos(\phi_\text{B}-\phi_I)/(cf)}\,
    {}_1 F_1\!\left[-i\gamma\,;1\,;-\frac{i\omega_\text{D}}{c}\left\{\sqrt{r_\text{B}^2+(z_\text{D}-Z_\text{M})^2}-(z_\text{D}-Z_\text{M})\right\}\right]\notag\\
    &=2\pi\int_0^\ell dr_\text{B}~
    r_\text{B}\,
    {}_1 F_1\!\left[-i\gamma\,;1\,;-\frac{i\omega_\text{D}}{c}\left\{\sqrt{r_\text{B}^2+(z_\text{D}-Z_\text{M})^2}-(z_\text{D}-Z_\text{M})\right\}\right]\,
    J_0\left[\frac{i\omega_\text{D} r_\text{B} r_I}{c f}\right]
    .
    \label{eq:CF_image_classical_2}
\end{align}
$J_\nu[z]$ is the Bessel function. 
The integral over $r_\text{B}$ is generally hard to evaluate directly, so we apply approximations to simplify the confluent hypergeometric function ${}_1 F_1$ as follows:
\begin{align}
    &{}_1 F_1\!\left[-i\gamma\,;1\,;-\frac{i\omega_\text{D}}{c}\left\{\sqrt{r_\text{B}^2+(z_\text{D}-Z_\text{M})^2}-(z_\text{D}-Z_\text{M})\right\}\right]\notag\\
    &\simeq {}_1 F_1\!\left[-i\gamma\,;1\,;
    -\frac{i\omega_\text{D}r_\text{B}^2}{2c(z_\text{D}-Z_\text{M})}\right]\notag\\
    &\simeq e^{-i\omega_\text{D} r_\text{B}^2/(4c (z_\text{D}-Z_\text{M}))}
    J_0\left[\sqrt{\frac{2\gamma\omega_\text{D}}{c(z_\text{D}-Z_\text{M})}}r_\text{B}\right]\notag\\
    &\simeq 
    \sqrt{\frac{2}{\pi}\sqrt{\frac{c (z_\text{D}-Z_\text{M})}{2\gamma \omega_\text{D}}}}\frac{e^{-i\omega_\text{D} r_\text{B}^2/(4c (z_\text{D}-Z_\text{M}))}}{\sqrt{r_\text{B}}}\cos\left[\sqrt{\frac{2\gamma\omega_\text{D}}{c (z_\text{D}-Z_\text{M})}}r_\text{B}-\frac{\pi}{4}\right]
    .
    \label{eq:confluentF_aprox}
\end{align}
In the second line, we have assumed $r_\text{B}\ll z_\text{D}-Z_\text{M}$ to simplify the argument.
For the third line, we utilized a property of the confluent hypergeometric function of the first kind: ${}_1 F_1\!\left[-\nu\,;1\,;\frac{z^2}{4\nu}\right]\to J_0(z)$ as $\nu\to\infty$ (for real $\nu$)\cite{Wolfram}. Although the first argument of the confluent hypergeometric function in the above equation is not a real number, we forcefully apply this property by approximation, verifying its accuracy through numerical consistency checks. Finally, in the last line, we used the condition $\sqrt{\frac{2\gamma\omega_\text{D}}{c(z_\text{D}-Z_\text{M})}}r_\text{B}\gg 1$, which allows us to approximate the Bessel function in cosine form.

Assuming $\omega_\text{D} r_\text{B} r_I/(cf)\gg 1$, we also approximate the Bessel function in the CF image expression Eq.~\eqref{eq:CF_image_classical_2} as a cosine function
\begin{align}\label{eq:Bessel_aprx}
    J_0\left[\frac{i\omega_\text{D} r_\text{B} r_I}{c f}\right]
    \simeq 
    \sqrt{\frac{2 c f}{\pi \omega_\text{D} r_\text{B} r_I}}\cos\left[\frac{\omega_\text{D} r_\text{B} r_I}{c f}-\frac{\pi}{4}\right]
    .
\end{align}
By substituting the approximations from Eqs.~\eqref{eq:confluentF_aprox} and \eqref{eq:Bessel_aprx} into our initial CF image expression Eq.~\eqref{eq:CF_image_classical_2}, we can perform the integral over $r_\text{B}$ to arrive at the following form:
\begin{align}
    &\tilde{\mathcal{Q}}_\text{CF}(\x_I)
    \simeq 
    \int_0^\ell dr_\text{B}~
    e^{-i\omega_\text{D} r_\text{B}^2/(4c  (z_\text{D}-Z_\text{M}))}
    \cos\left[\sqrt{\frac{2\gamma\omega_\text{D}}{c (z_\text{D}-Z_\text{M})}}r_\text{B}-\frac{\pi}{4}\right]\,
    \cos\left[\frac{\omega_\text{D}}{c}r_\text{B} \frac{r_I}{f}-\frac{\pi}{4}\right]\notag\\
    &\propto
    2\,e^{2i\sqrt{2\gamma\omega_\text{D} (z_\text{D}-Z_\text{M})}r_I/(cf)}\,
    \mathrm{erf}\!\left[(-1)^{1/4}\sqrt{\frac{\omega_\text{D} (z_\text{D}-Z_\text{M})}{c}}\left(r_I+\sqrt{\frac{2c\gamma}{\omega_\text{D} (z_\text{D}-Z_\text{M})}}\right)\right]\notag\\
    &\hspace{7mm}+
    \left\{
    e^{2i\sqrt{2\gamma\omega_\text{D} (z_\text{D}-Z_\text{M})}r_I/(cf)}
    \,
    \mathrm{erf}\!\left[(-1)^{1/4}\sqrt{\frac{\omega_\text{D} (z_\text{D}-Z_\text{M})}{c}}\left(r_I+\left(\sqrt{\frac{2c\gamma}{\omega_\text{D} (z_\text{D}-Z_\text{M})}}+\frac{\ell}{2z_\text{D}}\right)\right)\right]
    -i\left(r_I\to -r_I\right)
    \right\}\notag\\
    &\hspace{7mm}+\left\{
    e^{2i\sqrt{2\gamma\omega_\text{D} (z_\text{D}-Z_\text{M})}r_I/(cf)}
    \,
    \mathrm{erf}\!\left[(-1)^{1/4}\sqrt{\frac{\omega_\text{D} (z_\text{D}-Z_\text{M})}{c}}\left(r_I+\left(\sqrt{\frac{2c\gamma}{\omega_\text{D} (z_\text{D}-Z_\text{M})}}-\frac{\ell}{2z_\text{D}}\right)\right)\right]
    +i\left(r_I\to -r_I\right)
    \right\}
    .
    \label{eq:CF_image_classical_3}
\end{align}
Here, $\mathrm{erf}[\cdots]$ represents the error function.
Although the result may appear complicated, focusing on $r_I$ dependence reveals characteristic length scales in the image intensity. Specifically, we find that the error functions have roots around the following values of $r_I$:
\begin{align}
    r_I \sim \sqrt{\frac{2c\gamma}{\omega_\text{D} (z_\text{D}-Z_\text{M})}}\pm\frac{\ell}{2 (z_\text{D}-Z_\text{M})}
    .
\end{align}
This relationship highlights the primary length scales in the image intensity. Comparing these scales to the numerical plot of the image intensity, we find that the first term closely corresponds to the Einstein ring radius, while the second term explains the variance around it. This analysis yields the following analytic expressions for the Einstein ring radius and its variance, as also given in Eq.~\eqref{eq:Einstein_ring_radius}:
\begin{align}\label{eq:Einstein_ring_radius_apdx}
    R_I=\sqrt{\frac{2c\gamma}{(z_\text{D}-Z_\text{M})\omega_\text{D}}},
    \qquad
    \delta R_I = \frac{1}{z_\text{D}-Z_\text{M}}\left(\ell-\frac{3\pi c}{\omega_\text{D}}\right)
    .
\end{align}
We will demonstrate shortly that these approximations align well with the numerical CF image.

Next, we consider the case where the mass source is positioned at an arbitrary location, $\X_\text{M}=(X_\text{M},Y_\text{M},Z_\text{M})$. Here, with broken axisymmetry, it would be increasingly complicated to perform both $x_\text{B}$ and $y_\text{B}$ integrals directly. 
Thus, we only focus on the $x_\text{B}$ integral to anticipate how the Einstein ring image shifts as the mass source moves further from the $z$-axis.

Starting again from the CF image intensity on the classical spacetime provided in Eq.~\eqref{eq:CF_image_classical}, we proceed to simplify the integrand to perform the $x_\text{B}$ integral as follows:
\begin{align}
    \tilde{\mathcal{Q}}_\text{CF}(\x_I)
    &\propto 
        \iint_{\sqrt{x_\text{B}^2+y_\text{B}^2}\leq \ell} dx_\text{B} \, dy_\text{B} ~
        e^{i\omega_\text{D}(x_\text{B}x_I+y_\text{B}y_I)/(cf)}\,
        {}_1 F_1\!\left[-i\gamma\,;1\,;-\frac{i\omega_\text{D}}{c}\left\{|\x_\text{B}-\X_\text{M}|-(z_\text{D}-Z_\text{M})\right\}\right]\notag\\
    &\simeq\iint_{\sqrt{x_\text{B}^2+y_\text{B}^2}\leq \ell} dx_\text{B} \, dy_\text{B} ~
        e^{i\omega_\text{D}(x_\text{B}x_I+y_\text{B}y_I)/(cf)}\,
        e^{-i\omega_\text{D} \left\{(x_\text{B}-X_\text{M})^2+(y_\text{B}-Y_\text{M})^2+(z_\text{D}-Z_\text{M})^2\right\}/(4c (z_\text{D}-Z_\text{M}))}\notag\\
        &\hspace{20mm}\times 
        \left\{(x_\text{B}-X_\text{M})^2+(y_\text{B}-Y_\text{M})^2+(z_\text{D}-Z_\text{M})^2\right\}^{-1/4}\notag\\
        &\hspace{20mm}\times
        \cos\left[\sqrt{\frac{2\gamma\omega_\text{D}}{c (z_\text{D}-Z_\text{M})}}\sqrt{(x_\text{B}-X_\text{M})^2+(y_\text{B}-Y_\text{M})^2+(z_\text{D}-Z_\text{M})^2}-\frac{\pi}{4}\right]\notag\\
    &\simeq 
    \iint_{\sqrt{x_\text{B}^2+y_\text{B}^2}\leq \ell} dx_\text{B} \, dy_\text{B} ~
        e^{i\omega_\text{D}(x_\text{B}x_I+y_\text{B}y_I)/(cf)}
        e^{-i\omega_\text{D} \left\{(x_\text{B}-X_\text{M})^2+(y_\text{B}-Y_\text{M})^2+(z_\text{D}-Z_\text{M})^2\right\}/(4c (z_\text{D}-Z_\text{M}))}\notag\\
        &\hspace{20mm}\times (z_\text{D}-Z_\text{M})^{-1/2}
        \cos\left[\sqrt{\frac{2\gamma\omega_\text{D}}{c} (z_\text{D}-Z_\text{M})}
        \left(1+\frac{(x_\text{B}-X_\text{M})^2+(y_\text{B}-Y_\text{M})^2}{2(z_\text{D}-Z_\text{M})^2}\right)
        -\frac{\pi}{4}\right]
        .
\end{align}
In the second line, we approximate the confluent hypergeometric function following a similar procedure to Eq.~\eqref{eq:confluentF_aprox}. In the last line, we used $z_\text{D}-Z_\text{M}\gg \sqrt{(x_\text{B}-X_\text{M})^2+(y_\text{B}-Y_\text{M})^2}$, retaining only the leading term in the expansion of each integrand argument.
Integrating over $x_\text{B}$ yields a complex form involving multiple error functions similar to Eq.~\eqref{eq:CF_image_classical_3}. Notably, examining the $x_I$-dependence of these error functions reveals a consistent reliance on the form
\begin{align}
    x_I -\frac{X_\text{M}}{2(z_\text{D}-Z_\text{M})}.
\end{align}
This structure suggests that the $X_\text{M}$ dependence of the image intensity may be interpreted as a shift of $x_I\to x_I -\frac{X_\text{M}}{2(z_\text{D}-Z_\text{M})}$; a similar reasoning applies to the $Y_\text{M}$-dependence. Consequently, we propose that the center of the Einstein ring shifts according to the following relation, depending on the mass source position, as presented in Eq.~\eqref{eq:Einstein_ring_center}:
\begin{align}\label{eq:Einstein_ring_center_apdx}
    \bm{O}_I = \frac{1}{2(z_\text{D}-Z_\text{M})}\times (X_\text{M},Y_\text{M}).
\end{align}
Indeed, this ansatz aligns well with the numerical results for the CF image, confirming its validity.

We also discuss the period of the interference fringe observed around the peak of the Einstein ring in the image. This fringe pattern originates from the Fourier exponent used in the imaging process, as given in Eq.~\eqref{eq:imaging_formula}, leading to oscillations with the period:
\begin{align}\label{eq:Einstein_ring_fringe_apdx}
    \frac{\omega_\text{D} \ell}{c} \times \frac{|\x_I|}{f} =2\pi {n},
\end{align}
where $n$ is an arbitrary integer.

\begin{figure}[htbp]
    \centering
    \includegraphics[width=0.7\linewidth]{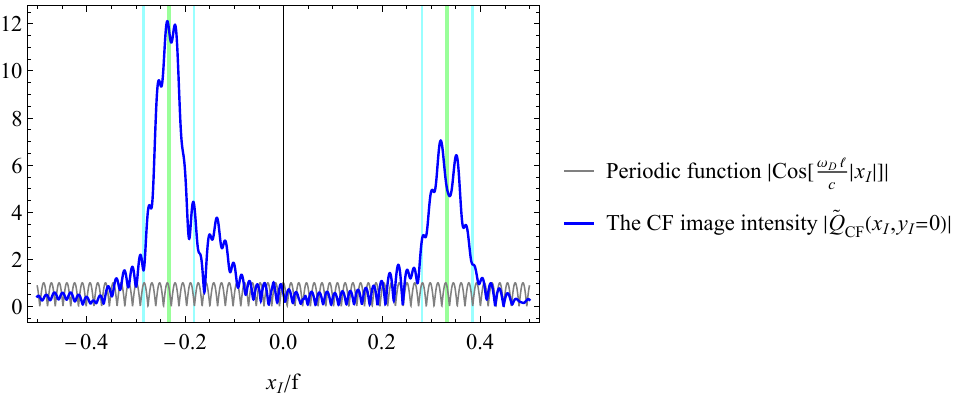}
    \caption{The numerical plot of the CF image intensity and its comparison with the analytically evaluated position of the Einstein ring.}
    \label{fig:compare_numerical_analytic}
\end{figure}

In Fig.~\ref{fig:compare_numerical_analytic}, we show a numerical plot of the CF image intensity and the approximate evaluations for the position, variance and interference fringe of the Einstein ring.
The horizontal axis represents the $x$-coordinate of the image, $x_I/f$, with $y_I$ fixed at zero. The parameters are set as $\gamma=100,~\ell=200$ and $z_\text{D}=2500$, normalized to a unit length scale of $c/\omega_\text{D}$. 
The blue line shows the cross-section of the CF image intensity $|\tilde{\mathcal{Q}}_\text{CF}(x_I,y_I=0)|$. Two vertical green grid lines indicate the position of the Einstein ring: $R_I+O_{I,x}$ for the right peak and $-R_I+O_{I,x}$ for the left peak. 
The cyan vertical grid lines show the variance of the Einstein ring position: $R_I+O_{I,x}\pm \delta R_I/2$ for the right peak and $-R_I+O_{I,x}\pm \delta R_I/2$ for the left peak. We observe that the evaluated positions and variances align well with the numerical data. Furthermore, we plot a cosine function with the period given by Eq.~\eqref{eq:Einstein_ring_fringe}: $\cos\left[\frac{\omega_\text{D}\ell}{c}|x_I|\right]$. This oscillatory pattern matches the numerical plot, confirming that the interference fringe is well described by Eq.~\eqref{eq:Einstein_ring_fringe}.

%%%%%------------------------------------------
\section{The Einstein ring images for various mass source configurations}
\label{apdx:more_images}

In the final part of this study, we present additional numerical results for the Einstein ring images in the curved spacetime induced by a superposed mass source.
Specifically, we provide a grid of the plots to illustrate the dependence on the configuration of the mass source.
Throughout this section, we assume that the mass source is in a \schrodinger~cat state, with the wave function given by Eq.~\eqref{eq:BMV_wave_function} as follows:
\begin{align}%\label{eq:BMV_wave_function}
    |\mu(\X)|^2
    = \frac{1}{2}\left( \delta^{(3)}(\X-\X_\text{L}) +\delta^{(3)}(\X-\X_\text{R}) \right).
\end{align}
The parameters are commonly set as $\gamma=100,~\ell=200$ and $z_\text{D}=2500$, normalized by the unit length scale of $c/\omega_\text{D}$. 

\begin{figure}[htbp]
    \centering
    \includegraphics[width=1\linewidth]{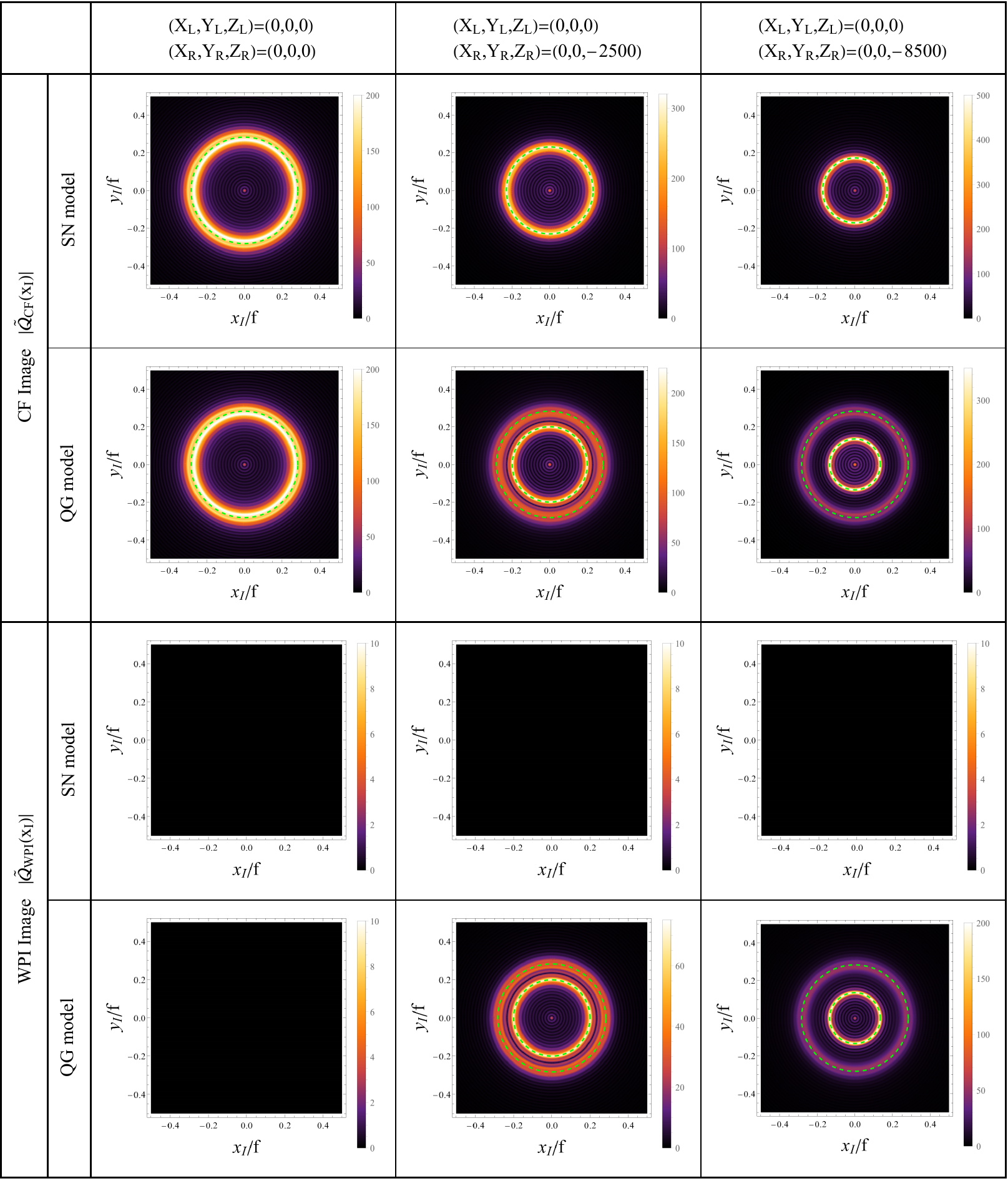}
    \caption{
    The Einstein ring images for the mass source superposed along the $z$-axis, aligned with the line of sight of the observer. The first and second rows show the CF image $|\tilde{Q}_\text{CF}(\x_I)|$ in the SN and QG spacetimes, respectively, while the third and fourth rows display the WPI image $|\tilde{Q}_\text{WPI}(\x_I)|$ in the SN and QG spacetimes, respectively. In the left, middle and right columns, we show the three different mass source configurations.
    }
    \label{fig:imageaxCFWPI_QG_SN_table}
\end{figure}

Let us first focus on the case where the mass source is superposed along the $z$-axis, aligned with the line of sight of the observer.  
In Fig.~\ref{fig:imageaxCFWPI_QG_SN_table}, we present the CF and WPI images for the QG and SN spacetimes. 
The first and second rows show the CF image $|\tilde{Q}_\text{CF}(\x_I)|$ in the SN and the QG spacetimes, respectively, while the third and fourth rows display the WPI image $|\tilde{Q}_\text{WPI}(\x_I)|$ in the SN and QG spacetimes, respectively. We show three mass source configurations in the left, middle, and right columns: $\X_\text{R}=(0,0,0),~(0,0,-2500)$ and $(0,0,-8500)$, with $\X_\text{L}$ fixed at the origin $(0,0,0)$. 
First, focusing on the CF images, we observe the following two key features:
(i) In the SN spacetime, each CF image reveals a single Einstein ring, which aligns with the Einstein ring for the ensemble-averaged mass source position $\langle \hat\X\rangle_\text{M}=(\X_\text{L}+\X_\text{R})/2$, as indicated by the green curve in each panel.
(ii) In the QG spacetime, the overlap of the two Einstein rings in the CF image varies depending on the mass source configuration. Specifically, a single Einstein appears in the left panel, where the localized states of the mass source $|\bm{L}\rangle_\text{M}$ and $|\bm{R}\rangle_\text{M}$ are positioned at the same point. In the middle and right panels, we observe the composition of two Einstein rings, each matching well with the classical Einstein ring positions for the mass source at $|\bm{L}\rangle_\text{M}$ and $|\bm{R}\rangle_\text{M}$, also indicated by green curves.
Next, we discuss the two key features observed in the WPI images.
(i) In the SN spacetime, the WPI image intensity vanishes in all cases, as explained in Eq.~\eqref{eq:WPI_image_SN}. This suggests that the SN spacetime does not produce gravity-induced entanglement, leading the WPI indicator to be zero, as given in Eq.~\eqref{eq:WPI_SN}.
(ii) In the QG model, the WPI image on the left panel also vanishes, as the mass source is not in a quantum superposition for this particular set of parameters. By contrast, the composition of two Einstein rings appears similarly to the CF images in the middle and the right panels.

\begin{figure}[htbp]
    \centering
    \includegraphics[width=1\linewidth]{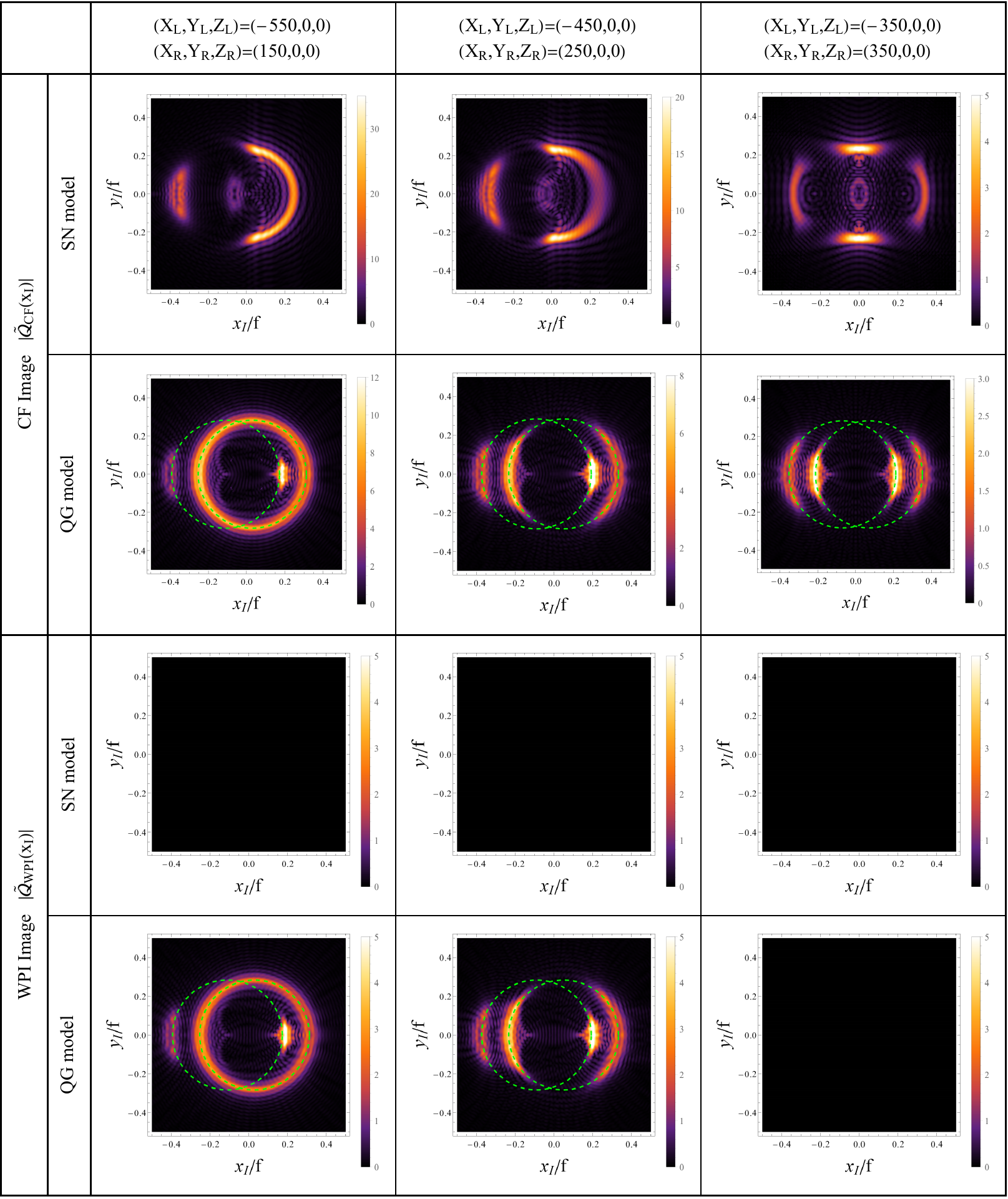}
    \caption{
    The Einstein ring images for the mass source superposed along the $x$-axis, perpendicular to the line of sight of the observer. The first and second rows show the CF image $|\tilde{Q}_\text{CF}(\x_I)|$ in the SN and the QG spacetimes, respectively, while the third and fourth rows display the WPI image $|\tilde{Q}_\text{WPI}(\x_I)|$ in the SN and QG spacetimes, respectively. In the left, middle, and right columns, we show three different mass source configurations.
    }
    \label{fig:imagenonsymCFWPI_QG_SNtable}
\end{figure}

Next, we consider the case where the mass source is superposed along the $x$-axis, perpendicular to the line of sight of the observer.  
Fig.~\ref{fig:imagenonsymCFWPI_QG_SNtable} presents the CF and WPI images for the QG and SN spacetimes. 
The first and second rows show the CF image $|\tilde{Q}_\text{CF}(\x_I)|$ in the SN and the QG spacetimes, respectively, while the third and fourth rows display the WPI image $|\tilde{Q}_\text{WPI}(\x_I)|$ in the SN and QG spacetimes, respectively.
Three different configurations of the mass source are shown in the left, middle and the right columns: $\X_\text{L}=(-550,0,0),~\X_\text{R}=(150,0,0)$, $\X_\text{L}=(-450,0,0),~\X_\text{R}=(250,0,0)$ and $\X_\text{L}=(-350,0,0),~\X_\text{R}=(350,0,0)$. As we move from the left to right columns, the separation between the two localized states of the mass source, $X_\text{R}-X_\text{L}=700$, remains constant, while the average position $\langle \hat\X\rangle_\text{M}=(X_\text{L}+X_\text{R})/2$ shifts from left to right according to the perspective of the observer.
The key characteristics are similar to those in Fig.~\ref{fig:imageaxCFWPI_QG_SN_table}.
We observe a single image in the CF images, representing the ensemble-averaged spacetime for the SN model, whereas in the QG spacetime, we see the composition of two classical images, each corresponding to the mass source at $\X_\text{L}$ and $\X_\text{R}$. The intensity of the WPI images vanishes consistently for the SN spacetime, while a composite image of two classical images appears for the QG model, except in the right panel.
Let us discuss two main differences from Fig.~\ref{fig:imageaxCFWPI_QG_SN_table}. 
First, each panel now exhibits either double arcs or a composite of them, instead of the ring structure. In addition, the SN model displays a single deformed image of double arcs. This results from the mass source being superposed along the $x$-axis, breaking the axisymmetry of the spacetime. 
Second, regarding the WPI images in the QG model, the image intensity vanishes in the right panel, despite the mass source being in a quantum superposition at distinct positions. This is because, for this particular configuration, the mass sources are symmetrically positioned from the perspective of the observer, making it impossible for detector A to discern the which-path information regarding whether the mass source is on the left or right. Further details are explained in the main context, specifically in Fig.~\ref{fig:zeroWPI}.

\begin{figure}[H]
    \centering
    \includegraphics[width=1\linewidth]{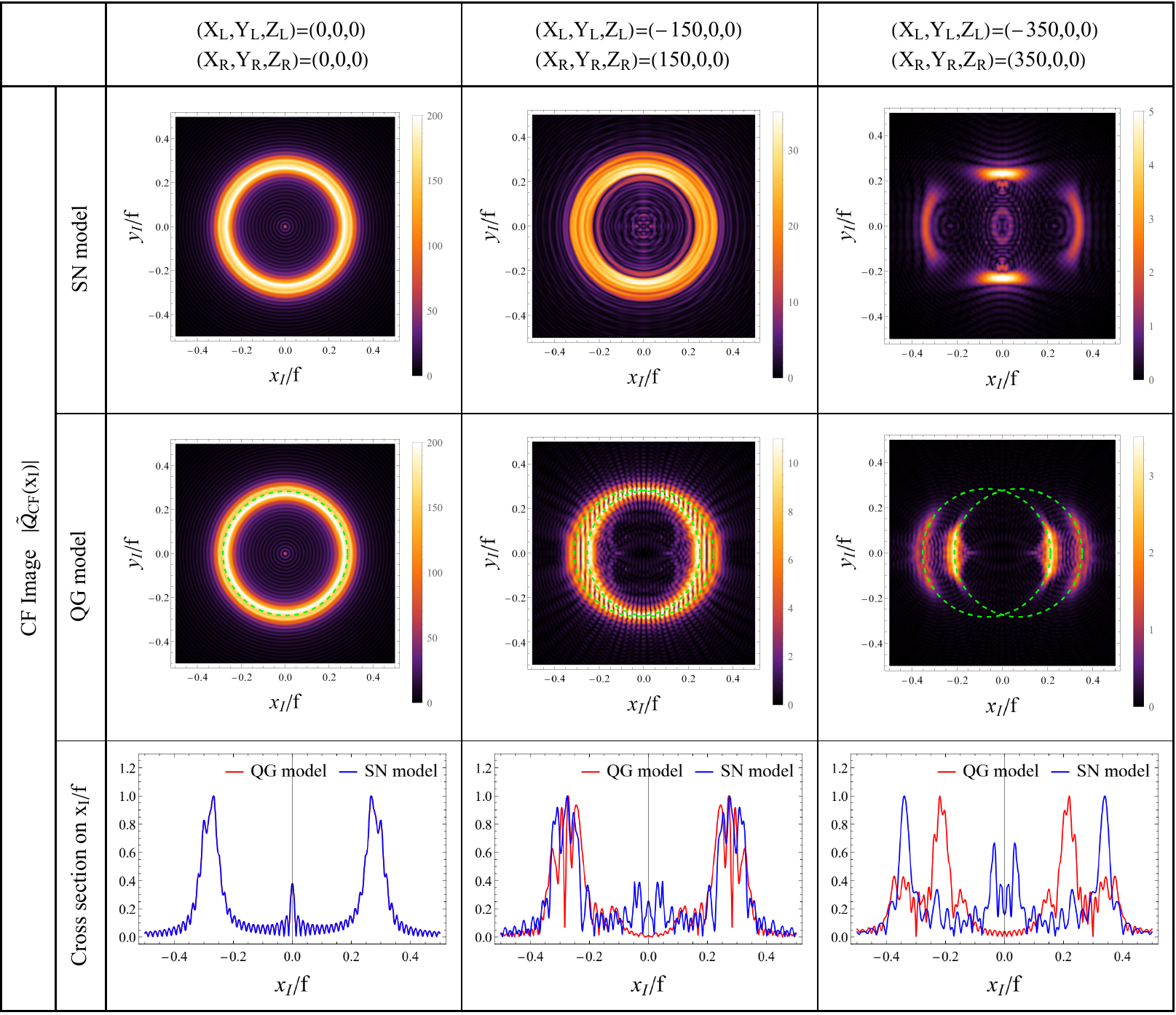}
    \caption{
    The Einstein ring images for the mass source that is superposed symmetrically along the $x$-axis from the perspective of the observer. The first and second rows show results for the SN and QG spacetimes, respectively. In the third row, we present the cross sections along the $x_I/f$ axis, where the SN model is plotted with a blue curve and the QG model with a red curve. 
    }
    \label{fig:imagesymCFWPI_QG_SN}
\end{figure}

Finally, we present the Einstein ring images for the case where the mass source is superposed symmetrically along the $x$-axis from the perspective of the observer; specifically $X_\text{L}=-X_\text{R}$, with $Y_\text{L}=Y_\text{R}=0$ and $Z_\text{L}=Z_\text{R}=0$.
In this parity symmetric configuration, detector A cannot acquire the which-path information regarding whether the mass source is on the left or right, causing the WPI images to consistently vanish, as described in Fig.~\ref{fig:zeroWPI}. Consequently, we focus only on the CF images, showing non-trivial behavior. 
In Fig.~\ref{fig:imagesymCFWPI_QG_SN}, we display the CF images $|\tilde{Q}_\text{CF}(\x_I)|$ for the SN and QG models. The first and second rows show results for the SN and QG spacetimes, respectively. In the third row, we present the cross sections along the $x_I/f$ axis, where the SN model is plotted with a blue curve and the QG model with a red curve. The vertical axis in each cross-sectional plot represents the CF image intensity $|\tilde{Q}_\text{CF}(\x_I)|$ normalized by its maximum value. 
The three columns, left, middle and right, illustrate different mass source configurations: $-\X_\text{L}=\X_\text{R}=(0,0,0)$, $-\X_\text{L}=\X_\text{R}=(150,0,0)$ and $-\X_\text{L}=\X_\text{R}=(350,0,0)$. From the left to right columns, the separation between the two localized states of the mass source, $X_\text{R}-X_\text{L}$, increases, while the average position $\langle \hat\X\rangle_\text{M}=(X_\text{L}+X_\text{R})/2=(0,0,0)$ remains fixed at the origin.

The figure reveals three notable features:
(i) For the SN model, a single deformed Einstein ring appears. In particular, the ring stretches horizontally and forms an elliptical shape in the right panel, where the two mass source states $|\bm{L}\rangle_\text{M}$ and $|\bm{R}\rangle_\text{M}$ are most separated. This deformation indicates that the light rays propagate through the ensemble-averaged spacetime of one classical configuration, reaching the observer after escaping from the mass source positioned at $|\bm{L}\rangle_\text{M}$ and $|\bm{R}\rangle_\text{M}$. 
(ii) For the QG model, the image forms a composition of two classical images resembling double arcs. In the middle panel, where the two mass source states $|\bm{L}\rangle_\text{M}$ and $|\bm{R}\rangle_\text{M}$ are moderately separated, the image shows two overlapping arcs and a vertical fringe pattern between them. 
This fringe pattern represents a beat effect resulting from the composing two images with a period given in Eq.~\eqref{eq:Einstein_ring_fringe}, each shifted by a distance of $\bm{O}_I|_{\X_\text{L}}$ and $\bm{O}_I|_{\X_\text{R}}$.
This image describes the interference between two propagations of light scattered by the mass source at $\X_\text{L}$ and $\X_\text{R}$, observable exclusively within the framework of wave optics, extending beyond the analysis of null geodesics.
In the right panel, where the two mass source states are well separated, the interference fringe largely vanishes. This suggests that the light scattered from $\X_\text{L}$ and $\X_\text{R}$ approaches nearly orthogonal, reducing the interference between them.
(iii) In the third row, we present the cross section of the normalized CF image for the SN and QG cases. Here, a central bright spot, or Poisson spot, consistently appears in the SN model but disappears in the QG model as the mass source achieves quantum superposition.
In wave optics, the Poisson spot results from the wave properties of light. The light waves arriving symmetrically from the left and right of the line of sight of the observer carry a common phase, which accumulates at the center of the screen, appearing as bright intensity.
Thus, the Poisson spot appears when the background spacetime maintains parity symmetry from the perspective of the observer. 
For the SN spacetime, the parity symmetry is always retained in the ensemble-averaged spacetime in the current case. By contrast, the QG model represents the quantum superposition of different spacetimes, which is not parity symmetric for now. This difference accounts for the Poisson spot observed only in the SN model. 
\newpage

%%%%%----------------------------------------------

\bibliographystyle{unsrt}
\bibliography{reference}

\end{document}